\documentclass[a4paper]{article} 
\usepackage{geometry}
\usepackage{tgpagella}
\usepackage{fontawesome5}
\usepackage[usenames,dvipsnames,svgnames,table]{xcolor}
\definecolor{mybg}{HTML}{282A36}
\definecolor{mycl}{HTML}{44475A}
\definecolor{myfg}{HTML}{F8F8F2}
\definecolor{mycomment}{HTML}{6272A4}
\definecolor{mycyan}{HTML}{8BE9FD}
\definecolor{mygreen}{HTML}{50FA7B}
\definecolor{myorange}{HTML}{FFB86C}
\definecolor{mypink}{HTML}{FF79C6}
\definecolor{mypurple}{HTML}{BD93F9}
\definecolor{myred}{HTML}{FF5555}
\definecolor{myyellow}{HTML}{F1FA8C}
\definecolor{VividPurple}{HTML}{3E0097}
\definecolor{SlateGrey}{HTML}{2E2E2E}
\definecolor{LightGrey}{HTML}{666666}
\definecolor{DarkPastelRed}{HTML}{450808}
\definecolor{PastelRed}{HTML}{8F0D0D}
\definecolor{GoldenEarth}{HTML}{E7D192}
\definecolor{awesome-emerald}{HTML}{00A388}
\definecolor{awesome-emerald-dark}{HTML}{00806A} 
\definecolor{awesome-skyblue}{HTML}{0395DE}
\definecolor{awesome-skyblue-dark}{HTML}{0376B0}
\definecolor{awesome-red}{HTML}{DC3522}
\definecolor{awesome-red-dark}{HTML}{B02A1C}
\definecolor{awesome-pink}{HTML}{EF4089}
\definecolor{awesome-pink-dark}{HTML}{EC136D}
\definecolor{awesome-orange}{HTML}{FF6138}
\definecolor{awesome-orange-dark}{HTML}{FF3300}
\definecolor{awesome-nephritis}{HTML}{27AE60}
\definecolor{awesome-nephritis-dark}{HTML}{219150}
\definecolor{awesome-concrete}{HTML}{95A5A6}
\definecolor{awesome-concrete-dark}{HTML}{74898B}
\definecolor{awesome-darknight}{HTML}{131A28}
\definecolor{awesome-darknight-dark}{HTML}{101623}
\definecolor{awesome-snowwhite}{HTML}{F9FBFD}
\definecolor{awesome-snowwhite-dark}{HTML}{F3F6FB}
\definecolor{awesome-blue-dark}{HTML}{0000FF}
\definecolor{awesome-golden}{HTML}{E1AD21}
\definecolor{awesome-silver}{HTML}{AAA9AD}
\definecolor{darktext}{HTML}{414141}
\definecolor{darktext-dark}{HTML}{262626}
\definecolor{text}{HTML}{333333}
\definecolor{graytext}{HTML}{5D5D5D}
\definecolor{lighttext}{HTML}{999999}
\definecolor{VividRed}{HTML}{7e2635}
\definecolor{DarkRed}{HTML}{a5402d}
\definecolor{SlateGrey}{HTML}{2E2E2E}
\definecolor{LightGrey}{HTML}{666666}
\UseRawInputEncoding
\usepackage[english]{babel}
\usepackage{lipsum}
\usepackage{amsmath, nccmath} 
\usepackage{gensymb}
\usepackage{amssymb}
\usepackage{graphics}
\usepackage{graphicx}
\usepackage{epsfig}
\usepackage{fancyhdr}
\usepackage{wasysym}
\usepackage{latexsym}
\usepackage{amsfonts}
\usepackage{float}
\usepackage{color,colortbl}
\usepackage{cite}
\usepackage{notoccite} 
\usepackage{soul}
\usepackage{multirow}
\usepackage{etoolbox}
\usepackage{authblk}
\usepackage{blindtext}
\usepackage{dcolumn}
\usepackage{bm}
\usepackage{qcircuit}
\usepackage{braket}
\usepackage{physics}
\usepackage{braket}
\usepackage{varwidth}
\usepackage{empheq}
\usepackage[]{algorithm2e}
						   
\usepackage{algorithmic}
\usepackage{framed}
\usepackage{enumitem}
\usepackage[utf8x]{inputenc} 
\usepackage[english]{babel}
\usepackage{url}
\usepackage{hyperref}
\usepackage[font={footnotesize}]{caption}
\usepackage{enumitem}
\usepackage{multirow}
\usepackage{textcomp}
\usepackage[T1]{fontenc}
\usepackage{listings}
\usepackage{subcaption}
\usepackage{textcomp}
\usepackage{mathrsfs}
\usepackage{caption}
\usepackage{xr}
\graphicspath{{./figures}}
\usepackage{xr-hyper}
\usepackage{hyperref}
\usepackage{cleveref}
\usepackage{amsthm}
\usepackage{enumerate}
\usepackage{array}
\usepackage[parfill]{parskip}
\usepackage{amscd}
\usepackage[]{units}
\usepackage{multicol}
\usepackage{tcolorbox}
\usepackage{media9}
\usepackage{pifont}
\usepackage{booktabs}
\usepackage[acronym]{glossaries}
\usepackage{alphalph}
\usepackage{mathtools}
\usepackage{blindtext}
\usepackage{subfiles}
\usepackage{fullpage}
\usepackage{makeidx}
\usepackage{fancyhdr}
\usepackage[calcwidth]{titlesec}
\usepackage[font={footnotesize},labelfont={bf}]{caption}
\usepackage{setspace}%
\providecommand{\U}[1]{\protect\rule{.1in}{.1in}}
\usepackage[symbol]{footmisc}
\usepackage{bookmark}
\usepackage[compat=1.1.0]{tikz-feynman}
\usepackage{tikz}
\usetikzlibrary{spy,calc}
\usetikzlibrary{matrix,backgrounds}
\usetikzlibrary{shapes.geometric, arrows}
\usetikzlibrary{decorations.pathmorphing}
\tikzstyle{startstop} = [rectangle, rounded corners, minimum width=3cm, minimum height=1cm,text centered, draw=black, fill=red!30]
\tikzstyle{io} = [trapezium, trapezium left angle=70, trapezium right angle=110, minimum width=3cm, minimum height=1cm, text centered, draw=black, fill=blue!30]
\tikzstyle{process} = [rectangle, minimum width=3cm, minimum height=1cm, text centered, text width=3cm, draw=black, fill=orange!30]
\tikzstyle{decision} = [diamond, minimum width=3cm, minimum height=1cm, text centered, draw=black, fill=green!30]
\tikzstyle{arrow} = [thick,->,>=stealth]

\tikzstyle{decision} = [diamond, draw, fill=blue!20, 
text width=4.5em, text badly centered, node distance=3cm, inner sep=0pt]
\tikzstyle{block} = [rectangle, draw, fill=blue!20, 
text width=5em, text centered, rounded corners, minimum height=4em]
\tikzstyle{line} = [draw, -latex']
\tikzstyle{cloud} = [draw, ellipse,fill=red!20, node distance=3cm,
minimum height=2em]		

\usetikzlibrary{shadows, patterns}
\tikzset{button/.style={
preaction={fill=blue,path fading=circle with fuzzy edge 20 percent,
opacity=.7,transform canvas={xshift=1mm,yshift=-1mm}},
preaction={pattern=#1,
path fading=circle with fuzzy edge 20 percent},
preaction={top color=white,
bottom color=red!50,
shading angle=180,
path fading=circle with fuzzy edge 20 percent,
opacity=0.4},
preaction={path fading=fuzzy ring 15 percent,
top color=black!5,
bottom color=black!80,
shading angle=180},
inner sep=2ex
},
button/.default=horizontal lines light blue,
circle}	
\newif\ifblackandwhitecycle
\gdef\patternnumber{0}

\pgfkeys{/tikz/.cd,
zoombox paths/.style={
draw=orange,
very thick
},
black and white/.is choice,
black and white/.default=static,
black and white/static/.style={ 
draw=white,   
zoombox paths/.append style={
draw=white,
postaction={
draw=black,
loosely dashed
}
}
},
black and white/static/.code={
\gdef\patternnumber{1}
},
black and white/cycle/.code={
\blackandwhitecycletrue
\gdef\patternnumber{1}
},
black and white pattern/.is choice,
black and white pattern/0/.style={},
black and white pattern/1/.style={    
draw=white,
postaction={
draw=black,
dash pattern=on 2pt off 2pt
}
},
black and white pattern/2/.style={    
draw=white,
postaction={
draw=black,
dash pattern=on 4pt off 4pt
}
},
black and white pattern/3/.style={    
draw=white,
postaction={
draw=black,
dash pattern=on 4pt off 4pt on 1pt off 4pt
}
},
black and white pattern/4/.style={    
draw=white,
postaction={
draw=black,
dash pattern=on 4pt off 2pt on 2 pt off 2pt on 2 pt off 2pt
}
},
zoomboxarray inner gap/.initial=5pt,
zoomboxarray columns/.initial=2,
zoomboxarray rows/.initial=2,
subfigurename/.initial={},
figurename/.initial={zoombox},
zoomboxarray/.style={
execute at begin picture={
\begin{scope}[
spy using outlines={%
zoombox paths,
width=\imagewidth / \pgfkeysvalueof{/tikz/zoomboxarray columns} - (\pgfkeysvalueof{/tikz/zoomboxarray columns} - 1) / \pgfkeysvalueof{/tikz/zoomboxarray columns} * \pgfkeysvalueof{/tikz/zoomboxarray inner gap} -\pgflinewidth,
height=\imageheight / \pgfkeysvalueof{/tikz/zoomboxarray rows} - (\pgfkeysvalueof{/tikz/zoomboxarray rows} - 1) / \pgfkeysvalueof{/tikz/zoomboxarray rows} * \pgfkeysvalueof{/tikz/zoomboxarray inner gap}-\pgflinewidth,
magnification=3,
every spy on node/.style={
zoombox paths
},
every spy in node/.style={
zoombox paths
}
}
]
},
execute at end picture={
\end{scope}
\node at (image.north) [anchor=north,inner sep=0pt] {\subcaptionbox{\label{\pgfkeysvalueof{/tikz/figurename}-image}}{\phantomimage}};
\node at (zoomboxes container.north) [anchor=north,inner sep=0pt] {\subcaptionbox{\label{\pgfkeysvalueof{/tikz/figurename}-zoom}}{\phantomimage}};
\gdef\patternnumber{0}
},
spymargin/.initial=0.5em,
zoomboxes xshift/.initial=1,
zoomboxes right/.code=\pgfkeys{/tikz/zoomboxes xshift=1},
zoomboxes left/.code=\pgfkeys{/tikz/zoomboxes xshift=-1},
zoomboxes yshift/.initial=0,
zoomboxes above/.code={
\pgfkeys{/tikz/zoomboxes yshift=1},
\pgfkeys{/tikz/zoomboxes xshift=0}
},
zoomboxes below/.code={
\pgfkeys{/tikz/zoomboxes yshift=-1},
\pgfkeys{/tikz/zoomboxes xshift=0}
},
caption margin/.initial=4ex,
},
adjust caption spacing/.code={},
image container/.style={
inner sep=0pt,
at=(image.north),
anchor=north,
adjust caption spacing
},
zoomboxes container/.style={
inner sep=0pt,
at=(image.north),
anchor=north,
name=zoomboxes container,
xshift=\pgfkeysvalueof{/tikz/zoomboxes xshift}*(\imagewidth+\pgfkeysvalueof{/tikz/spymargin}),
yshift=\pgfkeysvalueof{/tikz/zoomboxes yshift}*(\imageheight+\pgfkeysvalueof{/tikz/spymargin}+\pgfkeysvalueof{/tikz/caption margin}),
adjust caption spacing
},
calculate dimensions/.code={
\pgfpointdiff{\pgfpointanchor{image}{south west} }{\pgfpointanchor{image}{north east} }
\pgfgetlastxy{\imagewidth}{\imageheight}
\global\let\imagewidth=\imagewidth
\global\let\imageheight=\imageheight
\gdef\columncount{1}
\gdef\rowcount{1}

},
image node/.style={
inner sep=0pt,
name=image,
anchor=south west,
append after command={
[calculate dimensions]
node [image container,subfigurename=\pgfkeysvalueof{/tikz/figurename}-image] {\phantomimage}
node [zoomboxes container,subfigurename=\pgfkeysvalueof{/tikz/figurename}-zoom] {\phantomimage}
}
},
color code/.style={
zoombox paths/.append style={draw=#1}
},
connect zoomboxes/.style={
spy connection path={\draw[draw=none,zoombox paths] (tikzspyonnode) -- (tikzspyinnode);}
},
help grid code/.code={
\begin{scope}[
x={(image.south east)},
y={(image.north west)},
font=\footnotesize,
help lines,
overlay
]
\foreach \x in {0,1,...,9} { 
\draw(\x/10,0) -- (\x/10,1);
\node [anchor=north] at (\x/10,0) {0.\x};
}
\foreach \y in {0,1,...,9} {
\draw(0,\y/10) -- (1,\y/10);                        \node [anchor=east] at (0,\y/10) {0.\y};
}
\end{scope}    
},
help grid/.style={
append after command={
[help grid code]
}
},
}

\newcommand\phantomimage{%
\phantom{%
\rule{\imagewidth}{\imageheight}%
}%
}
\newcommand\zoombox[2][]{
\begin{scope}[zoombox paths]
\pgfmathsetmacro\xpos{
(\columncount-1)*(\imagewidth / \pgfkeysvalueof{/tikz/zoomboxarray columns} + \pgfkeysvalueof{/tikz/zoomboxarray inner gap} / \pgfkeysvalueof{/tikz/zoomboxarray columns} ) + \pgflinewidth
}
\pgfmathsetmacro\ypos{
(\rowcount-1)*( \imageheight / \pgfkeysvalueof{/tikz/zoomboxarray rows} + \pgfkeysvalueof{/tikz/zoomboxarray inner gap} / \pgfkeysvalueof{/tikz/zoomboxarray rows} ) + 0.5*\pgflinewidth
}
\edef\dospy{\noexpand\spy [
#1,
zoombox paths/.append style={
black and white pattern=\patternnumber
},
every spy on node/.append style={#1},
x=\imagewidth,
y=\imageheight
] on (#2) in node [anchor=north west] at ($(zoomboxes container.north west)+(\xpos pt,-\ypos pt)$);}
\dospy
\pgfmathtruncatemacro\pgfmathresult{ifthenelse(\columncount==\pgfkeysvalueof{/tikz/zoomboxarray columns},\rowcount+1,\rowcount)}
\global\let\rowcount=\pgfmathresult
\pgfmathtruncatemacro\pgfmathresult{ifthenelse(\columncount==\pgfkeysvalueof{/tikz/zoomboxarray columns},1,\columncount+1)}
\global\let\columncount=\pgfmathresult
\ifblackandwhitecycle
\pgfmathtruncatemacro{\newpatternnumber}{\patternnumber+1}
\global\edef\patternnumber{\newpatternnumber}
\fi
\end{scope}
}				
\definecolor{anti-flashwhite}{rgb}{0.95, 0.95, 0.96}
\definecolor{codegreen}{rgb}{0,0.6,0}
\definecolor{codepurple}{rgb}{0.58,0,0.82}
\lstdefinelanguage{NeMO}{
keywords={},
ndkeywords={solver},
keywordstyle=\color{blue},
ndkeywordstyle=\color{codepurple},
commentstyle=\color{codegreen},
stringstyle=\color{cyan},
sensitive=true
}
\hypersetup{colorlinks=true,
linktocpage=true,
colorlinks=true,
linkcolor=awesome-blue-dark,
filecolor=awesome-red,
citecolor=awesome-pink-dark,      
urlcolor=awesome-orange-dark,
pdftitle={Multi-scale Incoherent Electronic Transport Properties in Non-ideal CVD Graphene Devices},
bookmarks=true,
pdfauthor={Bhupesh~Bishnoi},
bookmarksopen=true,
pdfpagemode=UseOutlines,
pdfstartview={FitH},	
unicode   =true,
pdftoolbar=true,
pdfmenubar=true,
}
\makeatletter
\def\@fnsymbol#1{\ensuremath{\ifcase#1\or \pmb\ddagger \else\@ctrerr\fi}}
\makeatother
\title {\Huge {Multi-scale Incoherent Electronic Transport Properties in Non-ideal CVD Graphene Devices}}
\author{\bf{Bhupesh~Bishnoi\thanks{\href{mailto:bishnoi@computer.org}{\textcolor{PastelRed}{\texttt{\bf{bishnoi[At]computer[Dot]org}}}}~~~\href{mailto:bishnoi@ieee.org}{\textcolor{PastelRed}{\texttt{\bf{bishnoi[At]ieee[Dot]org}}}} }}}
\date{}

\begin{document}

\maketitle

\begin{abstract} 
In this research work, roll-to-roll chemical vapor deposited graphene device electronic transport properties are benchmarked to elucidate and comprehend mobility degradation in the real-world commercial application of graphene devices. Multifarious device design morphology in the graphene and two-dimensional material with diverse background materials compositions and processing recipes incorporate various scattering sources in the devices. However, the reported literature primarily discusses the best-manufactured device characteristics in careful laboratory experiments. Furthermore, most numerical calculate current in terms of coherent transport limit or the semi-classical Boltzmann treatment, with scattering mechanisms left residual for further treatment. Nevertheless, graphene being a two-dimensional sheet, a surface material with a high aspect ratio, the carrier goes through various scattering mechanisms. Therefore, to understand the nature of mobility degradation in roll-to-roll chemical vapor deposited graphene production devices, we employed multi-scale, multi-physics bottom-up, non-equilibrium Green's function-based quantum transport formalism. In this framework, we numerically incorporate various scattering mechanisms to deduce the measurand mobility at the last stage of computation to observe various scattering potential impacts on the production device performance. We have analyzed the variation in transmission, electronic charge density, electrostatic Poisson potential, energy-resolved flux density, and current-voltage characteristics and inferred the Drude mobility with different scattering potentials in various graphene devices. These scattering mechanisms treat scattering potentials as the first-order phonon Dyson self-energy term in the third loop of the two-looped self-consistent Poisson-Non-equilibrium Green's function iteration. Furthermore, multiple scattering scenarios implemented through a generalized contact self-energy scattering calculation ascribe the effect of contact scattering to the graphene device in quasi-ballistic transport limit. In this scheme, the effect of all the physical scattering mechanisms is lumped into one energy uncertainty or scattering rate parameter to include in the device's contact self-energy interaction term bound by the upper limit of Heisenberg uncertainty for the interacting quantum charged particles.
\end{abstract}

\section*{Introduction}
\label{section:Introduction}

In 2004 single-layer graphene was isolated in the laboratory with a promising adoption and application in the various fields of modern life. \cite{novoselov_electric_2004} To scale up the large-scale commercial production, the synthesis of graphene on metal catalysts by chemical vapor deposition (CVD) and plasma-enhanced chemical vapor deposition (PECVD) has been the most scalable process until now. The CVD/PECVD graphene synthesis process includes various growth kinetic chemistry steps. The adsorption of gas feedstock molecules, dehydrogenation of gas feedstock on metal catalysts, molecules surface diffusion, nucleation of the carbon atom, graphene flakes island growth, and island coalescence process to form a continuous sheet. In the overall graphene fabrication process, during the growth and transfer steps of the sheet, there can be polymer contamination on the transfer contact, wrinkles, tears damage, and folding of the sheet due to difference in thermal expansion during the cooling stage in the CVD/PECVD growth, incomplete transfer gaps, holes, and cracks. These nanoscale atomistic effects during the growth and transfer steps influence the sheet's background carrier density, carrier mobility, and sheet resistivity.

The most widely used method for electrical characterization of the sheet after the lithography step is the four-probe method and hall effect measurements, which are slow and one-time processes and permanently damage the graphene sheet. Extensive area high-density electrical mapping is essential for the roll-to-roll fabrication method's process optimization and quality control. Defect density is an essential measure of homogeneous and uniform graphene sheets. However, to quantify the effect, we have to consider both average defect density and distribution of defects in the graphene. \cite{cancado_quantifying_2011} The impact of microscopic non-uniformity on the output measurable electrical characteristic depends on the scale and size of non-uniformity in the devices. For example, a micrometer-scale device's microscopic non-uniformity of five hundred nanometers (nm) grain size gives an irregular current path. At the grain boundaries, current flow will see the high resistive path. However, a five hundred nanometers grain size appears to be homogenous for a millimeter-scale device, and many resistive current paths will be statistically averaged out. \cite{cummings_charge_2014} Unless the graphene sheet sample is very uniform in the distribution of doping variation, there is a significant deviation in estimating average carrier mobility. 
However, the four-terminal measurement method takes care of the geometrical aspect-ratio correction, If the sample size is an order of magnitude larger than probe point spacing. Nevertheless, for a microscopic non-uniform sample, the four-terminal measurement method has reliability issues. The dual configuration van der Pauw approach significantly reduces these errors in the mobility measurement for a carefully designed sample. \cite{pauw_method_1958} Graphene sheet resistance $R_{{{S}}}$ measured in terms of square resistance $R_{\square}$. Square resistance is the resistance of the device corrected for its geometrical shape and contact resistance value. Sheet conductance and sheet resistance are measured quantity of the devices.
In contrast, sheet conductivity and resistivity are material's inherent properties determined from the measured quantity with the inevitable spatial distribution error. Because, unlike metal in graphene, there is significant spatial variation in charge distribution and homogeneity. \cite{van_der_pauw_method_1958} Due to these reasons, there is no explicitly straightforward relationship between the measured device conductance and the inherent material conductivity of graphene. Also, graphene is a single atomic sheet; the bottom substrate, top oxide, passivation layers, and buildup contact potential influence the measured electrical properties. Further complexities arise in measurement due to hysteresis and time-dependent doping effect in the conductance measurement due to the realignment of polar water, gaseous molecules, and slow charge trap in the oxide near the graphene sheet surface. \cite{wang_hysteresis_2010} The sheet's quantitive values of electrical properties also depend upon the characterization scheme and their operating principle. For the contact-based measurement of electronic mobility, the Hall bar device, Van der Pauw device, and the top or bottom gate field-effect devices are employed in experiments. Each of these devices is structured by their operating principle and, by definition, gives different mobility concepts such as hall mobility, drift mobility, and field-effect mobility.
Furthermore, comparing these different mobility results is often not straightforward in the single atomic layer of graphene. Such as, in the top or bottom gate field-effect devices configuration, a direct current (DC) bias sweep on the gate electrode can change the Fermi energy level from the hole branch to the electron branch conduction in the graphene. We discussed the complete characterization of the graphene DC conductivity measurement formulation procedure for the roll-to-roll sample preparation section.

There are broadly three techniques to evaluate the quality of graphene sheets. The first method is spectrally resolved optical microscopy and inspection, revealing the extent of damage, contamination, wrinkle, coverage, and layer numbers in the sheet. \cite{pizzocchero_non-destructive_2015} The second widely used method is Raman spectroscopy, which provides extensive microscopic information about the quality of the graphene sheet. The Raman method, particularly $\mathrm{G}$-peak and narrow $\mathrm{2D}$-peak, gives extensive information about the quality of graphene sheets. A high ratio of $\mathrm{I_{2D}/I_{G}}$ and the absence of $\mathrm{D}$-peak indicate a low number of defects in the sheet, high uniformity, and low coupling to the substrate. All of these effects give rise to higher carrier mobility. \cite{suk_transfer_2011,nolen_high-throughput_2011} Graphene shows two characteristics peaks one the $\mathrm{2D}$-peak at 2675 cm${}^{-1}$ and another one the $\mathrm{G}$-peak at 1580 cm${}^{-}$${}^{1}$ in the Raman spectrum. However, due to the structural defect in the graphene sheet, there is a third $\mathrm{D}$-peak at 1350 cm${}^{-1}$ becomes visible, indicating the presence of $\mathrm{sp^3}$ bonds. Also, $\mathrm{I_{D}/I_{G}}$ intensity ratio is used to quantify the disorder in the graphene, which is related to the average distance between the defects. \cite{lucchese_quantifying_2010,rasappa_high_2015} A higher disorder in the system is represented by higher intensity $\mathrm{I_{D}/I_{G}}$ ratio and correlated with lower carrier mobility in the sample. \cite{ferrari_raman_2013} The full width at half maximum of the $\mathrm{2D}$-peak $({\mathrm{\Gamma_{2D}}})$, represent the strain variation in the high-quality graphene sheet such as exfoliated graphene on silicon dioxide $(\mathrm{Si}{\mathrm{O}}_2)$ substrate or epitaxial graphene on silicon carbide $(\mathrm{SiC})$ or chemical vapor deposited graphene on hexagonal boron nitride $(h\mathrm{-}\mathrm{BN})$. The minor deformation in the lattice correlated to the decrease in $({\mathrm{\Gamma_{2D}}})$, and hence enhanced the carrier mobility in the sheet. \cite{banszerus_ballistic_2016} Charge impurity scattering limited carrier mobility indicated by peak intensity $\mathrm{I_{2D}/I_{G}}$ ratio, because $\mathrm{G}$-peak and $\mathrm{2D}$-peak both are sensitive to doping level and doping type. \cite{casiraghi_raman_2007,newaz_probing_2012,das_monitoring_2008} The Third method for the quality assessment of the graphene sheet is an electrical measurement of the isolated devices by the lithography and hall bar test structure. The electrical characterization method will provide carrier density, sheet resistance, and carrier mobility related to the nanoscale atomistic quality of the sheet and the dominant origin of scattering effects in the devices. \cite{buron_graphene_2012,tan_measurement_2007} However, the definition of quality and norms of comparing the quality vary considerably in the different methods based on their operating principle, as discussed above briefly.  

There are four different, widely used methods for mapping the electrical properties of graphene. The first one is the fixed contact conventional lithography-based automated probe station. We can map the electrical conductivity of graphene sheet-based devices from the length scale of one micrometer to one-centimeter device length. The second method is the fixed contact dry laser lithography scheme, which has a resolution of the transport length scale of one-tenth millimeter to one hundred micrometers. \cite{mackenzie_fabrication_2015} The third method is a moveable contact micro-four point probe, where the scale of the probed region is one micrometer to one hundred micrometers. \cite{buron_electrically_2014,thorsteinsson_accurate_2009} The fourth and most advanced method is the non-contact terahertz time-domain spectroscopy method, where the length scale of resolution is ten to one hundred nanometers. \cite{buron_terahertz_2015}
The disadvantages of the first three contact-based probing schemes are that device fabrication process-chemistry influences the sheet's intrinsic electrical quality, the long turnaround time to make test devices, and the destructive nature of lithography contact, which permanently damages and prevents future use of graphene sheets. In addition, though graphene shows very high mechanical strength due to the atomic thinness of the graphene sheet, it is very fragile, and physical contact will damage the sheet permanently for reuse application.

In the previous article from the author, we numerically probed the above-mentioned microscopic variation and non-idealities in the macroscopic CVD graphene device in the coherent transport regime. The multi-scale non-equilibrium Green's function (NEGF) framework details its incoherent extension, implementation,  approximation, and bottom-up atomistic tight-binding (TB) models to treat these non-ideal CVD graphene devices discussed in the aforementioned article. In this article, we will incorporate the effect of each scattering mechanism in a multi-scale simulation framework, critical assumptions, and their combined effect as self-energy incorporating explicitly in the non-equilibrium Green's function loop as self-consistent Dyson's equation, and numerical truncation for the two-dimensional graphene sheet devices. The article's \hyperref[section:Graphene DC conductivity measurement]{graphene DC conductivity measurement section}  describes the graphene DC conductivity measurement formulation procedure for the roll-to-roll sample preparation. Next, we will discuss the \hyperref[section:Graphene tight-binding atomistic modeling]{Graphene atomistic tight-binding modeling} used in non-equilibrium Green's function formalism. The article's \hyperref[section:framework]{Multi-scale theoretical framework section} provides a detailed description of the theoretical framework. We have discussed the underline physical theory and complete evolutionary derivation of the theoretical framework of Non-equilibrium Green's function formalism. We will derive, evaluate, and in-depth discuss the multi-scale non-equilibrium Green's function formalism used to obtain subsequent section results. For the complete cumulative description of the theoretical framework, we encourage the readers to further deep dive into the relevant section of the article. Next, we will present the semi-classical treatment of the scattering rate through \hyperref[section:Fermi's Golden rule]{Fermi's Golden rule section} for the non-equilibrium Green's function formalism truncation procedure. Next, we present The multi-scale non-equilibrium Green's function formalism simulation \hyperref[section:implementation]{implementation section} on the supercomputer cluster. The following \hyperref[section:Result and Discussion]{result and discussion section} elucidate the variability in the measurand electrical properties for benchmarking purposes. We will discuss and recapitulate the main results from the multi-scale non-equilibrium Green's function formalism numerical simulation results and their physical interpretations for the graphene device's various aspects of variabilities in the device simulation result. Finally, in the last \hyperref[section:Summary]{summary section}, we conclude the article by summarizing the results, observations, and future work in the two-dimensional material field.

\section*{Graphene DC Conductivity Measurement}
\label{section:Graphene DC conductivity measurement}

The ultra-flat intrinsic graphene sheet is a zero-bandgap semiconductor, and at the edge of the Brillouin zone, the conduction and valence band coincide on the six Dirac points. Graphene lattices consist of two inequivalent triangular sublattices. Therefore, the six Dirac points are labeled as two sets of $K$ and $K^{'}$-points. Within the $\mathrm{1eV}$ of the Dirac points, the energy dispersion around the $K$ and $K^{'}$-points show linear dependence. The electrical current is mainly carried out in the graphene by this low-energy, mass-less quasi-particles, the Dirac fermions around the $K$ and $K^{'}$-points. The graphene sheet carrier mobility $\mu $, sheet carrier density $n$, and sheet conductivity ${\sigma }_s$ related by the Drude model is, \cite{drude_zur_1900,drude_zur_1900-1}

\begin{equation} \label{eq1} 
\sigma_{s}=qn\mu  
\end{equation}

From the semiclassical Boltzmann transport theory, sheet conductivity $\sigma_{s}$ of graphene is, \cite{sarma_electronic_2011,stauber_conductivity_2008,neto_electronic_2009} 

\begin{equation} \label{eq02}
\sigma_{s}=\frac{2q^{2}k_{f}l_{mfp}}{h}
\end{equation}

Where the Fermi wave-number $k_{f}=\sqrt{(\pi n)}$, The elastic mean free path $l_{mfp}=\upsilon_{f}\tau$ for the momentum relaxation time $\tau$. Fermi velocity $\upsilon_{f}\approx 10^{6}ms^{-1}$ in the graphene sheet. The elastic mean free path is the smallest length scale in the material where we can define the resistivity, and below this length scale, transport is non-local and ballistic. Also, DC Drude's conductivity represents as,

\begin{equation} \label{eq03}
\sigma_{s}=\frac{nq^{2}\tau}{m^*}
\end{equation}

Putting the effective mass of Dirac Fermions in the \cref{eq03}, \cite{sarma_electronic_2011,stauber_conductivity_2008,neto_electronic_2009}

\begin{equation} \label{eq04} 
m^*=\frac{\hbar k_{f}}{\upsilon_{f}}
\end{equation}

The DC Drude Conductivity for graphene is,

\begin{equation} \label{eq05}
\sigma_{s}=\frac{nq^{2}\tau\upsilon_{f}}{\hbar k_{f}}=\ \frac{nq^{2}l_{mfp}}{\hbar k_{f}}=\frac{2q^{2}(\pi n)l_{mfp}}{hk_{f}}=\frac{2q^{2}k_{f}l_{mfp}}{h} 
\end{equation}

And, the carrier mobility $\mu$ is,

\begin{equation} \label{eq06}
\mu=\frac{\sigma_{s}}{nq}=\frac{2q\upsilon_{f}\tau}{h}\sqrt{\frac{\pi}{n}}=\frac{ql_{mfp}}{\hbar \sqrt{(\pi n)}} 
\end{equation}

In a rectangular or quadratic graphene device geometry, the field-effect measurements and the hall measurement determine the residual carrier density $n$, field-effect carrier mobility$ \mu $. The carrier density $n$ varied in the field-effect measurement devices by applying the $V_G$ gate voltage as follows,

\begin{equation} \label{eq07}
n=\frac{C_{OX}V_{G}}{q} 
\end{equation}

The gate electrode and graphene sheet separated by thickness $t$, of the dielectric permittivity of $\epsilon_{r}$, form the parallel plate capacitor $C_{OX}$, of capacitance per unit area,

\begin{equation} \label{eq08}
C_{OX}=\frac{\epsilon_{o}\epsilon_{r}}{t}
\end{equation}

This is valid only if graphene and gate-electrode overlap capacitance area's linear dimensions are significantly larger than the spacing $t$ between them, and the effect of electric fringe fields is negligible. \cite{van_der_pauw_method_1958,pauw_method_1958,swartzendruber_four-point_1964,miccoli_100th_2015} Also, as graphene is a two-dimensional material and has a low state density, a very thin dielectric layer can contribute to the quantum of the capacitance effect. However, for the thicker dielectric layer in the range of 90 nm to 300 nm, quantum capacitance is negligible and can omit out. \cite{lee_is_2014, giannazzo_screening_2009} For the electrically homogenous device sample, from the measured conductance slope value as a function of gate voltage, the field-effect mobility is, \cite{zhong_comparison_2015,venugopal_effective_2011} 

\begin{equation} \label{eq09}
\mu \approx \frac{1}{C_{OX}}\frac{L}{W}\frac{\Delta G}{\Delta V_G} 
\end{equation}

To accurately estimate the graphene mobility values from the conductance curve, it is necessary to exclude the metal contact resistance, comparable to the channel resistance of the field-effect device. The total measured resistance $R_{tot}=\frac{V_{DS}}{I_{DS}},$ for a channel length $L$ and a channel width $\mathrm{W}$ is,

\begin{equation} \label{eq10}
R_{tot}=R_{contact}+R_{channel}=R_{contact}+\frac{L}{Wnq\mu} 
\end{equation}

Where $V_{DS}$ is the applied voltage between the drain and source terminal and  $I_{DS}$ is the measured current between the drain and source terminal of the graphene field-effect device. And the total carrier density $n$ in the graphene field-effect device is, \cite{adam_self-consistent_2007}

\begin{equation} \label{eq11}
n=\sqrt {n_{i}^{2}+n_{g}^{2}} 
\end{equation}

Where $n_i$ is the residual charge density in the sheet, and $n_g$ is the back gate that induces charge density in the device. The back gate induces charge density $n_g$ is,

\begin{equation} \label{eq12}
n_g=\frac{\epsilon_{o}\epsilon_{r}(V_{G}-\Delta V_{dirac})}{qt}
\end{equation}

The gate voltage value corresponding to the $G_{min}$ minimum conductance is the gate voltage offset $\Delta V_{dirac}$ can deduce the residual charge density $n_{i}$ as, \cite{gon_lee_influence_2013}

\begin{equation} \label{eq13}
n_{i}=\frac{\epsilon_{o}\epsilon_{r}\Delta V_{dirac}}{qt}
\end{equation}

The positive voltage shift offset in the current-voltage curve indicates the presence of positive charge impurity on the graphene surface. The local density of charge impurity changes from point to point on the graphene surface. The charge impurity density strongly affects the electron mean free path $l_{mfp}$ on the low dielectric substrate material where the substrate permittivity does not adequately screen out. Moreover, the charge neutrality point in the transfer characteristic exhibits an asymmetry between electron and hole branches in the conductance curve. Due to the charge transfer between the graphene and metal contact interface, a potential barrier developed, and hence conductance in the electron branch is mostly suppressed. The carrier density $n$ in the graphene expressed in Fermi energy $E_{F}$ at a temperature $T$ as, \cite{neto_electronic_2009}

\begin{equation} \label{eq14}
n=\frac{2{(k_{B}T)}^2}{\pi {\hbar v_f}^2}\Gamma(2)\bigg[{F}_1\Big(\frac{E_F}{k_{B}T}\Big)-{F}_1\Big(-\frac{E_F}{k_{B}T}\Big)\bigg]
\end{equation}

Where $\Gamma(N)$ is the Gamma function of order $(N)$, $F_N\big(\frac{E_F}{k_{B}T}\big)$ is Fermi-Dirac integral of order $(N)$ in energy, $k_B$ is the Boltzmann constant, and $v_f$ is the Fermi velocity in the graphene. For the $T=0 K$ temperature approximation the Fermi energy $E_F$ is, \cite{das_monitoring_2008,pisana_breakdown_2007}

\begin{equation} \label{eq15}
E_F=sgn(n)\hbar v_f\sqrt{\pi|n|}
\end{equation}

Where $sgn(n)$ is the sign function of $(n)$. The zero-temperature approximation for calculating the Fermi energy level gives the relatively higher energy level at room temperature operation $T=300K$ because the Fermi-Dirac distribution is relatively broadened. However, this difference is small; hence, this approximation holds valid at room temperature. \cite{wan_enhanced_2013} Next, we will discuss the graphene atomistic tight-binding model used in non-equilibrium Green's function formalism. 

\section*{Graphene Tight-binding Atomistic Modeling}
\label{section:Graphene tight-binding atomistic modeling}

At the nano-scale atomic composition, crystal symmetry and the spatial disorder affect the material's bulk properties as quantum mechanics effects come into the picture and modify the electronic-phononic structure. Therefore, atomistic simulations are more appropriate to model the quantum device's electronic properties within the range of a few meV in the entire Brillouin zone. \cite{vogl_semi-empirical_1983} The electronic band structure derives from the tight-binding method. The method is similar to a linear combination of atomic orbitals (LCAO) used to construct molecular orbitals. \cite{slater_simplified_1954,podolskiy_compact_2004} However, in the tight-binding approximation, electron-electron interactions of the orbital are neglected, but it gives an excellent approximation to the electronic band structure. Furthermore, for more rigorous treatment, the Hubbard model was employed. Graphene is a two-dimensional sheet of carbon atoms arranged in a hexagonal lattice. $2\mathrm{S}, 2\mathrm{P_{x}}$, and $2\mathrm{P_{y}}$ orbitals of carbon atom in graphene are $\mathrm{S}\mathrm{P}^{2}$ hybridize, resulting in strong $\sigma$ bonds where each carbon atom is bonded to its three neighbors carbon atoms. The $\mathrm{P_{z}}$ or $\pi$ orbital of the carbon atoms defines the low-energy electronic bandstructure in graphene.

The primitive lattice vectors in graphene are,

\begin{equation}\label{eq67}
\begin{aligned}
\vec{a}_1 = \frac{\sqrt{3}a}{2} \widehat{x}+ \frac{a}{2}\widehat{y} \\
\vec{a}_2 = \frac{\sqrt{3}a}{2} \widehat{x}- \frac{a}{2}\widehat{y} 
\end{aligned}
\end{equation} 

The graphene primitive unit cell comprises two carbon atoms, and each atom has one valence $ 2\mathrm{P_{z}} $ orbital. At the C1 atoms site valence $ \phi_{\text{2p}_{z1}} $ orbital is centered and at the C2 atoms site valence $ \phi_{\text{2p}_{z2}} $ orbital is centered, The tight-binding wavefunction for graphene is,

\begin{equation}\label{eq68}
\begin{aligned}
\psi_{\vec{k}}(\vec{r})=\frac{1}{\sqrt{N}}\sum\limits_{h,j}e^{i(h\vec{k}\cdot\vec{a}_1 + j\vec{k}\cdot\vec{a}_2)} &\big[ c_1\phi_{\text{2p}_{z1}}(\vec{r}-h\vec{a}_1-j\vec{a}_2) + c_2 \phi_{\text{2p}_{z2}}(\vec{r}-h\vec{a}_1-j\vec{a}_2) \big] 
\end{aligned}
\end{equation}	

After multiplying from the left by both of $ \mathrm{P_{z}} $ orbitals to time-independent Schr\"{o}dinger equation and integrating  overall space, the dispersion relation is,

\begin{equation}\label{eq69}
\begin{aligned}
\langle\phi_{\text{2p}_{z1}}|\widehat{H}|\psi_{k}\rangle = E\langle\phi_{\text{2p}_{z1}}|\psi_{k}\rangle, \\ \langle\phi_{\text{2p}_{z2}}|\widehat{H}|\psi_{k}\rangle = E\langle\phi_{\text{2p}_{z2}}|\psi_{k}\rangle 
\end{aligned}
\end{equation}

In tight-binding approximation, on-site and nearest-neighbor matrix elements retain, and all the other terms are assumed to be too small enough to ignore the equation,

\begin{equation}\label{eq70}
\begin{aligned}
\epsilon c_1 -tc_2\bigg[1+e^{-i\vec{k}\cdot\vec{a_1}} +  e^{-i\vec{k}\cdot\vec{a_2}}\bigg] = Ec_1 \\
\epsilon c_2 -tc_1\bigg[1+e^{i\vec{k}\cdot\vec{a_1}} +  e^{i\vec{k}\cdot\vec{a_2}}\bigg] = Ec_2 
\end{aligned}
\end{equation}

Where,

\begin{equation*}
\begin{aligned}
\epsilon & = \big\langle\phi_{\text{2p}_{z1}}(\vec{r})|\widehat{H}|\phi_{\text{2p}_{z1}}(\vec{r})\big\rangle \;\; 
\mathrm{and} \;\; t & = - \big\langle\phi_{\text{2p}_{z1}}(\vec{r})|\widehat{H}|\phi_{\text{2p}_{z2}}(\vec{r})\big\rangle= - \bigg\langle\phi_{\text{2p}_{z1}}(\vec{r})|\widehat{H}|\phi_{\text{2p}_{z1}}(\vec{r}-\frac{a}{\sqrt{3}}\widehat{x})\bigg\rangle 
\end{aligned}
\end{equation*}

Graphene electronic band structure is defined in the tight-binding approximation while considering interactions up to third-nearest neighbors as, \cite{white_hidden_2007, gunlycke_tight-binding_2008}

\begin{equation}\label{eq-C1 }
\widehat{H}=\widehat{H}_{1}+\widehat{H}_{3}
\end{equation}

Where $\widehat{H}_{1}$ is the first nearest neighbor Hamiltonian and $\widehat{H}_{3}$ is the third nearest neighbor Hamiltonian. In the second quantization language, Hamiltonians are in the real space representation is expressed as creation and annihilation operator acting on the $\pi$ state on each carbon atom site as follow,

\begin{equation}\label{eq-C2 }
\begin{aligned}
\widehat{H}_{1}&=\sum_{\langle n,l\rangle}t_{n,l}\widehat{c}_{n}^{\dagger}\widehat{c}_{l}\\ \widehat{H}_{3}&=\sum_{\langle n,m\rangle}t_{n,m}\widehat{c}_{n}^{\dagger}\widehat{c}_{m}
\end{aligned}	
\end{equation}

Where $c^{\dagger}$ creation and  $c$ annihilation operators, and summation runs over entire  $n$ lattice point, and $l$ is first and $m$ is third nearest neighbor site of $n$ lattice point. First nearest-neighbor hopping parameter $t_{n,l}\approx-3.2\mathrm{e}\mathrm{V}$ and third nearest neighbor $t_{n,m}\approx-0.3\mathrm{e}\mathrm{V}$. In the device simulation, we have passivated the dangling bond at the device edge by hydrogen passivation treatment. \cite{boykin_accurate_2011} In the tight-binding framework dangling bonds at the surface or edge are passivated by primarily two numerical methods. In the first strategy, passivation atoms are implicitly incorporated without distinguishing passivation atom types and add a passivation potential to the dangling bonds' orbital energies. This method works well with a relatively large system, arbitrary crystal structures, and hybridization symmetries. Furthermore, with appropriate parameters, it is applied to any passivation scenario. \cite{lee_boundary_2004} The second method is explicit, including the passivation atoms and their coupling to the surface or edge atoms Hamiltonian matrix and limited to small molecules and systems. \cite{tan_tight-binding_2015} In this explicit treatment, ab-initio results for different passivation atoms fit targets. \cite{podolskiy_compact_2004} The unsaturated dangling bonds at the edge or surface will result in edge or surface states at the electronic band structure. These unwanted states iron out by coupling the hydrogen passivation atoms to the surface's or edge's unsaturated dangling bonds in the device. We have used the $ \mathrm {P-D} $ orbital tight-binding model, which represents the edge effects by explicitly including the passivated hydrogen in the Hamiltonian matrix. The carbon atom is represented by three $\mathrm{P_{z}} $, $ \mathrm{D_{yz}} $, and $ \mathrm{D_{zx}} $ orbitals. The simple single orbital $\mathrm{P_{z}}$ tight-binding model works well for the two-dimensional graphene sheet.\cite{boykin_accurate_2011} Furthermore, with appropriate parameters, it is applied to any passivation scenario. \cite{lee_boundary_2004} The second method is explicit, including the passivation atoms and their coupling to the surface or edge atoms Hamiltonian matrix and limited to small molecules and system. \cite{tan_tight-binding_2015} In the explicit treatment, ab-initio results for different passivation atoms fit targets. \cite{podolskiy_compact_2004} The unsaturated dangling bonds at the edge or surface will result in edge or surface states at the electronic band structure. These unwanted states were ironed out by coupling the hydrogen passivation atoms to the device's surface or edge's unsaturated dangling bonds. We have used the $ \mathrm {P-D} $ orbital tight-binding model, which represents the edge effects by explicitly including the passivated hydrogen in the Hamiltonian matrix. The carbon atom is represented by three $ \mathrm {P_z} $, $ \mathrm {D_{yz}} $, and $ \mathrm {D_{zx}} $ orbitals. The simple single orbital $ \mathrm {P_z} $ tight-binding model works well for the two-dimensional graphene sheet. \cite{boykin_accurate_2011} Another method to solve large-scale Atomistic devices is the Wannier function approach. The Wannier function's advantage is matching the higher energy band's band structure data with the density functional theory, which is sometimes overlooked by the tight-binding approach. \cite{marzari_maximally_2012} The wannier formalism of quantum mechanics involves representing the physical quantity in the phase space coordinates in the quasi-distribution function. Therefore, working with this approach for the transport problem is more intuitive. In the Wannier function approach, Ab-initio calculations are performed on a small homogeneous section of the actual device. We extract the hamiltonian and a basis set from the DFT calculation, which is the input of the Wannier function approach. In Wannier formalism, overlapping Bloch wave functions are appropriately chosen with a phase factor to localize them maximally. The objective is to get a low-rank hamiltonian and the basis set compared to the original DFT input. This intermediate step reduces the numerical load for the subsequent transport calculation on the device hamiltonian. The efficiency of the Wannier formalism relies upon the transferability of the parameterization. The critical property of the maximum localized Wannier function representation is that they are orthogonal basis like the empirical tight-binding used for subsequent transport calculation. However, the hamiltonian's non-vanishing elements and sparsity pattern are denser in the Wannier function than the tight-binding. Hence, the transport calculation is four to five-time numerically more expensive than the tight-binding approach, but they are still two to three times faster than the DFT calculation. However, the Wannier approach's parametrization transferability is as good as DFT, especially in low-dimensional material. The interlayer bonding is Vander wall, and electronic coupling between layers is less dense than the intralayer strong covalent bonding and much denser coupling hamiltonian. Therefore variability in the band structure data from the multilayer to a single-layer device is much less than tight-binding. Linear interpolation uses to bind the low-energy DFT data point more efficiently. \cite{stieger_winterface_2020} The range of localized Wannier functions depends upon the efficiency of the wannierization step. A range of 3 - 4 {\AA} angstrom or five to ten atoms is reasonable to estimate, but it also depends upon the relevant material layers and individual coupling strength. Next, we will discuss the complete evolutionary derivation of the theoretical framework of non-equilibrium Green's function formalism. 

\section*{Multi-scale Theoretical Framework}
\label{section:framework}

The non-equilibrium Green's function formalism was developed by Keldysh, Kadanoff, and Baym \textit{et al.} in their seminal work in 1960. \cite{baym_conservation_1961,keldysh_diagram_1964} Lake and Datta \textit{et al.} first demonstrated the adaption of non-equilibrium Green's function formalism to semiconductor devices in 1992 and later in 2002 by Wacker \textit{et al.} \cite{lake_nonequilibrium_1992,lake_single_1997,wacker_semiconductor_2002} In semiconductors, electrons, phonons, and spin constitute the many-body quantum fields. Their time evolution in thermodynamical equilibrium and non-equilibrium states investigates through non-equilibrium Green's function formalism. \cite{fanchenko_generalized_1983, danielewicz_quantum-I_1984,danielewicz_quantum-II_1984,lake_single_1997,gebauer_current_2004,burke_density_2005,frederiksen_inelastic_2007}
In the non-equilibrium Green's function formalism, The Schr\"{o}dinger-Poisson equation solved with the open boundary conditions under the non-equilibrium, Fermi contact potentials with the coupling to the contacts and energy dissipative scattering processes. \cite{landauer_spatial_1957,beenakker_quantum_1991,weinmann_quantum_1994,datta_electronic_1997,datta_nanoscale_2000,luisier_quantum_2006,anantram_modeling_2008,lundstrom_fundamentals_2009,kubis_quantum_2009,hirsbrunner_review_2019} 

\subsection*{Non-equilibrium Green's Function Formalism}

By this formalism, quantum mechanical effects such as quantum mechanical tunnelling, quantization of density of states, quantum mechanical transmission, reflections, and resonance states, discretization of energy levels due to spatial confinement, metal-induced gap states, edge states effects investigated. Furthermore, we can also incorporate various scattering mechanisms, such as electron-electron scattering electron-phonon scattering in the formalism. In the Schr\"{o}dinger representation of the quantum system, the ground state $|\Psi_{0}\rangle$ of an interacting device is determined by the time-independent $H$ Hamiltonian,

\begin{equation}\label{eq-B1}
H|\Psi_{0}\rangle=E|\Psi_{0}\rangle 
\end{equation}

Where Hamilton operator $H=H_{0}+V$ constitutes a non-interacting Hamiltonian $H_{0}$ part and $V$ an interacting perturbation part. $H_{0}$ solved exactly with its eigenvalues and eigenvectors as a solution. The solution of Schr\"{o}dinger equation gives the time dependence of Schr\"{o}dinger wave-functions $|\Psi_{S}(t)\rangle$. The interacting part contains all the many-body effects, e.g., phonon-carrier, ionized dopant, impurity atom, carrier-carrier interaction, interface roughness effects, and evaluated by the respective self-energy calculation. The interaction part $V$ is treated as a perturbation of system $H$ as both wave-functions and operators are time-dependent in the interaction picture. In the alternative Heisenberg representation, the interacting quantum mechanical system is solved using the operator's time-dependent property while the wave functions are now time-independent. The time-dependent Heisenberg $\widehat{\psi}_{H}^{\dagger}(x',t')$ and $\widehat{\psi}_{H}(x, t)$ evolve as per,

\begin{equation}\label{eq-B2}
\widehat{\psi}_{H}(x, t)=e^{iHt/\hbar}\widehat{\psi}(x)e^{-iHt/\hbar}
\end{equation}

Where $\widehat{\psi}(x)$ destroy a fermion at place $x$, In the second quantization language, hamiltonian in the real space representation presents a relatively intuitive picture,

\begin{equation}\label{eq-B3}
\widehat{\psi}(x)=\sum_{k}\phi_{k}(x)c_{k}
\end{equation}

Where $c_{k}^{\dagger}$ is the creation and $c_{k}$ is the annihilation operators of a fermion in orbital state $\phi_{k}$. Such a quantum interaction picture is described by one-particle Green's functions in an equilibrium system. The causal/time-ordered zero-temperature single-fermion Green's function is defined as, \cite{haug_quantum_2008,hirsbrunner_review_2019}

\begin{equation}\label{eq-B4}
G(x,t;x',t')=-\frac{i}{\hbar}\frac{\Big\langle\Psi_{0}\big|T^{c}\big\{\widehat{\psi}_{H}(x,t)\widehat{\psi}_{H}^{\dagger}(x',t')\big\}\big|\Psi_{0}\Big\rangle}{\Big\langle\Psi_{0}\big|\Psi_{0}\Big\rangle}
\end{equation}

Where a time-ordering operator $T^{c}$ moves the earlier time-argument to the right, and the resulting expression changes its sign whenever two fermion operators interchanged. By inserting \cref{eq-B2} into \cref{eq-B4}, the Green's functions physical interpretation is defined as, \cite{rammer_quantum_1986}

\begin{equation}\label{eq-B5}
G(x,t;x',t')=-\frac{i}{\hbar}\frac{\Big\langle\Psi_{0}\big|T^{c}\big\{e^{iHt/\hbar}\widehat{\psi}(x)e^{-iH(t-t')/\hbar}\widehat{\psi^\dagger}(x')e^{-iHt'/\hbar}\big\}\big|\Psi_{0}\Big\rangle}{\Big\langle\Psi_{0}\big|\Psi_{0}\Big\rangle}
\end{equation} 

The phenomenological explanation of the above equation interpreted as for the $t>t'$, The probability that a fermion created in the quantum system at time $t'$ and place $x'$ in real space moves to another time $t$ and another place $x$, is represented by Green's function $G(x,t;x',t')$. $G(x,t;x',t')$ is defined at initial zero-time, quantum system is in ground state $|\Psi_{0}\rangle$ after that to a time $t'$ system evolves with the factor $e^{-iHt/\hbar}$. At reaching the place $x'$, at that time $t'$, $\widehat{\psi}^{\dagger}(x')$ a fermion is created. Next, from time $t'$ to $t$, system continuously evolving with a factor $e^{-iH(t-t')/\hbar}$. After reaching at place $x$, fermion is destroyed by annihilation $\widehat{\psi}(x)$ and system return to $\langle\Psi_{0}|$ initial ground state by evolving $e^{iHt/\hbar}$. For the $t'>t$ contrary process, hold true. The ground state expectation value represents by a bracket $\langle\cdots\rangle$ at the zero temperature. At the finite non-zero temperature, the quantum system is no longer in the ground state. Therefore bracket $\langle\cdots\rangle$ represents the grand canonical ensemble of thermodynamic average. The quantum device is in contact with a reservoir with a temperature $\mathrm{T}$, and with the reservoir, the device might exchange heat and fermions. To represent the zero and finite temperature equilibrium and non-equilibrium quantum system interaction, real-time and imaginary Green's functions and advanced $G^{A}$ and retarded $G^{R}$ Green's functions are defined similarly. As well to completely describe the evolution of the system, two additional Green's functions, the greater $G^{>}$ and the lesser $G^{<}$ Green's functions are also established as, \cite{martin_theory_1959,baym_conservation_1961,kadanoff_theory_1961,keldysh_diagram_1964}

\begin{equation}\label{eq-B6}
\begin{aligned}
G^{R}(x,t;x^{\prime}, t^{\prime}) & = -\frac{i}{\hbar}\theta(t-t^{\prime})\Big\langle\big[\widehat{\psi}(x, t),\widehat{\psi}^{\dagger}(x^{\prime}, t^{\prime})\big]_{+}\Big\rangle \\
G^{A}(x, t;x^{\prime}, t^{\prime}) &= \frac{i}{\hbar}\theta(t^{\prime}-t)\Big\langle\big[\widehat{\psi}(x, t),\widehat{\psi}^{\dagger}(x^{\prime}, t^{\prime})\big]_{+}\Big\rangle	
\end{aligned}
\end{equation}
\begin{equation}\label{eq-B8}
\begin{aligned}
G^{<}(x, t;x',t') &=\frac{i}{\hbar}{\Big\langle\widehat{\psi}^{\dagger}(x',t')\widehat{\psi}(x,t)\Big\rangle} \\
G^{>} (x, t;x',t') &=-\frac{i}{\hbar}\Big\langle\widehat{\psi}(x, t)\widehat{\psi}^{\dagger}(x',t')\Big\rangle
\end{aligned}
\end{equation}

The lesser $G^{<}$, greater $G^{>}$, advanced $G^{A}$, retarded $G^{R}$, and $G$ Green's functions are not uniquely independent but interrelated by following relationships. \cite{martin_theory_1959,baym_conservation_1961,kadanoff_theory_1961,keldysh_diagram_1964}

\begin{equation}\label{eq-B10}
\begin{aligned}
G(x,t;x',t') & = \theta(t-t')G^{>}(x, t;x', t')+\theta(t'-t)G^{<}(x, t;x',t') \\
G^{R,A}(x,t;x', t') & = \pm\theta(\pm t\mp t')\big[G^{>}(x, t;x', t')-G^{<}(x, t;x', t')\big] \\
G^{R}(x, t;x', t') & -G^{A}(x, t;x', t') = G^{>}(x, t;x', t')-G^{<}(x, t;x', t')
\end{aligned}
\end{equation}

These equalities hold for both equilibrium and non-equilibrium pictures, though the fluctuation-dissipation theorem linked all these properties in equilibrium. One subtle difference between the equilibrium and non-equilibrium pictures is in the derivation of the perturbation assumption. In equilibrium, zero-temperature Green's functions scenario system is guaranteed to return its initial state after an asymptotically large time. However, In the non-equilibrium picture, this is not true, as at time $t$ equal to $+\infty$, the final state will be very distinct from the initial state at time $t$ equal to $-\infty$. Consequently, operator expectation values are built through the Feynman diagrams technique for contour integration, \cite{feynman_space-time_1948,schwinger_on_1951,mattuck_guide_1976} and by Wick's decomposition theorem \cite{wick_the_1950,binder_nonequilibrium_1995} and linked-graph theorem techniques for non-equilibrium situations. \cite{fanchenko_generalized_1983, fetter_quantum_1971,wagner_expansions_1991} The definition of non-equilibrium Green's function defined as,

\begin{equation}\label{eq-B13}
G(x, t;x', t') =  -\frac{i}{\hbar}\Big\langle T^{c}\big\{\widehat{\psi}_{\mathcal{H}}(x, t)\widehat{\psi}_{\mathcal{H}}^{\dagger}(x', t')\big\}\Big\rangle	
\end{equation}

Where now the field operators are expressed in the Heisenberg representation as $\widehat{\psi}_{\mathcal{H}}(x, t)$ and $\widehat{\psi}_{\mathcal{H}}^{\dagger}(x', t')$ and are correspond to the total Hamiltonian $\mathcal{H}(t)$. By using the 
\cref{eq-B8}, which holds for non-equilibrium situation, The non-equilibrium Green's function \cref{eq-B13} is,

\begin{equation}\label{eq-B14}
G(x, t;x^{\prime}, t^{\prime}) = \theta(t, t^{\prime})G^{>}(x, t;x^{\prime}, t^{\prime})+\theta(t^{\prime}, t)G^{<}(x, t;x^{\prime}, t^{\prime})	
\end{equation}

Where on the contour the definition of the function $\theta(t, t^{\prime})$ is,

\begin{equation}\label{eq-B15}
\theta(t,t^{\prime}) = \left\{\begin{array}{l}
0,\ \mathrm{if}\ t\ \mathrm{is}\ \mathrm{earlier}\ \mathrm{on}\ \mathrm{a}\ \mathrm{contour}\ \mathrm{than}\ t^{\prime} \\
1,\ \mathrm{if}\ t\ \mathrm{is}\ \mathrm{later}\ \mathrm{on}\ \mathrm{a}\ \mathrm{contour}\ \mathrm{than}\ t^{\prime}
\end{array}\right.	
\end{equation}

The system is defined by total Hamiltonian $\mathcal{H}(t)$,

\begin{equation}\label{eq-B16}
\mathcal{H}(t) = H+H^{ext}(t)=H_{0}+V+H^{ext}(t)	
\end{equation}

Where again non-interacting part of Hamiltonian is $H_{0}$, and the external perturbation $H^{ext}(t)$ tries to drive the system out of equilibrium, and carrier-carrier interactions and Poisson potential is contained in $V$. The external perturbation $H^{ext}(t)$ defined as,

\begin{equation}\label{eq-B17}
H^{ext}(t) = \int \mathrm{d}x\widehat{\psi}^{\dagger}(x)U(x, t)\widehat{\psi}(x)
\end{equation}

Where the external potential due to interactions is $U(x, t)$. When treating Green's functions in the device domain, changing the variable from time and real space basis to energy and momentum basis is convenient. The $G(x, t;x^{\prime}, t^{\prime})$ is the functions of the continuous space variables ($\boldsymbol{r}$) and ($\boldsymbol{r}^{\prime}$). The bold mathematical symbol is used throughout the work to represent vector position unless otherwise stated. Using the Fourier transformation, Green's functions represent the momentum and energy space. Furthermore, to solve for the real space finite element two-dimensional/three-dimensional devices by the Green's functions method, the mathematical equations are discretized on the grid of Green's functions $G_{nm}(tt^{\prime})$. For the $z$-directional current transport solution, the potential is homogeneous in the transverse $x$ and $y$ direction. The creation and annihilation operators $\widehat{\psi}^{\dagger}(x', t')$ and $\widehat{\psi}(x, t)$  from \cref{eq-B3} are expanded into a series of eigenfunctions as $G(x, t;x^{\prime}, t^{\prime})$, \cite{binder_nonequilibrium_1995,Schafer_semiconductor_2002}

\begin{equation}\label{eq-B18}
G(x, t;x^{\prime}, t^{\prime}) = \sum_{n_{1},m_{1}}\sum_{\boldsymbol{k'},\boldsymbol{k''}} \phi_{n_{1},\boldsymbol{k}^{\prime}}(\boldsymbol{r})\cdot G_{n_{1}m_{1}}(\boldsymbol{k}^{\prime}\boldsymbol{k}^{\prime\prime};tt^{\prime})\cdot\phi_{m_{1},\boldsymbol{k''}}^{*}(\boldsymbol{r}^{\prime})  
\end{equation}

The wave functions $\phi$ are factorized in the transport $z$ direction and homogeneous confined in $x$ and $y$ directions.

\begin{equation}\label{eq-B19}
\phi_{\boldsymbol{k}}(\boldsymbol{r}) = \frac{1}{\sqrt{A}}e^{i\boldsymbol{k}_{\mathrm{t}}\cdot \boldsymbol{r_t}}\phi_{\boldsymbol{k}_{\mathrm{t}}}(z) 
\end{equation}

Where xy-surface area is $A$, $\boldsymbol{r}_{\mathrm{t}}=(x, y)$  and $\boldsymbol{k}_{\mathrm{t}}=(k_{x}, k_{y})$. Due to $x$ and $y$ directions homogeneous condition, $G(x, t; x^{\prime}, t^{\prime})$ Green's function only depend upon difference $\boldsymbol{r}_{\mathrm{t}}-\boldsymbol{r}_{\mathrm{t}}^{\prime}$ and  $\boldsymbol{k}_{\mathrm{t}}^{\prime}=\boldsymbol{k}_{\mathrm{t}}^{\prime\prime}$. The discretized wave-functions $\phi$ are very localized functions. The atomic orbitals' wave function is tightly bound to the atoms and is usually assumed to be one lattice point in the grid. In the tight-binding approximation, this is a fundamental assumption. Hence lattice wave functions expanded on the basis of atomic orbitals. \cite{lake_single_1997,svizhenko_two-dimensional_2002} Therefore Green's function discretized as,

\begin{equation}\label{eq-B20}
G(x, t;x^{\prime}, t^{\prime}) = \frac{1}{A}\sum_{n_{1},m_{1}}\sum_{\boldsymbol{k}_{\mathrm{t}}^{\prime}}\phi_{n_{1},\boldsymbol{k}_{\mathrm{t}}^{\prime}}(z)\cdot G_{n_{1}m_{1}}(\boldsymbol{k}_{\mathrm{t}}^{\prime} ; tt^{\prime})\cdot\phi_{m_{1},\boldsymbol{k}_{\mathrm{t}}^{\prime}}^{*} (z^{\prime})\cdot e^{i\boldsymbol{k}_{\mathrm{t}}^{\prime}\cdot(\boldsymbol{r}_{\mathrm{t}}-\boldsymbol{r}_{\mathrm{t}}^{\prime})} 	
\end{equation}

Self-energies for electron-phonon and other interactions can also discretize by applying similar procedures. The time evolution of an interacting non-equilibrium quantum system is derived by solving the equations of motion of Green's functions $G(x, t;x^{\prime}, t^{\prime})$ from the time $t$ to the  $t^{\prime}$, for the evolution of non-equilibrium Green's function $\frac{\mathrm{d}}{\mathrm{d}t'}G(x, t;x^{\prime}, t^{\prime})$ and $\frac{\mathrm{d}}{\mathrm{d}t}G(x, t;x^{\prime}, t^{\prime})$ is derived with respect to $G(x, t;x^{\prime}, t^{\prime})$ as represented by \cref{eq-B13}, the equation of motion for $G(x, t;x^{\prime}, t^{\prime})$ at the time $t_{1}$ is,

\begin{equation}\label{eq-B21}
\begin{aligned}
\Bigg[i\hbar\frac{\mathrm{d}}{\mathrm{d}t_{1}}+\frac{\hbar^{2}\nabla^{2}(x_{1},t_{1})}{2m}-U(x_{1},t_{1})\Bigg]G(x_{1}, t_{1};x_{1}^{\prime}, t_{1}^{\prime})  &=\delta(t_{1},t_{1}')\delta(x_{1}-x_{1}') +\\& \int_{C} d(x_{3},t_{3}) \Sigma^{tot}(x_{1}, t_{1};x_{3}, t_{3})G(x_{3}, t_{3};x_{1}^{\prime}, t_{1}^{\prime})
\end{aligned}
\end{equation}

Where $ \delta(t,t') = \frac{\mathrm{d}}{\mathrm{dt}}\theta(t,t') $ and $ \Sigma^{tot} $ total self-energy. $ \Sigma^{tot} $ is derived by decomposition of 2-fermion Green's function into single fermion Green's function via variational derivation and Wick's decomposition theorem, and it corresponds to the Dyson equations. \cite{dyson_the_1949,schwinger_greens-I_1951,schwinger_greens-II_1951,hedin_new_1965} The computation of the self-energy matrices is based on the treatment of the Dyson equation. The solution of non-equilibrium equations of motion is achieved by dividing the contour integrals over a time-loop $\int_{C}$ into the time-ordered integrals. To solve these modified equations of motion, all types of $G^{R, A}$ and $G^{<,>}$ Green's functions are required. In the device simulation, the carrier densities distribution $\mathfrak{N}(x,t)$ and $\mathfrak{J}(x,t)$ the current densities are the two most important physical observable quantities and which can also be measured by various experimental technique as discussed in the previous section. By solving the equations of motion, lesser Green's functions $G^{<}(x,t;x^{\prime},t^{\prime})$ are calculated, which is only possible because Green's function $G(x_{1},t_{1};x_{2},t_{2})$ and the self-energy $\Sigma(x_{1},t_{1};x_{2},t_{2})$ has the same symmetry properties, and that is required for the calculation of carrier densities. Green's function $G(x_{1},t_{1};x_{2},t_{2})$ and the self-energy $\Sigma(x_{1},t_{1};x_{2},t_{2})$ symmetry properties evaluated by Craig ansatz et.al. \cite{craig_perturbation_1968} and proved by Danielewicz et.al.\cite{danielewicz_quantum_1984} for the formal solution of NEGF equations. By using the Langreth theorem, and expansion of eigenfunction, for ($t-t^{\prime}$) time difference by taking the Fourier transform of the Green's functions. \cite{haug_quantum_2008} The closed set of equations of motion is defined as,

\begin{equation}\label{eq-B22}
\begin{aligned}
{\it ih} \frac{\mathrm{d}}{\mathrm{d}t}G_{nm}^{<}(\boldsymbol{k}_{\mathrm{t}};tt^{\prime})-\sum_{l}h_{nl}G_{lm}^{<}(\boldsymbol{k}_{\mathrm{t}};tt^{\prime}) &= \sum_{l}\int_{t_{0}}^{\infty}\mathrm{d}t_{1}\Sigma_{nl}^{R}(\boldsymbol{k}_{\mathrm{t}};tt_{1})G_{lm}^{<}(\boldsymbol{k}_{\mathrm{t}};t_{1}t^{\prime}) \\&
+\sum_{l}\int_{t_{0}}^{\infty}\mathrm{d}t_{1}\Sigma_{nl}^{<}(\boldsymbol{k}_{\mathrm{t}};tt_{1})G_{lm}^{A}(\boldsymbol{k}_{\mathrm{t}};t_{1}t^{\prime})
\end{aligned}
\end{equation}
\begin{equation}\label{eq-B23}
\begin{aligned}
-ih\frac{\mathrm{d}}{\mathrm{d}t^{\prime}}G_{nm}^{<}(\boldsymbol{k}_{\mathrm{t}};tt^{\prime})-\sum_{l}G_{nl}^{<}(\boldsymbol{k}_{\mathrm{t}};tt^{\prime})h_{lm}\ & =\ \sum_{l}\int_{t_{0}}^{\infty}\mathrm{d}t_{1}G_{nl}^{R}(\boldsymbol{k}_{\mathrm{t}};tt_{1})\Sigma_{lm}^{<}(\boldsymbol{k}_{\mathrm{t}};t_{1}t^{\prime}) \\&
+\sum_{l}\int_{t_{0}}^{\infty}\mathrm{d}t_{1}G_{nl}^{<}(\boldsymbol{k}_{\mathrm{t}};tt_{1})\Sigma_{lm}^{A}(\boldsymbol{k}_{\mathrm{t}};t_{1}t^{\prime})\
\end{aligned}
\end{equation}

Where hamiltonian is hermitian $h_{ml}^{*}=h_{lm}$ and $h_{nm}=\int \mathrm{d}\boldsymbol{r}\phi_{n}^{*}(\boldsymbol{r})H_{0}(\boldsymbol{r})\phi_{m}(\boldsymbol{r})$. Retarded and advanced Green's function $(G^{R,A})$, as well as lesser self-energy $(\Sigma^{<})$, and retarded and advanced self-energy $(\Sigma^{R,A})$ is required to solve the close-set of equations of motion \cref{eq-B22} and \cref{eq-B23}. These Green's functions and self-energy are calculated by solving the equation of motion. Similarly, $R's$ replaced by $A's$, and by similarly procedure equations for $G^{A}$ are obtained.   

\begin{equation}\label{eq-B24}
{i\hbar} \frac{\mathrm{d}}{\mathrm{d}t}G_{nm}^{R}(\boldsymbol{k}_{\mathrm{t}};tt^{\prime})-\sum_{l}h_{nl}G_{lm}^{R}(\boldsymbol{k}_{\mathrm{t}};tt^{\prime}) = \delta_{nm}(tt^{\prime})+\sum_{l}\int_{t_{0}}^{\infty}\mathrm{d}t_{1}\Sigma_{nl}^{R}(\boldsymbol{k}_{\mathrm{t}};tt_{1})G_{lm}^{R}(\boldsymbol{k}_{\mathrm{t}};t_{1}t^{\prime})
\end{equation}

Where $ \delta_{nm}(tt^{\prime}) $ is defined as, 

\begin{equation}\label{eq-B25}
\delta_{nm}(tt^{\prime}) =
\sum_{l}\int_{t_{0}}^{\infty}\mathrm{d}t_{1}[G_{nl}^{R}(\boldsymbol{k}_{\mathrm{t}};tt_{1})]^{-1}G_{lm}^{R}(\boldsymbol{k}_{\mathrm{t}};t_{1}t^{\prime})
\end{equation}

After simplification by Langreth theorem, The lesser Green's function, the central equation of motion is with the coupling between $G^{<}$ and $G^{R,A}$ as,

\begin{equation}\label{eq-B26}
G_{nm}^{<}(\boldsymbol{k}_{\mathrm{t}};tt^{\prime}) =\ \sum_{l,v}\int_{t_{0}}^{\infty}\mathrm{d}t_{1}\int_{t_{0}}^{\infty}\mathrm{d}t_{2}G_{nl}^{R}(\boldsymbol{k}_{\mathrm{t}};tt_{1})\Sigma_{lv}^{<}(\boldsymbol{k}_{\mathrm{t}};t_{1}t_{2})G_{vm}^{A}(\boldsymbol{k}_{\mathrm{t}};t_{2}t^{\prime})
\end{equation}

Green's functions $G_{nm}(\boldsymbol{k}_{\mathrm{t}};tt^{\prime})$ usually depend upon time ($t$) and ($t^{\prime}$), However once non-equilibrium system reach to a stationary state solution the Green's functions depend on the time difference $(t-t^{\prime})$. As per the Langreth theorem, the system of evolution, ($t$) and ($t^{\prime}$) upon reaching the stationary state, no longer reside on the imaginary time contour. Furthermore, using the advantage of Fourier transform Green's functions for the time difference $(t-t^{\prime})$ are modified in the energy domain. Comparable relationship holds for $G^{R,A}$ also,

\begin{equation}\label{eq-B27}
G_{nm}^{<}(\boldsymbol{k}_{\mathrm{t}};E) = \int \mathrm{d}(t-t^{\prime})e^{iE(t-t^{\prime})/\hbar}G_{nm}^{<}(\boldsymbol{k}_{\mathrm{t}};t-t^{\prime}) 
\end{equation}

Consequently, finally, in the stationary state, equations of motion of the quantum system are simplified as follows in the coupled set of the equations to describe the NEGF formalism,

\begin{equation}\label{eq-B28}
\begin{aligned}
EG_{nm}^{<}(\boldsymbol{k}_{\mathrm{t}};E)-\sum_{l}h_{nl}G_{lm}^{<}(\boldsymbol{k}_{\mathrm{t}};E) & = \sum_{l}\Sigma_{nl}^{R}(\boldsymbol{k}_{\mathrm{t}};E)G_{lm}^{<}(\boldsymbol{k}_{\mathrm{t}};E)+\sum_{l}\Sigma_{nl}^{<}(\boldsymbol{k}_{\mathrm{t}};E)G_{lm}^{A}(\boldsymbol{k}_{\mathrm{t}};E) \\ 
EG_{nm}^{<}(\boldsymbol{k}_{\mathrm{t}};E)-\sum_{l}G_{nl}^{<}(\boldsymbol{k}_{\mathrm{t}};E)h_{lm} &  = \sum_{l}G_{nl}^{R}(\boldsymbol{k}_{\mathrm{t}};E)\Sigma_{lm}^{<}(\boldsymbol{k}_{\mathrm{t}};E)+\sum_{l}G_{nl}^{<}(\boldsymbol{k}_{\mathrm{t}};E)\Sigma_{{lm}}^{A}(\boldsymbol{k}_{\mathrm{t}};E) \\
G_{nm}^{<}(\boldsymbol{k}_{\mathrm{t}};E)\ & =\ \sum_{l,v}G_{nl}^{R}(\boldsymbol{k}_{\mathrm{t}};E)\Sigma_{lv}^{<}(\boldsymbol{k}_{\mathrm{t}};E)G_{vm}^{A}(\boldsymbol{k}_{\mathrm{t}};E) \\ 
G_{nm}^{<}(\boldsymbol{k}_{\mathrm{t}};E)\ & =\ -[G_{mn}^{<}(\boldsymbol{k}_{\mathrm{t}};E)]^{\dagger} \\
EG_{nm}^{R}(\boldsymbol{k}_{\mathrm{t}};E)-\sum_{l}h_{nl}G_{lm}^{R}(\boldsymbol{k}_{\mathrm{t}};E) & = \delta_{nm}+\sum_{l}\Sigma_{nl}^{R}(\boldsymbol{k}_{\mathrm{t}};E)G_{lm}^{R}(\boldsymbol{k}_{\mathrm{t}};E) \\ 
G_{nm}^{A}(\boldsymbol{k}_{\mathrm{t}};E)\ & =\ [G_{mn}^{R}(\boldsymbol{k}_{\mathrm{t}};E)]^{\dagger} \\ 
G_{mn}^{R}(\boldsymbol{k}_{\mathrm{t}};E)-G_{mn}^{A}(\boldsymbol{k}_{\mathrm{t}};E)\ & =\ G_{mn}^{>}(\boldsymbol{k}_{\mathrm{t}};E)-G_{mn}^{<}(\boldsymbol{k}_{\mathrm{t}};E) \\ 
A(\boldsymbol{k}_{\mathrm{t}};E) & = i\Big[G_{mn}^{R}(\boldsymbol{k}_{\mathrm{t}};E)-G_{mn}^{A}(\boldsymbol{k}_{\mathrm{t}};E)\Big]
\end{aligned}
\end{equation}

Where $G^{R}$, $G^{<}$, and for the relevant different interactions self-energies $\Sigma_{nm}^{R}(\boldsymbol{k}_{\mathrm{t}};E)$, $\Sigma_{nm}^{<}(\boldsymbol{k}_{\mathrm{t}};E)$ calculated in coupled system of \cref{eq-B28}. \cite{haug_quantum_2008,wacker_semiconductor_2002,lake_single_1997,mahan_many-particle_1990,lee_nonequilibrium_2002,hirsbrunner_review_2019} The couple set of equations is computational intensive to solve, for a device with tight-binding model Hamiltonian $h$, matrix element is $N_{H}\times N_{H}$, if the device contain lattice $N_{L}$ points, $k$ wavevector $N_{k}$ points, and energy $N_{E}$ points. The functions $G^{R}, G^{<}, \Sigma^{R}$, and $\Sigma^{<}$ size to calculate and store is $N_{L}\times N_{L}\times N_{k}\times N_{E}\times N_{H}\times N_{H}$. The size of the matrices is $N \times N$, as $N$ denotes the total number of atoms in the device. The retarded Green's function computed by the Recursive Green's Function (RGF) algorithm of complexity $\mathcal{O}(N)$, exploiting the property of block tri-diagonal matrix structure with minimal computational resources compared to the massive matrix inversion operation of complexity $\mathcal{O}(N^3)$. \cite{haydock_recursive_1980,teichert_improved_2017,thouless_conductivity_1981,mackinnon_calculation_1985} The algorithm to calculate the couple set of equation start with an initial value of $G^{<}$ and $G^{R}$. The system with no interactions is taken as the initial value of free Green's function $G^{0R}$ and $G^{0<}$. For all the lattice, $H$ $E$, $k$, points, self-energies $\Sigma^{<}$and $\Sigma^{R}$ derived by calculating the actual $G^{<}$ and $G^{R}$. The calculation of new $G^{R}$ and $G^{<}$ is performed by using the self-energies $\Sigma^{<}$ and $\Sigma^{R}$ values from the previous iteration. This loop continues until Jacobi iterations reach convergence. With this final $G^{<}$, the actual carrier density is obtained and solved in the Poisson equation loop. The carrier-carrier interactions will be approximately treated on the mean-field level with the Hartree self-energies as part of the Poisson potential. The algorithm restarted with the newly calculated Poisson potential and continued to run till convergence was achieved. In Green's function formalism, calculating carrier and current density is computationally expensive. Speed up is achieved by parallelizing the self energies computation and neglecting some parts of the self-energies. \cite{landauer_spatial_1957,datta_electronic_1997,lake_single_1997,datta_nanoscale_2000,lundstrom_fundamentals_2009,anantram_modeling_2008,rahman_theory_2003,cauley_distributed_2011}

\subsection*{Density of State $ \mathcal{D}(E) $}

The stationary state solution of non-equilibrium Green's functions from  \cref{eq-B20}  and \cref{eq-B27}, eigenfunction expansion in $(\mathrm{z})$ direction as eigenstate $ \phi_{\boldsymbol{k}_{\mathrm{t}}n}(z) $,
$ \phi_{\boldsymbol{k}_{\mathrm{t}}m}^{*}(z) $ are orthonormal, the Density of state $ \mathcal{D}(E) $ by taking trace of,

\begin{equation}\label{eq-B42-A}
\mathcal{D}(E) = -\frac{1}{\pi}\mathrm{Tr} \Bigg[\sum_{\boldsymbol{k}_{\mathrm{t}}}\sum_{nm}G_{nm}^{<}(\boldsymbol{k}_{\mathrm{t}};E)\Bigg]
\end{equation}

\subsection*{Carrier Density $\mathfrak{N}(r,t)$}

In the non-equilibrium Green's functions formalism, the carrier density $\mathfrak{N} (\boldsymbol r, t)$ is,

\begin{equation}\label{eq-B42}
\mathfrak{N} (\boldsymbol r, t) = -i\hbar G^{<}(\boldsymbol r, t;\boldsymbol r, t)
\end{equation}

In the stationary regime of non-equilibrium Green's functions solution in the one-dimension transport direction eigenfunction expansion $(\mathrm{z})$, the carrier density from \cref{eq-B20}  and \cref{eq-B27} is,

\begin{equation}\label{eq-B43}
\mathfrak{N}(z) = -\frac{i}{A}\sum_{\boldsymbol{k}_{\mathrm{t}}}\sum_{nm}\int\frac{\mathrm{d}E}{2\pi}G_{nm}^{<}(\boldsymbol{k}_{\mathrm{t}};E)\phi_{\boldsymbol{k}_{\mathrm{t}}n}(z)\phi_{\boldsymbol{k}_{\mathrm{t}}m}^{*}(z)
\end{equation}

\subsection*{Self-Energy Interaction $\Sigma^{\lessgtr,\mathrm{A},\mathrm{R}}(\boldsymbol{k}_{\mathrm{t}};E)$}

This section will discuss different scattering mechanisms, their Hamiltonian, eigenfunction expansion, and corresponding self-energy in lesser, greater, retarded, and advanced $\Sigma^{\lessgtr,\mathrm{A},\mathrm{R}}(\boldsymbol{k}_{\mathrm{t}};E)$ forms. The self-energy  calculated by Feynman diagrams, \cite{feynman_space-time_1948,schwinger_on_1951,mattuck_guide_1976}  Wick's decomposition and  variational derivation. \cite{binder_nonequilibrium_1995} As calculating carrier density, the current density is computationally expensive in Green's function formalism, parallelizing the self energies computation and neglecting some parts of the self-energies, speed up is achieved. \cite{rahman_theory_2003,cauley_distributed_2011}

\subsubsection*{Acoustic-Phonon Scattering $\Sigma^{\lessgtr,\mathrm{A},\mathrm{R};\mathcal{AC-PH}}(\boldsymbol{k}_{\mathrm{t}};E)$}

The Hamiltonian for carrier acoustic-phonon interaction is,

\begin{equation}\label{eq-B29}
H^\mathcal{AC-PH} = \int \mathrm{d}\boldsymbol{r}\widehat{\psi}^{\dagger}(\boldsymbol{r})\sum_{\boldsymbol{q}}e^{i\boldsymbol{qr}}|\boldsymbol{q}|[U_{\mathcal{AC}}a_{\boldsymbol{q}}+U_{\mathcal{AC}}^{*}a_{-\boldsymbol{q}}^{{\dagger}}]\widehat{\psi}(\boldsymbol{r})
\end{equation}
Where $a_{-\boldsymbol{q}}^{{\dagger}}$ creation phonon operator and $a_{\boldsymbol{q}}$ annihilation operator. The coupling for acoustic phonons is,

\begin{equation}\label{eq-B30}
|U_{\mathcal{AC}}|^{2}=\frac{\hbar D_{\mathcal{AC}}^{2}}{2V\rho v_{s}}	
\end{equation}
$V$ the volume, $\rho$ the material density, $D_{\mathcal{AC}}$ is the material acoustic deformation potential, and $v_{s}$ the velocity of sound in the crystal. Hence, the form factor $M_{nn_{1}}(\boldsymbol{q})$ for the carrier-acoustic-phonon interaction self-energies is,

\begin{equation}\label{eq-B31}
M_{nn_{1}}(\boldsymbol{q}) = |U_{\mathcal{AC}}||\boldsymbol{q}|\int \mathrm{d}z\phi_{n}^{*}(z)e^{iq_{z}z}\phi_{n_{1}}(z) = |U_{\mathcal{AC}}||\boldsymbol{q}|M_{nn_{1}}^{\prime}(q_{z})
\end{equation}

With the high temperature and low energy elastic scattering assumptions $\omega_{\boldsymbol{q}}=v_{s}|\boldsymbol{q}|$, $n_{\boldsymbol{q}}\approx n_{\boldsymbol{q}}+1\approx k_{B}T/\hbar\omega_{\boldsymbol{q}}=k_{B}T/\hbar v_{s}\boldsymbol{q}$, and $E\pm h\omega_{\boldsymbol{q}}=E^{\prime}$ then the carrier acoustic-phonons interaction self-energy is, 

\begin{equation}\label{eq-B32}
\Sigma_{nm}^{\lessgtr,\mathrm{A},\mathrm{R};\mathcal{AC-PH}}(\boldsymbol{k}_{\mathrm{t}};E)=\frac{D_{\mathcal{AC}}^{2}k_{B}T}{V\rho v_{s}^{2}}\sum_{n_{1}m_{1}}\sum_{\boldsymbol{q}}M_{nn_{1}}^{\prime}(q_{z})M_{mm_{1}}^{\prime*}(q_{z})G_{nm}^{\lessgtr,\mathrm{A},\mathrm{R};\mathcal{AC-PH}}({\boldsymbol{k_t}}-{\boldsymbol{q_t}};E)	
\end{equation}

For further details, we described the semi-classical treatment of scattering rate through Fermi's Golden rule for the non-equilibrium Green's function formalism truncation in the next section.

\subsubsection*{Optical-Phonon Scattering $ \Sigma^{\lessgtr;\mathcal{OP-PH}}(\boldsymbol{k}_{\mathrm{t}};E)$}

The carrier-optical-phonon interaction is inelastic primarily in nature, and the optical phonon's energy is larger than  $\boldsymbol{k}_{\mathrm{B}}T$. The optical phonons scattering matrix elements are assumed independent of the wave vector. However, in optical phonons, neighboring atoms oscillate in the opposite direction. Consequently, long-wavelength optical phonons may affect electronic energy directly. The coupling for optical phonons is,							

\begin{equation}\label{eq-B33}
|U_{\mathcal{OP}}|^{2}=\frac{\hbar D_{\mathcal{OP}}^{2}}{2V\rho \omega_{\boldsymbol{q}}}	
\end{equation}

$D_{\mathcal{OP}}$ is the material acoustic deformation potential, $\rho$ the material density, $V$ the volume, and $\omega_{\boldsymbol{q}}$ is the phonon wave vector in a phonon branch. The carrier optical-phonon interaction self-energies is, \cite{wacker_semiconductor_2002} 

\begin{equation}\label{eq-B34}
\begin{aligned}
\Sigma_{nm}^{\lessgtr;\mathcal{OP-PH}}(\boldsymbol{k}_{\mathrm{t}};E) & =\frac{\hbar D_{\mathcal{OP}}^{2}}{2V\rho \omega_{\boldsymbol{q}}}\sum_{n_{1}m_{1}}\sum_{\boldsymbol{q}}M_{nn1}(\boldsymbol{q})M_{mm1}^{*}(\boldsymbol{q}) \Big\{N_{\boldsymbol{q}}G_{nm}^{\lessgtr;\mathcal{OP-PH}}({\boldsymbol{k_t}}-{\boldsymbol{q_t}};E-\hbar\omega_{\boldsymbol{q}})\\ &+(N_{\boldsymbol{q}}+1)G_{nm}^{\lessgtr;\mathcal{OP-PH}}({\boldsymbol{k_t}}-{\boldsymbol{q_t}};E+\hbar\omega_{\boldsymbol{q}})\Big\}	
\end{aligned}
\end{equation}

The retarded self-energy evaluates by the lesser and the greater self-energies. The retarded self-energy's real part is discarded and approximated only with the imaginary part to reduce computation time. For further details, we described the semi-classical treatment of scattering rate through Fermi's Golden rule for the non-equilibrium Green's function formalism truncation in the next section.

\subsubsection*{Polar Optical-Phonon Scattering $\Sigma^{\lessgtr;\mathcal{POP-PH}}(\boldsymbol{k}_{\mathrm{t}};E)$}

The carrier polar-optical phonon interaction Hamiltonian is given by, \cite{riddoch_scattering_1983}

\begin{equation}\label{eq-B35}
H^\mathcal{POP-PH} = \int \mathrm{d}\boldsymbol{r}\widehat{\psi}^{\dagger}(\boldsymbol{r})\sum_{\boldsymbol{q}}\frac{e^{i\boldsymbol{q}\boldsymbol{r}}}{|\boldsymbol{q}|}[U_{\mathcal{POP}}a_{\boldsymbol{q}}+U_{\mathcal{POP}}^{*}a_{-\boldsymbol{q}}^{{\dagger}}]\widehat{\psi}(\boldsymbol{r})
\end{equation}

Where the phonon annihilation $a_{\boldsymbol{q}}$ and creation $a_{-\boldsymbol{q}}^{\dagger}$ operators, and $U_{\mathcal{POP}}$ is the Frohlich \textit{et al.} coupling constant. \cite{frohlich_interaction_1952} 

\begin{equation}\label{eq-B36}
U_{\mathcal{POP}}=i\sqrt{\frac{e^{2}}{V}\frac{\hbar\omega_{lo}}{2}\bigg[\frac{1}{\epsilon_{\infty}}-\frac{1}{\epsilon_{0}}\bigg]}
\end{equation}

Using the Feynman \textit{et al.} diagrams, \cite{feynman_space-time_1948,schwinger_on_1951,mattuck_guide_1976} and Wick's \textit{et al.} decomposition, The form factor $M_{nn_{1}}(\boldsymbol{q})$ and $D(\boldsymbol{q};t_{1}t_{2})$ for the carrier polar-optical phonon interaction self-energies is,

\begin{equation}\label{eq-B37}
\begin{aligned}
M_{nn_{1}}(\boldsymbol{q}) &= \Bigg|\frac{U_{\mathcal{POP}}}{\boldsymbol{q}}\Bigg|\int \mathrm{d}z\phi_{n}^{*}(z)e^{iq_{z}z}\phi_{n_{1}}(z) \\ D(\boldsymbol{q};t_{1}t_{2}) &= D(\boldsymbol{q},-\boldsymbol{q};t_{1}t_{2})
\end{aligned}
\end{equation}

Finally, the carrier polar-optical phonon interaction self-energies is,

\begin{equation}\label{eq-B38}
\begin{aligned}
\Sigma_{nm}^{\lessgtr;\mathcal{POP-PH}}(\boldsymbol{k}_{\mathrm{t}};E) &= \sum_{n_{1}m_{1}}\sum_{\boldsymbol{q}}M_{nn_{1}}(\boldsymbol{q})M_{mm_{1}}^{*}(\boldsymbol{q})  \Big\{N_{\boldsymbol{q}}G_{nm}^{\lessgtr;\mathcal{POP-PH}}({\boldsymbol{k_t}}-{\boldsymbol{q_t}};E-\hbar\omega_{\boldsymbol{q}})\\
&+(N_{\boldsymbol{q}}+1)G_{nm}^{\lessgtr;\mathcal{POP-PH}}({\boldsymbol{k_t}}-{\boldsymbol{q_t}};E+\hbar\omega_{\boldsymbol{q}})\Big\} \\
\Sigma_{nm}^{\mathrm{A},\mathrm{R};\mathcal{POP-PH}}(\boldsymbol{k}_{\mathrm{t}};E) & = \sum_{n_{1}m_{1}}\sum_{\boldsymbol{q}}M_{nn_{1}}(\boldsymbol{q})M_{mm_{1}}^{*}(\boldsymbol{q})  \Bigg[(N_{\boldsymbol{q}}+1)G_{nm}^{\mathrm{A},\mathrm{R};\mathcal{POP-PH}}({\boldsymbol{k_t}}-{\boldsymbol{q_t}};E-\hbar\omega_{\boldsymbol{q}})\\
&+N_{\boldsymbol{q}}G_{nm}^{\mathrm{A},\mathrm{R};\mathcal{POP-PH}}({\boldsymbol{k_t}}-{\boldsymbol{q_t}};E+\hbar\omega_{\boldsymbol{q}})\\
&+\frac{1}{2}\Big(G_{nm}^{<;\mathcal{POP-PH}}({\boldsymbol{k_t}}-{\boldsymbol{q_t}};E-\hbar\omega_{\boldsymbol{q}})-G_{nm}^{<;\mathcal{POP-PH}}({\boldsymbol{k_t}}-{\boldsymbol{q_t}};E+\hbar\omega_{\boldsymbol{q}})\Big)\\
&+i\mathcal{P}\bigg\{\int\frac{\mathrm{d}E^{\prime}}{2\pi}\Big(\frac{G_{nm}^{<;\mathcal{POP-PH}}({\boldsymbol{k_t}}-{\boldsymbol{q_t}};E-E^{\prime})}{E^{\prime}-\hbar\omega_{\boldsymbol{q}}}-\frac{G_{nm}^{<;\mathcal{POP-PH}}({\boldsymbol{k_t}}-{\boldsymbol{q_t}};E-E^{\prime})}{E^{\prime}+\hbar\omega_{\boldsymbol{q}}}\Big)\bigg\}\Bigg]
\end{aligned}
\end{equation}

The retarded self-energy's imaginary part obtains by evaluating the lesser and greater self-energies. The Hilbert transforming the imaginary part gives the real part. However, to reduce the computation overhead, we have discarded it. The physical significance of the self-energy imaginary part is to give a finite lifetime of the state. In contrast, the real part signifies an energy shift. This energy shift due to scattering is neglected compared to electrostatic potential in the device. Polar optical phonon is long-range scattering due to its coulomb-based nature. However, local polar optical phonon approximations are often used to reduce computational resources. Nevertheless, nonlocal polar optical phonon becomes significant at the high-temperature regime, and local polar optical phonon approximation does not hold. In the self-consistent Born approximation, electron-phonon self-energies evaluate by full electron and phonon Green's functions. For the self-consistent solution, the first influence on the phonons by the bare electrons compute, and then the renormalized phonon state's influence on the electrons is evaluated. \cite{wagner_expansions_1991} The complete set of equations should be solved for the exact solution in the many-body quantum system. Therefore, Dyson's equation for the phonon Green's function $D(\boldsymbol{q}; E)$ solved in a coupled way which is very expansive. \cite{dyson_the_1949} The first-order phonon renormalization process was neglected at a price to miss to capture a possible phonon lifetime reduction. According to the Migdal \textit{et al.} theorem, \cite{migdal_interaction_1958} phonon induced renormalization process of the electron-phonon vertex scales with the ratio of electron mass to ion mass. Hence it is safe to omit the renormalization process at the first level. \cite{fetter_quantum_1971} Therefore, We have assumed the phonon bath is in thermal equilibrium and full phonon Green's function $D(\boldsymbol{q};E)$ approximated to the non-interacting free phonon Green's functions $D^{(0)}(\boldsymbol{q};E)$. The Bose distribution for the phonons  is $N_{\boldsymbol{q}}$ with phonon frequency $\omega_{\boldsymbol{q}}$. For further details, we described the semi-classical treatment of scattering rate through Fermi's Golden rule for the non-equilibrium Green's function formalism truncation in the next section.

\subsubsection*{Charge Impurity Scattering $\Sigma^{\lessgtr,\mathrm{A},\mathrm{R};\mathcal{C-IMP}}(\boldsymbol{k_t};E)$}

An interaction between a moving charge carrier and a fixed ionized atom describes the impurity scattering mechanism as, \cite{Brooks_Scattering_1951,Lindhard_Dan_1954,ridley_reconciliation_1977}

\begin{equation}\label{eq-B39}
H^\mathcal{C-IMP} = \sum_{i}\int \mathrm{d}\boldsymbol{r}\;\widehat{\psi}^{\dagger}(\boldsymbol{r})U_{\mathcal{C-IMP}}(\boldsymbol{r}-\boldsymbol{R}_{\mathrm{i}})\widehat{\psi}(\boldsymbol{r})
\end{equation}

Where creation $\widehat{\psi}^\dagger$ and annihilation operators $\widehat{\psi}$, and the potential $U_{\mathcal{C-IMP}}(\boldsymbol{r}-\boldsymbol{R}_{\mathrm{i}})$ describe the interaction between an impurity at site ($\boldsymbol{R}_{\mathrm{i}}$) and moving carrier at site ($\boldsymbol{r}$). The inverse Fourier transform of the potential is,

\begin{equation}\label{eq-B40}
U_{\mathcal{C-IMP}}(\boldsymbol{r}-\boldsymbol{R}_{\mathrm{i}}) = \frac{1}{A}\sum_{\boldsymbol{q}}U_{\mathcal{C-IMP}}(\boldsymbol{q})\exp^{i\boldsymbol{q}(\boldsymbol{r}-\boldsymbol{R}_{\mathrm{i}})}
\end{equation}

Where $A$ is area and potential $U_{\mathcal{C-IMP}}(\boldsymbol{r})$ Fourier transform is $U_{\mathcal{C-IMP}}(\boldsymbol{q})$. The Coulomb potential for charge impurity scattering in the momentum space is, \cite{visscher_dielectric_1971}

\begin{equation} \label{eq28}
U_{\mathcal{C-IMP}}({q})=\frac{1}{{2\epsilon}_o\epsilon}{\frac{1}{q}\rho}^{ind}(q)+\frac{Ze}{{2\epsilon}_o\epsilon}{\frac{1}{q}\exp}^{-q|Z_c|}
\end{equation}

Where $\epsilon $ is the average of the permittivity of substate ${\epsilon }_{sub}$ and vacuum permittivity $({\epsilon }_{vac}=1)$. ${\rho }^{ind}(q)$ induced charge density, $Z_c$ is the shortest distance between the two-dimensional graphene sheet and the external charge impurity atom. The screened potential at the graphene devices deduced by the Thomas-Fermi approach as,

\begin{equation}\label{eq29}
U_{\mathcal{C-IMP}} (q)=\frac{Ze}{{2\epsilon }_o\epsilon}\frac{\exp^{-q|Z_c|}}{q+\rho(E_F)e^2/{2\epsilon }_o\epsilon}
\end{equation}

Finally, the charge impurity interaction self-energy is,

\begin{equation}\label{eq-B41}
\Sigma_{nm}^{\lessgtr,\mathrm{A},\mathrm{R};\mathcal{C-IMP}}(\boldsymbol{k_t};E) = \frac{\rho}{V}\sum_{n_{1}m_{1}}\sum_{\boldsymbol{q}}U_{\mathcal{C-IMP}}(\boldsymbol{q})U_{\mathcal{C-IMP}}(-\boldsymbol{q})M_{nn_{1}}^{\prime}(q_{z})M_{mm_{1}}^{\prime*}(q_{z})G_{n_{1}m_{1}}^{\lessgtr,\mathrm{A},\mathrm{R};\mathrm{Imp}}({\boldsymbol{k_t}}-{\boldsymbol{q_t}};E)
\end{equation}

Where $\rho$ is the impurity density present in the device. The matrix elements $M_{nn_{1}}'(q_{z})$ and $M_{mm_{1}}^{\prime*}(q_{z})$ are defined in \cref{eq-B31}. The interface roughness scattering potential and disorder potential scattering treatment on the same mathematical footing as the impurity scattering can be incorporated into non-equilibrium Green's function formalism in further study. In this proposed approach, the statistically averaged scattering potential $U(\boldsymbol{r},\boldsymbol{R})$ depends upon the roughness or dopant potential amplitude at position $\boldsymbol{R}$ and carrier position at $\boldsymbol{r}$.

\subsubsection*{Electron-Electron Scattering $ \Sigma^{\lessgtr;\mathbb{E-E}}(\boldsymbol{k_t};E)$}

The exact treatment of thermalizing electron-electron scattering is computationally challenging with explicit non-equilibrium Green's function simulations. Assuming elastic scattering and momentum conserving process, self-energy due to electron-electron interaction $ \Sigma_{nm}^{\lessgtr;\mathbb{E-E}} $ matrix elements defined as, \cite{goodnick_effect_1988}

\begin{equation} \label{eqR1}
\Sigma_{nm}^{\lessgtr;\mathbb{E-E}}(\boldsymbol{k_t};E)=D_{nm}^{\mathbb{E-E}}G_{nm}^{\lessgtr;\mathbb{E-E}}({\boldsymbol{k_t}}-{\boldsymbol{q_t}};E)
\end{equation}

Where the electron-electron interaction strength is represented by $ D_{nm}^{\mathbb{E-E}} = D_{p} $ for the momentum conserving interaction. \cite{golizadeh-mojarad_nonequilibrium_2007} In graphene electron-electron scattering rate is order of 60 pico-second, \cite{li_Influence_2010} by using $D_{p} = 10^{-3}$ as,

\begin{equation} \label{eqR2}
\frac{\hbar}{\tau_{\mathbb{E-E}}} = \Im {\Sigma_{nm}^{\lessgtr;\mathbb{E-E}}(\boldsymbol{k_t};E)}
\end{equation}

In principle, electron-electron scattering is non-local in energy/momentum space, physical space, and many-body nature. For the realistic device, complete full matrices of electron-electron interaction are too large to invert and solve precisely. On the other hand, for the high electron density contact or device region, electron-electron interactions are essential for carrier occupancy's thermalization process. Nevertheless, up to now, there is no efficient, formal scattering self-energy model available that can manage the full thermalization in the realistic high carrier density device regions. For further details, we described the semi-classical treatment of scattering rate through Fermi's Golden rule for the non-equilibrium Green's function formalism truncation in the next section.

\subsection*{Current Density Flux $ \mathfrak{J}(z,t)$}

The current density flux calculation is more computationally expensive compared to carrier density. The current density $\mathfrak{J}(z,t)$ is related to carrier density $\rho(z,t)$ by the continuity equation,

\begin{equation}\label{eq-B44}
\frac{\mathrm{d}}{\mathrm{d}t}\rho(z, t)+\mathrm{d}\mathrm{i}\mathrm{v}\mathfrak{J}(z, t) = 0
\end{equation}

The carrier density $\rho(z, t)$ is derived from the lesser Green's functions $G^{<}(x, t;x^{\prime}, t^{\prime})$ as,

\begin{equation}\label{eq-B45}
\frac{\mathrm{d}}{\mathrm{d}t}\rho(z,t) = \lim_{t'\rightarrow t}(-i\hbar)e\Big[\frac{\mathrm{d}}{\mathrm{d}t}G^{<}(z,t,z,t^{\prime})+\frac{\mathrm{d}}{\mathrm{d}t'}G^{<}(z,t,z,t^{\prime})\Big]
\end{equation}

In the case of the $z$-directional current transport and assuming the  eigenfunctions are centered around one lattice point, \cite{lake_single_1997} by using \cref{eq-B44},

\begin{equation}\label{eq-B46}
\begin{aligned}
\frac{\mathrm{d}}{\mathrm{d}t}\rho_{n}(t) & = \frac{e}{A\triangle}\sum_{\boldsymbol{k}_{\mathrm{t}}}\frac{\mathrm{d}}{\mathrm{d}t}\Big\langle a_{n,\boldsymbol{k}_{\mathrm{t}}}^{\dagger}(t)a_{n,\boldsymbol{k}_{\mathrm{t}}}(t)\Big\rangle \\ &
=\lim_{t'\rightarrow t}(-i\hbar)\frac{e}{A\triangle}\sum_{\boldsymbol{k}_{\mathrm{t}}}\Big[\frac{\mathrm{d}}{\mathrm{d}t}G_{nn}^{<}(\boldsymbol{k}_{\mathrm{t}};tt^{\prime})+\frac{\mathrm{d}}{\mathrm{d}t'}G_{nn}^{<}(\boldsymbol{k}_{\mathrm{t}};tt^{\prime})\Big]=  -\frac{\mathfrak{J}_{n}(t)-\mathfrak{J}_{n-1}(t)}{\triangle}
\end{aligned}
\end{equation}

Where $\rho_{n}(t)$ is charge density and $\mathfrak{J}_{n}(t)$  is the current density at place $z_{n}$, $a_{n,\boldsymbol{k}_{\mathrm{t}}}(t)$ annihilates an electron at position $z_{n}$, with state $\boldsymbol{k}_{\mathrm{t}}$, at time $t$ with in a volume $V=A\Delta$, $e$ is negative charge for electrons and the positive charge for holes transport, $A$ the area in the {\it xy} plane, $\Delta=z_{n}-z_{n-1}$, $a_{n,\boldsymbol{k}_{\mathrm{t}}}^{\dagger}$ creates an electron at position $z_{n}$, with state $\boldsymbol{k}_{\mathrm{t}}$, at time $t$ with in a volume $V=A\Delta$, 
$\mathfrak{J}_{n}(t)$ is the current density between point $n$ and $n+1$. The current density $\mathfrak{J}_{n}(t)$ is calculated by taking the two derivatives of the lesser Green's function $G_{nn}^{<}(\boldsymbol{k}_{\mathrm{t}}; {\it tt}')$  from equation \cref{eq-B22}  and \cref{eq-B23}, and inserting into \cref{eq-B46} yield,

\begin{equation}\label{eq-B47}
\begin{aligned}
\frac{\mathrm{d}}{\mathrm{d}t}\rho_{n}(t) & = -\frac{e}{A\triangle}\sum_{\boldsymbol{k}_{\mathrm{t}}}\sum_{m}\Big\{(h_{nm}G_{mn}^{<}(\boldsymbol{k}_{\mathrm{t}};tt)-G_{nm}^{<}(\boldsymbol{k}_{\mathrm{t}};tt)h_{mn})\Big\} 
=-\frac{\mathfrak{J}_{n}(t)-\mathfrak{J}_{n-1}(t)}{\triangle}
\end{aligned}
\end{equation}

By decomposing \cref{eq-B47}, $\mathfrak{J}_{n}$ and $\mathfrak{J}_{n-1}$ are separated. An ansatz has been given by Caroli \textit{et al.}\cite{caroli_direct_1971}. The current $\mathfrak{J}_{n}$ between point $n-1$ and point $n$ define as the difference between the flow of fermions from right to left and from left to right. Therefore, for stationary as well as for non-stationary cases when scattering mechanisms are present, the current $\mathfrak{J}_{n}(t)$ is given by,

\begin{equation}\label{eq-B48}
\mathfrak{J}_{n}(t) = -\frac{e}{A}\sum_{l\geq n+1}\sum_{m\leq n}\sum_{\boldsymbol{k}_{\mathrm{t}}}\Big[h_{lm}G_{ml}^{<}(\boldsymbol{k}_{\mathrm{t}};tt)-G_{lm}^{<}(\boldsymbol{k}_{\mathrm{t}};tt)h_{ml}\Big] 
\end{equation}

for $\mathfrak{J}_{n}$ \cref{eq-B48} with similar expression for $\mathfrak{J}_{n-1}$ satisfies the \cref{eq-B47}. The current is everywhere the same in the stationary state of the device. Hence, we can choose where to compute the current, assuming that contacts are big and in thermal equilibrium. We have assumed that in-between active parts of the device and contacts, no scattering occurs. Current calculated at the interface of the active region and contact. By choosing this place, the index $l$ corresponds to points in the active region, and the index $m$ corresponds to the contact points. By using equation \cref{eq-B27}, \cref{eq-B48} $\mathfrak{J}_{n}$ is simplified as,

\begin{equation}\label{eq-B49}
\begin{aligned}
\mathfrak{J}_{n} & = -\frac{e}{\hbar A}\sum_{l\geq n+1}\sum_{m\leq n}\sum_{\boldsymbol{k}_{\mathrm{t}}}\int\frac{\mathrm{d}E}{2\pi}\Big[h_{lm}G_{ml}^{<}(\boldsymbol{k}_{\mathrm{t}};E)-G_{lm}^{<}(\boldsymbol{k}_{\mathrm{t}};E)h_{ml}\Big]  \\ &
=-\frac{e}{\hbar A}\sum_{l\geq n+1}\sum_{m\leq n}\sum_{\boldsymbol{k}_{\mathrm{t}}}\int\frac{\mathrm{d}E}{2\pi}2Re\Big\{h_{lm}G_{ml}^{<}(\boldsymbol{k}_{\mathrm{t}};E)\Big\}
\end{aligned}
\end{equation}

The second line in the above equation is evaluated using \cref{eq-B28}. $G_{ml}^{<}(\boldsymbol{k}_{\mathrm{t}};E)$ is simplified by using the appropriate boundary conditions. Where $l$ belongs to the active part in a device,  $m$ belongs to any point in the contacts with defined carriers Fermi distribution, and $n$ belongs to the interface between both regions using the corresponding boundary conditions for $h_{ml, lm}$ and with equilibrium contacts assumption. The Fermi distribution is in equilibrium in the contacts using the fluctuation-dissipation theorem. \cite{Kubo1966Jan} The current density \cref{eq-B49} simplified to,

\begin{equation}\label{eq-B50}
\mathfrak{J}_{n}=\frac{e}{\hbar A}\sum_{l\geq n+1}\sum_{l_{1}\geq n+1}\sum_{\boldsymbol{k}_{\mathrm{t}}}\int\frac{\mathrm{d}E}{2\pi}\Gamma_{ll_{1}}^{\mathrm{Contact}}(\boldsymbol{k}_{\mathrm{t}};E)\Big[f^{\mathrm{Contact}}(E)A_{l_{1}l}(\boldsymbol{k}_{\mathrm{t}};E)+iG_{l_{1}l}^{<}(\boldsymbol{k}_{\mathrm{t}};E)\Big]
\end{equation}

Where $f^{Cont}(E)$ is contact Fermi-distribution function as,

\begin{equation}\label{eq-B51}
\Gamma_{ll_{1}}^{\mathrm{Contact}}(\boldsymbol{k}_{\mathrm{t}};E) = \sum_{m\leq n}\sum_{m_{1}\leq n}h_{lm}{A_{mm_{1}}}(\boldsymbol{k}_{\mathrm{t}};E)h_{m_{1}l_{1}}\end{equation}

With many-body interactions and scattering processes in the device's active region, \cref{eq-B50} is still valid. Only one assumption was made to derive the \cref{eq-B50} that self-energies between active region and contacts disappear. This equation corresponds to equation (5) of the Landauer \textit{et al.} formula for the current through an interacting electron region. \cite{meir_landauer_1992} The carrier density from \cref{eq-B43}, and current density from \cref{eq-B50} computed in the non-equilibrium Kadanoff-Keldysh-Martin \textit{et al.} formalism. \cite{martin_theory_1959,kadanoff_theory_1961,keldysh_diagram_1964}

\subsubsection*{Coherent Current}

Ballistic current in the non-interacting device evaluates by assuming no interaction self-energy in the active region. The \cref{eq-B50} simplified by appropriate boundary conditions by expressing lesser Green's function $G_{l_{1}l}^{<}(\boldsymbol{k}_{\mathrm{t}};E)$, and the spectral function $A_{l_{1}l}(\boldsymbol{k}_{\mathrm{t}};E)$ and using mathematical algebra on the running indices $m_{1}$ and $m_{2}$ to the left contact, non-interacting active part indices $l_{2}$ and $l_{3}$, and $m_{3}$ and $m_{4}$ for the right contact. Carrier distribution within the equilibrated contacts represent by $g_{m_{1}m_{2}}^{<}$ and $g_{m_{3}m_{4}}^{<}$ which enables the use of the fluctuation-dissipation theorem.\cite{Kubo1966Jan} Furthermore, recalling the definition \cref{eq-B51} leads to the following equations, that is two-terminal non-interacting Landauer \textit{et al.} formula as, \cite{meir_landauer_1992}

\begin{equation}\label{eq-B52}
\mathfrak{J} = \frac{e}{\hbar A}\sum_{ll_{1}}\sum_{l_{2}l_{3}}\sum_{\boldsymbol{k}_{\mathrm{t}}}\int\frac{\mathrm{d}E}{2\pi}\Big\{\Gamma_{ll_{1}}^{L}G_{l_{1}l_{2}}^{R}\Gamma_{l_{2}l_{3}}^{R}G_{l_{3}l}^{A}\Big\}(\boldsymbol{k}_{\mathrm{t}};E)\Big[f^{L}(E)-f^{R}(E)\Big] 
\end{equation}

Where in the non-interacting active part of the device indices $l, l_{1}, l_{2}, l_{3}$ run covering all the points. 

\subsubsection*{Transmission $ \mathcal{T}(\boldsymbol{k}_{\mathrm{t}};E) $}

Now, the Transmission $ \mathcal{T} $ at $ (\boldsymbol{k}_{\mathrm{t}};E) $ is defined as,

\begin{equation}\label{eq-B52-A}
\mathcal{T}(\boldsymbol{k}_{\mathrm{t}};E) = \Big\{\Gamma_{ll_{1}}^{L}G_{l_{1}l_{2}}^{R}\Gamma_{l_{2}l_{3}}^{R}G_{l_{3}l}^{A}\Big\}(\boldsymbol{k}_{\mathrm{t}};E)
\end{equation}

Where in the non-interacting active part of the device indices $l, l_{1}, l_{2}, l_{3}$ run covering all the points. And, $ \Gamma_{ll_{1}}^{L} $ is connected towards the left contact and $ \Gamma_{l_{2}l_{3}}^{R} $ connected through the right contact.

\subsubsection*{Mode Density $ \mathcal{M}(E) $}

The total number of the transmissible propagating mode of the wave-function in the device define as Density of Mode or Mode Density $ \mathcal{M} $ at energy $(E)$ as,

\begin{equation}\label{eq-B52-B}
\mathcal{M}(E) = \frac{1}{ A}\sum_{ll_{1}}\sum_{l_{2}l_{3}}\sum_{\boldsymbol{k}_{\mathrm{t}}}\int\Big\{\Gamma_{ll_{1}}^{L}G_{l_{1}l_{2}}^{R}\Gamma_{l_{2}l_{3}}^{R}G_{l_{3}l}^{A}\Big\}(\boldsymbol{k}_{\mathrm{t}};E){\mathrm{d}E}
\end{equation}

\subsubsection*{Interacting Current}

Interacting current in the device, where the central part interacts with self-energy and contacts or leads are non-interacting, is defined by \cref{eq-B50}. However, the interacting part of the device has a different set of expressions for $A_{l_{1}l}$ and $G_{l_{1}l}^{<}$ is,

\begin{equation}\label{eq-B53}
\begin{aligned}
G_{l_{1}l}^{<}& = \sum_{l_{2},l_{3}}G_{l_{1}l_{2}}^{R}\Big[if^{L}\Gamma_{l_{2}l_{3}}^{L}+if^{R}\Gamma_{l_{2}l_{3}}^{R}+\Sigma_{l_{2}l_{3}}^{<}\Big]G_{l_{3}l}^{A} \\
A_{l_{1}l} & = \sum_{l_{2},l_{3}}G_{l_{1}l_{2}}^{R}\Big[\Gamma_{l_{2}l_{3}}^{L}+\Gamma_{l_{2}l_{3}}^{R}+i\big[\Sigma_{l_{2}l_{3}}^{>}-\Sigma_{l_{2}l_{3}}^{<}\big]\Big]G_{l_{3}l}^{A}  
\end{aligned}
\end{equation}

The self-energies due to interactions e.g. carrier-phonon scattering are incorporated into $\Sigma_{l_{2}l_{3}}^{<}$ and $\Sigma_{l_{2}l_{3}}^{>}$. In the \cref{eq-B53} $G_{l_{1}l}^{<}$ Green's function and spectral function $A_{l_{1}l}$ produce two-part current density $\mathfrak{J}$ as,

\begin{equation}\label{eq-B54}
\mathfrak{J} = \mathfrak{J}_{coh}+\mathfrak{J}_{in}
\end{equation}

Coherent current density $\mathfrak{J}_{coh}$ is same as \cref{eq-B52}, however, $G_{l_{1}l_{2}}^{R}$ and $G_{l_{3}l}^{A}$ in it is calculated in the presence of interaction and hence gives different value from \cref{eq-B52}, The interaction current density $\mathfrak{J}_{in}$ is, 

\begin{equation}\label{eq-B55}
\begin{aligned}
\mathfrak{J}_{in} & = \frac{ie}{\hbar A}\sum_{ll_{1}}\sum_{l_{2}l_{3}}\sum_{\boldsymbol{k}_{\mathrm{t}}}\int\frac{\mathrm{d}E}{2\pi}\Gamma_{ll_{1}}^{L}(\boldsymbol{k}_{\mathrm{t}};E)G_{l_{1}l_{2}}^{R}(\boldsymbol{k}_{\mathrm{t}};E)\Big\{\Sigma_{l_{2}l_{3}}^{>}(\boldsymbol{k}_{\mathrm{t}};E)f^{L}(E)  
+\Sigma_{l_{2}l_{3}}^{<}(\boldsymbol{k}_{\mathrm{t}};E)\big[1-f^{L}(E)\big]\Big\}G_{l_{3}l}^{A}(\boldsymbol{k}_{\mathrm{t}};E) \\
& = \frac{ie}{\hbar A}\sum_{ll_{1}l_{2}l_{3}}\sum_{m_{1}m_{2}}\sum_{\boldsymbol{k}_{\mathrm{t}}}\int\frac{\mathrm{d}E}{2\pi}h_{lm_{1}}A_{m_{1}m_{2}}(\boldsymbol{k}_{\mathrm{t}};E)h_{m_{2}l_{1}}G_{l_{1}l_{2}}^{R}(\boldsymbol{k}_{\mathrm{t}};E)G_{l_{3}l}^{A}(\boldsymbol{k}_{\mathrm{t}};E) \\&
\times\Big\{\Sigma_{l_{2}l_{3}}^{>}(\boldsymbol{k}_{\mathrm{t}};E)if^{L}(E)+\Sigma_{l_{2}l_{3}}^{<}(\boldsymbol{k}_{\mathrm{t}};E)i\big[1-f^{L}(E)\big]\Big\} \\
& = \frac{e}{\hbar A}\sum_{ll_{1}l_{2}l_{3}}\sum_{m_{1}m_{2}}\sum_{\boldsymbol{k}_{\mathrm{t}}}\int\frac{\mathrm{d}E}{2\pi}h_{lm_{1}}h_{m_{2}l_{1}}G_{l_{1}l_{2}}^{R}(\boldsymbol{k}_{\mathrm{t}};E)G_{l_{3}l}^{A}(\boldsymbol{k}_{\mathrm{t}};E) \\&
\times\Big\{\Sigma_{l_{2}l_{3}}^{>}(\boldsymbol{k}_{\mathrm{t}};E)g_{m_{1}m_{2}}^{<}(\boldsymbol{k}_{\mathrm{t}};E)-\Sigma_{l_{2}l_{3}}^{<}(\boldsymbol{k}_{\mathrm{t}};E)g_{m_{1}m_{2}}^{>}(\boldsymbol{k}_{\mathrm{t}};E)\Big\}
\end{aligned}
\end{equation}

Where in left lead or contact, indices $m_{1}, m_{2}$ are situated and within the interacting central part of the device, indices $l, l_{1}, l_{2}, l_{3}$ run covering all the points. In \cref{eq-B55} last equality evaluated employing fluctuation-dissipation theorem.\cite{Kubo1966Jan} In the self-consist born approximation, the carrier density from \cref{eq-B43} and current density from \cref{eq-B50} is computed. The algorithm drives as follows, at the start of the simulation, at the first step  $G_{0}^{<,>}$  and $G_{0}^{R, A}$ non-interacting Green's functions evaluated to calculate the first iterated self-energies $\Sigma^{<,>}$ and $\Sigma^{R, A}$. In the second step, the $G^{R, A}$ matrix equation calculate to get the actual values of $\Sigma^{<,>}$ and $\Sigma^{R, A}$ self-energies. Actual self-energies use for the computation of $G^{<,>}$. In the third step, updated values of $G^{<,>}$ and $G^{R,A}$  adopt to estimate new scattering self-energies $\Sigma^{<,>}$ and $\Sigma^{R,A}$. The scattering self-energies utilizes to determine the new Hartree potential, which through the $V(z)$ part directly updates the Hamiltonian $H_{0}$. It is equivalent to finding the solution of potential in Poisson's equation with the carrier density from \cref{eq-B43} and iteratively updating the device potential. The self-consistent iterative loop between the self-energies and Green's functions will run continuously until convergence. Once the convergence achieves, the algorithm proceeds in the last step. In the fourth step, definitive device potential obtained from self-consistent self-energies and Green's functions loop is used in \cref{eq-B50} to calculate the current density. Truncation to the self-energies and Green's functions is used in the self-consistent calculation to reduce the computation time when incorporating scattering processes. Also, in \cref{eq-B38} the principle integral is neglected for calculating carrier-optical-phonon interaction. The generalized contact method implements multiple scattering scenarios. A single-scattering rate represents the effect of multiple scattering events and correlates with momentum relaxation time and the mean free path of the carriers. The contacts thermalize in the generalized contact method, and the contact Green's function's diagonal elements compute by the Recursive Green's Function (RGF) algorithm. \cite{klimeck_quantum_1995} Furthermore, the drift-diffusion equation solve in the thermalized contacts to determine the quasi-Fermi levels in the contact. \cite{klimeck_quantum_2003} The contacts are also affecting electrical characteristics in real devices. Furthermore, there are uncertainties about the dielectric constant of low-dimensional materials. Further details of multi-scale non-equilibrium Green's function formalism implementation on the supercomputer cluster are provided in the Multi-scale non-equilibrium Green's function simulation implementation section. Here for the brevity of time, Next, we will present the semi-classical treatment of the scattering rate through Fermi's Golden rule for the non-equilibrium Green's function formalism truncation procedure.

\section*{Scattering Semi-classical Treatment Fermi's Golden Rule}
\label{section:Fermi's Golden rule}

After growth, chemical vapor deposited graphene is wet transferred on the polymer support substrate either through the copper etching process or delamination process. \cite{li_large-area_2009,pizzocchero_non-destructive_2015} During the growth and transfer process, various factors can influence the electrical property of the final graphene sheet. Adsorbent gas and water molecule on the surface or trapped between graphene and substrate can change the residual carrier density of the graphene sheet. \cite{wang_hysteresis_2010,piazza_graphene_2015} Also, polymer residues can contaminate the device during the transfer and lithography process. Moreover, plasma treatment of the surface can also influence the final device property. \cite{lin_graphene_2012,goniszewski_correlation_2016} Water and other polar molecule and trap charges in oxide can dynamically vary the total career density in the capacitive getting and charge transfer process. \cite{wang_hysteresis_2010} All of these factors influence graphene mobility through various nanoscale atomistic scattering mechanisms. These scattering mechanisms dominate the device operating at room temperature and ambient conditions. Therefore, the mobility values obtained in such a scenario are orders of magnitude lower than the intrinsic graphene mobility. The charge particles described above can contribute to the charge impurity scattering mechanism. Charge impurity scattering shows the square root dependence on the carrier density, where $n_{C-Imp}$ is charge impurity density in the sheet, \cite{adam_self-consistent_2007,chen_charged-impurity_2008}

\begin{equation} \label{eq16}
\tau_{\mathcal{C-IMP}}\propto \frac{\sqrt{n_s}}{n_{\mathcal{C-IMP}}}
\end{equation}

Charge impurity scattering on the macroscopic length scale shifts the conductance-gate voltage characteristic along the gate voltage axis. On the other hand, charge impurity scattering creates electron-hole puddles at the charge neutrality point on the atomistic nanoscale length scale. \cite{martin_observation_2008,zhang_origin_2009} Nevertheless, at the nanoscale, the formation of electron-hole puddles is also greatly influenced by the surface roughness of the substrate, and in the ultra-flat single-crystal hexagonal boron nitride $(h\mathrm{-}\mathrm{BN})$ substrate, such formation is actively suppressed. \cite{xue_scanning_2011} Furthermore, the electrical permittivity of the environment strongly influences the long-range coulomb scattering, which originated from the charged impurity scattering. \cite{ando_screening_2006} However, with the increase of electrical permittivity of different substrates for the graphene, there is a slight increment in the mobility values, suggesting that the charge impurity scattering suppresses with increasing dielectric screening. However, other scattering mechanisms in the room temperature condition become dominant in such a scenario. \cite{peres_electronic_2006,fratini_substrate-limited_2008} In the graphene sheet, atomic-scale defects, naturally occurring vacancy in graphene, adsorbate gaseous and fluid atoms, molecules in the fabrication process, grain boundaries, cracks, fold, and tears give rise to point-like defects which will be the center of robust and short-range scattering potential, \cite{sarma_electronic_2011,peres_electronic_2006,ostrovsky_electron_2006,wehling_resonant_2010,ni_resonant_2010}

\begin{equation} \label{eq17}
\tau_{\mathcal{N-IMP}}=\frac{2q^2}{\pi h}\frac{n}{n_{\mathcal{N-IMP}}}{{\mathrm{ln}}^{\mathrm{2}}(k_fR)} 
\end{equation}

Where $n_{\mathcal{N-IMP}}$ is neutral defect density in the sample, and $R$ is the radius of the point-like scattering potential. Besides charge contamination and defects, the electrons in the graphene are also scattered by the intrinsic longitudinal acoustic phonon. Depending on the substrate can also scatter from the substrate polar optical phonon. The conductance versus gate voltage curve shows the linear dependence in the charged impurity scattering-dominated sample. The exceptionally clean sample will become sublinear where short-range scattering or ballistic transport dominates. The practical impact of various nanoscale atomistic scattering mechanisms in the semiconductor expressed by the Matthiesen rule and total scattering time is, \cite{jacoboni_monte_1983,lake_single_1997,datta_electronic_1997,datta_quantum_2005,chen_charged-impurity_2008,lundstrom_fundamentals_2009,steiger_nemo5_2011}

\begin{equation} \label{eq18}
\tau^{-1}_{\mathcal{TOT}}=\tau^{-1}_{\mathcal{AC-PH}}+\tau^{-1}_{\mathcal{POP-PH}}+\tau^{-1}_{\mathcal{C-IMP}}+\tau^{-1}_{\mathcal{N-IMP}}+\dots
\end{equation}

The contribution through different scattering mechanisms to the mean free path is,

\begin{equation} \label{eq19}
\ell^{-1}_{\mathcal{TOT}}=\ell^{-1}_{\mathcal{AC-PH}}+\ell^{-1}_{\mathcal{POP-PH}}+\ell^{-1}_{\mathcal{C-IMP}}+\ell^{-1}_{\mathcal{N-IMP}}+\dots
\end{equation}

The defects, cracks, adsorbates, and grain boundaries in the graphene cause the resonant scattering rate $\tau^{-1}_{\mathcal{N-IMP}}=\upsilon_f\ell^{-1}_{\mathcal{N-IMP}}$.The contribution through different scattering mechanisms to the mean free path is, \cite{lundstrom_fundamentals_2009,mahan_many-particle_1990} The resonant scattering rate $\tau^{-1}_{\mathcal{N-IMP}}$, via the phase shift induces by the scattering center, assuming only the elastics scattering events and considering the s-wave scattering, the transition rate $\tau^{-1}_{\mathcal{N-IMP}}$ is,

\begin{equation} \label{eq20}
\hbar\tau^{-1}_{\mathcal{N-IMP}}=\frac{8N_{\mathcal{N-IMP}}}{\pi\rho (E_k)}{\sin}^2(\delta_k)
\end{equation}

Where $N_{\mathcal{N-IMP}}$ is the density of short-range resonant scatters, $ \delta_k$ is the phase shift in the s-wave channel due to the scattering center. The phase shift, $\delta_k$ is, \cite{hentschel_orthogonality_2007}

\begin{equation} \label{eq21}
\delta_k=-\frac{\pi}{2}\frac{1}{\ln(kR)}
\end{equation}

Where $R\approx2a$ is the effective radius of resonant scatters, $a\approx 0.14nm$ in graphene bond length. \cite{chen_defect_2009} Assuming the $kR\ll 1$ for the graphene, putting phase shift, $\delta_k$ from \cref{eq21} in to the \cref{eq20}, the resonant scattering relaxation time is, \cite{stauber_electronic_2007}

\begin{equation} \label{eq22}
\tau_{\mathcal{N-IMP}}=\frac{\hbar\rho(E_k)}{2\pi N_{\mathcal{N-IMP}}}{\ln^2 (kR)}
\end{equation}

And, the resonant scattering length $\ell_{\mathcal{N-IMP}}$ is, \cite{stauber_electronic_2007}

\begin{equation} \label{eq23}
\ell_{\mathcal{N-IMP}}(n)=\frac{\sqrt{\pi n}}{{\pi}^2N_{\mathcal{N-IMP}}}{{\ln}^2(\sqrt{\pi n}R)} 
\end{equation}

The charge impurities reside in or on the substrate $(\mathrm{SiO}_2)$ or $(h-\mathrm{BN})$ layer screened the conduction electrons of the graphene sheet and gave rise to the charge impurity scattering rate $\tau^{-1}_{\mathcal{C-IMP}}={\upsilon_f\ell}^{-1}_{\mathcal{C-IMP}}$ \cite{lundstrom_fundamentals_2009,mahan_many-particle_1990}

\begin{equation} \label{eq24}
\tau^{-1}_{\mathcal{C-IMP}}=N_{\mathcal{C-IMP}}\sum_{k^\prime}\Gamma(k,k^\prime)(1-\cos\theta_{k,k^\prime})
\end{equation}

Where $N_{\mathcal{C-IMP}}$ is the charge impurity density, from Fermi's golden rule for the transition probability of the scattering for the electrons,

\begin{equation} \label{eq25}
{\Gamma}(k,k^{\prime})=\frac{2\pi}{\hbar}{|\langle k\mathrel{|\vphantom{k V_{Imp} k^{\prime}}\kern-\nulldelimiterspace}V_{\mathcal{C-IMP}}\mathrel{|\vphantom{k V_{scatt} k^{\prime}}\kern-\nulldelimiterspace}k^{\prime}\rangle |}^2\delta(E_k-E_{k^{\prime}}) 
\end{equation}

Where taking Fourier transform of $V_{\mathcal{C-IMP}}$ scattering potential, $V_{\mathcal{C-IMP}}(q)=eU_{\mathcal{C-IMP}}(q)$, and putting in the scattering rate $\tau^{-1}_{\mathcal{C-IMP}}$ expression, \cite{stauber_electronic_2007}

\begin{equation} \label{eq26}
\hbar {\tau }^{-1}_{\mathcal{C-IMP}}= \frac{N_{\mathcal{C-IMP}} }{8}\rho (E_F) \int{d\theta }{{dq}}{|eU_{\mathcal{C-IMP}} (q)|}^2(1- {{\cos}^2 \theta }) 
\end{equation}

Where for graphene $q= |(k^{'}-k)|=2k\sin(\theta/2)$, and the density of the state-defined as,

\begin{equation} \label{eq27}
\rho (E_F)=\frac{2|E_F|}{\pi{(\hbar v_f)}^2}
\end{equation}

And from \cref{eq-B28}, Coulomb potential for charge impurity scattering in the momentum space is, \cite{visscher_dielectric_1971}

\begin{equation} \label{eq28-a}
U_{\mathcal{C-IMP}}({q})=\frac{1}{{2\epsilon}_o\epsilon}{\frac{1}{q}\rho}^{ind}(q)+\frac{Ze}{{2\epsilon}_o\epsilon}{\frac{1}{q}\exp}^{-q|Z_c|}
\end{equation}

Where $\epsilon $ is the average of the permittivity of substate ${\epsilon}_{sub}$ and vacuum permittivity $({\epsilon}_{vac}=1)$. ${\rho}^{ind}(q)$ induced charge density, $Z_c$ is the shortest distance between the two-dimensional graphene sheet and the external charge impurity atom. And again, from equation \cref{eq29} the screened potential inside the graphene sheet deduced by the Thomas-Fermi approach, \cite{lundstrom_fundamentals_2009,mahan_many-particle_1990}

\begin{equation} \label{eq29-a}
U_{\mathcal{C-IMP}} (q)=\frac{Ze}{{2\epsilon }_o\epsilon}\frac{\exp^{-q|Z_c|}}{q+\rho(E_F)e^2/{2\epsilon }_o\epsilon}
\end{equation}

By plugging \cref{eq29-a} in \cref{eq26} with the assumption of $Z_c\approx 0$, and using \cref{eq15}, The charge impurity scattering time,  

\begin{equation} \label{eq30}
{\tau}_{\mathcal{C-IMP}}=\frac{16{{\epsilon}_o}^2{\epsilon }^2{\hbar }^2{\upsilon}_f}{{Z}^2q^4N_{\mathcal{C-IMP}}}{\bigg(1+\frac{q^2}{\pi \hbar {\upsilon}_f{\epsilon}_o\epsilon}\bigg)}^2\sqrt{\pi n} 
\end{equation}

And, the charge impurity scattering length ${\ell}_{\mathcal{C-IMP}}$ is, \cite{stauber_electronic_2007}

\begin{equation} \label{eq31}
{\ell}_{\mathcal{C-IMP}}(n)=\frac{16{{\epsilon}_o}^2{\epsilon}^2{\hbar}^2{{\upsilon}_f}^2}{{Z}^2q^4N_{\mathcal{C-IMP}}}{\bigg(1+\frac{q^2}{\pi\hbar{\upsilon}_f{\epsilon}_o\epsilon}\bigg)}^2\sqrt{\pi n} 
\end{equation}

Where $\hbar$ is reduced plank constant, $Z=1$ is the net charge of impurities. The carrier in the graphene on the polar substrate like $(\mathrm{Si}{\mathrm{O}}_2)$ electrostatically coupled to the long-range polarization field created at the interface by the polar molecules, and this surface polar phonon scattering strongly depends upon the dielectric function of the substrate. However, for the single-crystal hexagonal boron nitride $(h\mathrm{-}\mathrm{BN})$ substrate, surface polar phonon scattering is massively reduced. The surface polar phonon scattering rate ${\tau}^{-1}_{\mathcal{POP-PH}}={{\upsilon}_f\ell}^{-1}_{\mathcal{POP-PH}}$ is related to the scattering by surface polar phonon mode of energy $E_i= \hbar{\omega}^i_q$ and the transition probability ${\tau}^{-1}_{\mathcal{POP-PH}}$ is, \cite{lundstrom_fundamentals_2009,mahan_many-particle_1990}

\begin{equation} \label{eq32}
{\tau}^{-1}_{\mathcal{POP-PH}}= \sum_q{{\Gamma}(k,k+q)(1-{\cos{\theta}_{k,k+q}})}
\end{equation}

Where the transition rate is, 

\begin{equation}\label{eq33}
\begin{split}
{\Gamma}(k,k+q)=&\frac{2\pi}{\hbar}{|\langle k\mathrel{|\vphantom{k V_{\mathcal{POP-PH}}k+q}\kern-\nulldelimiterspace}V_{\mathcal{POP-PH}}\mathrel{|\vphantom{k V_{SPP}k+q}\kern-\nulldelimiterspace}k+q\rangle |}^2 \times [N_{\mathcal{POP-PH}}\delta (E_k-E_{k+q}+\hbar {\omega }^i_q)+(N_{\mathcal{POP-PH}}+1)\delta(E_k-E_{k+q}-\hbar {\omega }^i_q)] 
\end{split}
\end{equation}

Where $N_{\mathcal{POP-PH}}$ is the surface polar phonon occupation number given by Bose-Einstein statistics. The surface polar phonon coupling in the graphene is,

\begin{equation} \label{eq34}
{|\langle k\mathrel{|\vphantom{k V_{\mathcal{POP-PH}} k+q}\kern-\nulldelimiterspace}V_{\mathcal{POP-PH}}\mathrel{|\vphantom{k V_{\mathcal{POP-PH}} k+q}\kern-\nulldelimiterspace}k+q\rangle |}^2=e^2U_{\mathcal{POP-PH},i}^2\ \bigg(\frac{e^{-2qz_o}}{q}\bigg){cos}^2\bigg(\frac{{\theta }_{k,k+q}}{2}\bigg)
\end{equation}

Where $z_o\approx 0.35nm$ distance between the polar substrate and the graphene layer. The magnitude of the polarization field given by the Frohlich coupling constant ${U_{\mathcal{POP-PH}, i}^2}$ and depends upon substrate permittivity as, \cite{frohlich_interaction_1952,wang_electron_1972} 

\begin{equation} \label{eq35}
U_{\mathcal{POP-PH},i}^2=\frac{\hbar {\omega }^i_q}{2A\epsilon_o}\bigg(\frac{1}{\epsilon^{\infty }_{ox}+1}-\frac{1}{\epsilon^0_{ox}+1}\bigg)
\end{equation}

Where $A$ is the area of graphene sheet, $\epsilon^0_{ox}$ and $\epsilon^{\infty}_{ox}$ is low and high frequency dieletric constant of the substrate, respectively. By using \cref{eq33}, \cref{eq34}, \cref{eq35} in the \cref{eq32}, The surface polar phonon scattering length ${\ell}_{\mathcal{POP-PH}}$ of Energy $E_{i}$ is, \cite{fratini_substrate-limited_2008,perebeinos_inelastic_2010,sonde_role_2010}

\begin{equation} \label{eq36}
{\ell }_{\mathcal{POP-PH},i}= \sqrt{\frac{\beta}{E_i}} \frac{\hbar{\upsilon}_f4\pi \epsilon_o}{q^2} \frac{q{\upsilon}_f}{U_{\mathcal{POP-PH},i}^2} \frac{exp(\epsilon_oz_o)}{N_{\mathcal{POP-PH},i}}\frac{\hbar \sqrt{\pi}}{q}
\end{equation}

Where $\epsilon_o\approx{[{({2E}_i{\hbar }^{-1}{{\upsilon }_f}^{-1})}^2+\chi n]}^{\frac{1}{2}},$ $\chi \approx 10.5$, $\beta \approx 0.153\times {10}^{-4}eV$, and for $(\mathrm{Si}{\mathrm{O}}_2)$ substrate $E_1=58.9meV$, ${E}_2=156.4meV$, $U_{\mathcal{POP-PH},1}^2=0.237meV$, and $U_{\mathcal{POP-PH},2}^2=1.612meV$. \cite{perebeinos_inelastic_2010,giannazzo_mapping_2011} Next, we present The multi-scale non-equilibrium Green's function formalism simulation implementation on the supercomputer cluster.

\section*{Multi-scale Non-equilibrium Green's Function Simulation Implementation}
\label{section:implementation}

Non-equilibrium Green's function formalism is computationally intensive to solve and has simulation overhead compared to classical and semi-classical approaches. Also, non-equilibrium Green's function formalism is limited by the single-particle charging energy and mean-field assumption. One should explore the Hartree energy approximation or ab-initio calculation to more accurately explore the potential energy Hamiltonian. In the non-equilibrium Green's function simulation, the most computationally expensive part is the evolution of the retarded Green's function, which the matrix inversion operation requires at every energy grid point. However, the problem is partially simplified in the ballistic transport regime as only a few of the Green's function columns matrix are required. The actual metal contact connected to the active device is incorporated in the simulation through the contact metal work function and injection of the continuum density of states near the Fermi level. For the calculation of current, the local density of state, and transmission spectra through the finite element calculation, We have employed $\mathrm{Nano-Electronics\ Modeling\ Tools}$ (\texttt{NEMO5}), which is an all-purpose multiscale simulation toolbox for nano-electronic device modeling. \cite{fonseca_efficient_2013, steiger_nemo5_2011} Its modular architecture parallelized with 5-level Message Passing Interface \texttt{(MPI)} in the position space, momentum space, energy space, bias space, and random seeding space. \texttt{NEMO5} has the advantage that physical models are added and extended due to the modular architecture of the kernel. \texttt{NEMO5} used various numerical packages for the scalability and computation of mathematical functions. For the Finite element method \texttt{(FEM)} discretization of Poisson's equation, \texttt{libmesh} library~package is used. The eigenvalue solution \texttt{SLEPc} package and \texttt{PETSc} package is used for linear and nonlinear matrices equation solution—the \texttt{Boost} package used for the file-system and input deck operations. \texttt{Silo} and  \texttt{VTK} packages are used for an output file format visualization. We used the empirical tight-binding parameter to describe the device material property. We are not tailoring new material at the nano-scale from scratch, and we do not need a first principle calculation and ab-initio density-functional theory (DFT) simulator. Also, due to the applied potential bias condition device being in a non-equilibrium state, the applicability of ab-initio is limited in the off-equilibrium and excited state condition. As in the nano-scale devices, non-ideal effects, non-parabolicity in bands, strain, crystal orientation, energy quantization, size confinement, band coupling, band-to-band tunneling, local disorders, valley splitting, and valley mixing, two-dimensional/three-dimensional spatial variations, and potential variations in device starts arising into the microscopic landscape. Through the \texttt{NEMO5} kernel, we can solve the atomistic tight-binding contacted-Schr\"{o}dinger equation, Poisson's equation, and coupled Schr\"{o}dinger-Poisson loop for the relatively large size of atomistic devices. We can calculate the transmission spectra, the local density of state, and current density through wave-function formalism or non-equilibrium Green's function formalism in the two-dimensional geometries. The non-linear Poisson's equation solves by the Newton-Ralphson iteration method, and the Lanczos eigenvalue solver algorithm is employed to provide the Schr\"{o}dinger equation solution. The Self-Energy of the contacts calculate through surface Green's function method. An iterative solution based on Sancho-Rubio algorithms is employed to calculate surface Green's function. \cite{steiger_nemo5_2011,anantram_modeling_2008, fonseca_efficient_2013} In the non-equilibrium Green's function formalism, calculation of the retarded Green's function is the most computationally intensive step, and the recursive algorithm computes it. To reduce the complexity time, a nested dissection approach in which the graph partitioning approach uses to calculate the electron density at a reduced complexity compared to the Recursive Green's Function (RGF) algorithm was developed. \cite{hetmaniuk_nested_2013,hetmaniuk_reduced-order_2015} In this approach from the full-order model, the unitary matrices construct a small energy subset and contact grid point. The reduced-order method calculates charge and current density by combining two techniques. First, find an approximate solution of Green's functions based on the moment matching scheme at all the energy points. They further sampled a small subset of energy, and spatial points are computed Green's functions in every lead. The reduced-order method demonstrates three to seven times the speedup for different device geometry. Also, Fast Inverse using the Nested Dissection (FIND) algorithm was employed to evaluate Green's functions in the non-equilibrium Green's function formalism. In this method, the FIND algorithm calculates the specific entries of the inverse of a sparse matrix. To calculate the inverse of a dense matrix the Recursive Green's Function (RGF) method have a run time of the order of $\mathcal{O}({{\mathrm{N}}_{\mathrm{x}}}^3\mathrm{\times }{\mathrm{N}}_{\mathrm{y}})$, where ${\mathrm{N}}_{\mathrm{x}}$ is number of points after discretization in the $\mathrm{x}$ direction and ${\mathrm{N}}_{\mathrm{y}}$ in $\mathrm{y-}$direction. Whereas in the FIND algorithm computation complexity scale down to the order of $\mathcal{O}({{\mathrm{N}}_{\mathrm{x}}}^2\mathrm{\times }{\mathrm{N}}_{\mathrm{y}})$. \cite{li_extension_2012} The following result and discussion section elucidate the variability in the measurand electrical properties for benchmarking purposes. We will discuss and recapitulate the main results from the multi-scale non-equilibrium Green's function formalism numerical simulation results and their physical interpretations for the graphene device's various aspects of variabilities in the device simulation result. 

\section*{Result and Discussion}
\label{section:Result and Discussion}

We have employed multi-scale, multi-physics-based bottom-up, non-equilibrium Green's function mechanism based quantum transport simulation techniques to investigate various device aspects and deduce the observable and measurable at the final stage in the computationally simulated devices. The scattering nature and origin depend on graphene's different fabrication process steps, the device orientation and substrate, and surface encapsulation in the device preparation step. In a recent room-temperature operational demonstration of graphene nanoribbon tunnel field-effect transistors, electrostatic doped gating is used to make the tunnel junction. \cite{hwang_room-temperature_2019} We have used the $ \mathrm {P-D} $ orbital tight-binding model, which represents the edge effects by explicitly including the passivated hydrogen in the Hamiltonian matrix. The carbon atom is represented by three $\mathrm{P_{z}}$, $ \mathrm{D_{yz}} $, and $ \mathrm {D_{zx}} $ orbitals. The simple single orbital $ \mathrm {P_z} $ tight-binding model works well for the two-dimensional graphene sheet. \cite{boykin_accurate_2011} The quantum transmitting boundary method (QTBM) is a purely ballistic charge transport model in the space of quantum propagating lead modes. \cite{ting_multiband_1992} In this method, the calculations numerical load is less than the ballistic non-equilibrium Green's function formalism or Recursive Green's Function (RGF) algorithm where all the modes consider. Furthermore, a small rank of original tight-binding achieves by incorporating incomplete spectral transformations of non-equilibrium Green's function equations into the Hilbert space. \cite{zeng_low_2013} In all ballistic non-equilibrium Green's function scenario a low rank approximation based contact block reduction (CBR) method, \cite{mamaluy_efficient_2003} and mode space approach, \cite{wang_three-dimensional_2004} are formulated to reduce the numerical load. Poisson's equation discretizes through the finite-difference method. For the source and drain contact boundary condition modeling, the Von Neumann scheme used the initial electric field value to be zero in the normal direction. Standard Newton–Rapson calculates and distributes new potential to the contact in the iteration. For the gate contact Dirichlet, the open boundary condition is employed, and potential value is kept fixed to $ V_\mathrm{GS} - \Phi_\mathrm{MS} $ value, here $ V_\mathrm{GS} $ is Gate to source voltage and $ \Phi_\mathrm{MS} $ is the metal work-function of the gate electrode, The value for $ \Phi_\mathrm{MS} $ is 4.2 eV. The channel is assumed ballistic in this calculation. In reality, a phonon-assisted current path is always available in the devices. Therefore, the subsequent section performs a more accurate current calculation incorporating electron-phonon scattering through self-consistent Born approximation (SCBA). Also, The multi-scale non-equilibrium Green's function simulation implementation on the supercomputer cluster was discussed there. We encourage the reader to deep dive into the relevant section of the article. Here for the brevity of time, first, we will discuss the main results from numerical calculation.

\subsection*{Quantum Transmitting Boundary Transport}

\begin{figure}[!htbp]
\centering
\begin{subfigure}{0.19\textwidth}\centering \includegraphics[width=\textwidth]{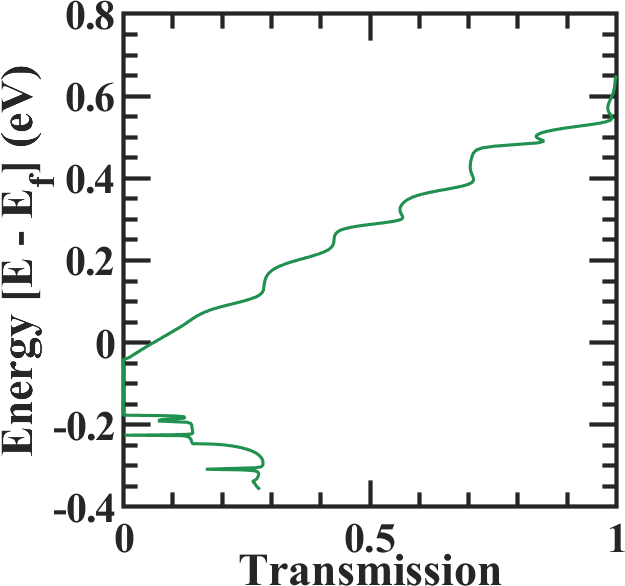}\subcaption{Transmission profile 1}\label{fig-QTBM-2-k}\end{subfigure}
\begin{subfigure}{0.19\textwidth}\centering \includegraphics[width=\textwidth]{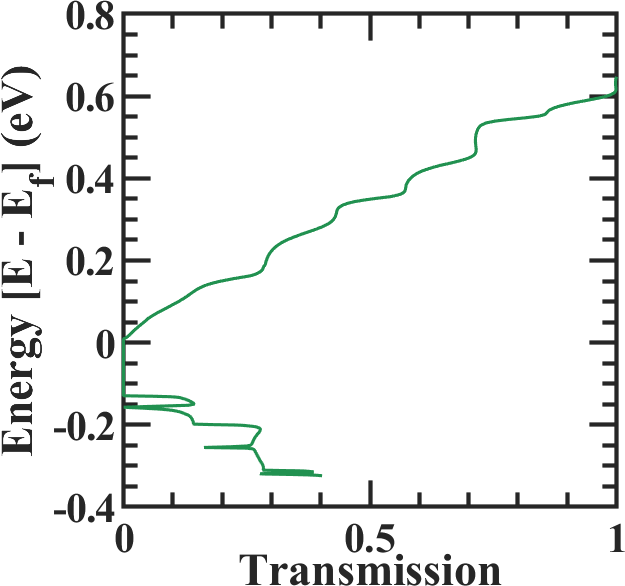}\subcaption{Transmission profile 2}\label{fig-QTBM-2-o}\end{subfigure}
\begin{subfigure}{0.19\textwidth}\centering \includegraphics[width=\textwidth]{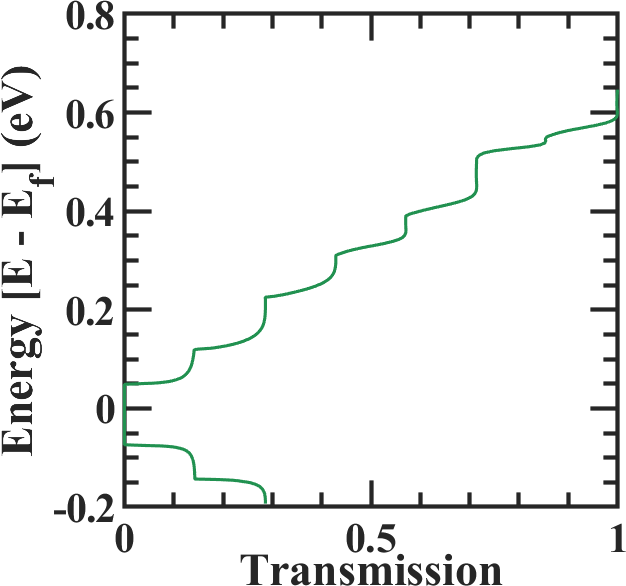}\subcaption{Transmission profile 3}\label{fig-QTBM-2-l}\end{subfigure}
\begin{subfigure}{0.19\textwidth}\centering \includegraphics[width=\textwidth]{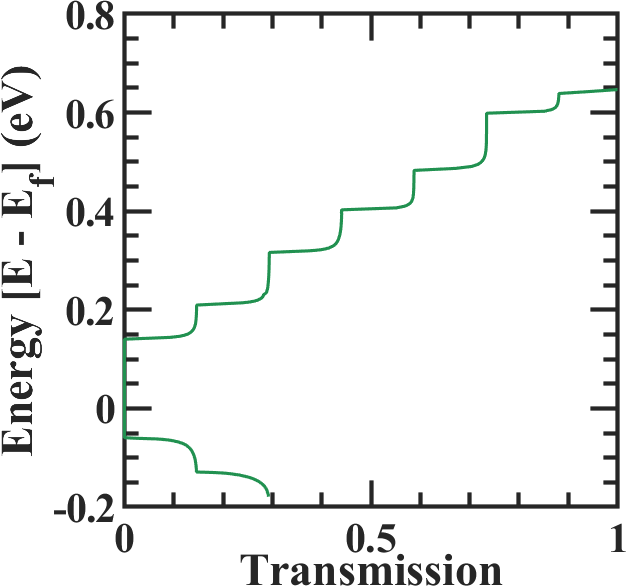}\subcaption{Transmission profile 4}\label{fig-QTBM-2-m}\end{subfigure}
\begin{subfigure}{0.19\textwidth}\centering \includegraphics[width=\textwidth]{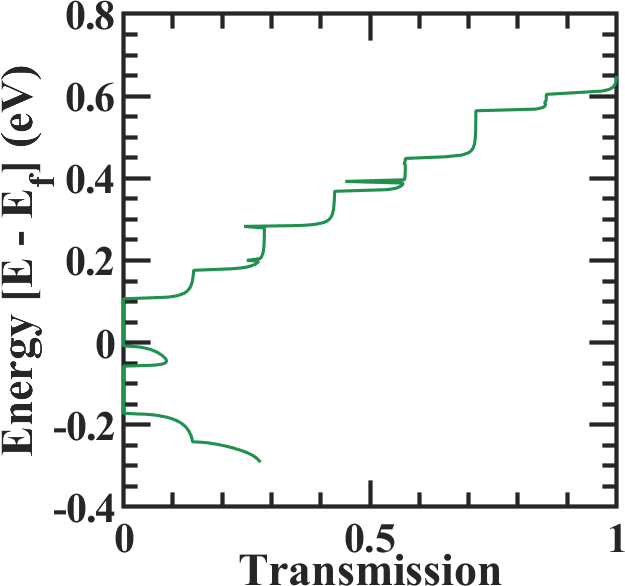}\subcaption{Transmission profile 5}\label{fig-QTBM-2-n}\end{subfigure}
\begin{subfigure}{0.19\textwidth}\centering \includegraphics[width=\textwidth]{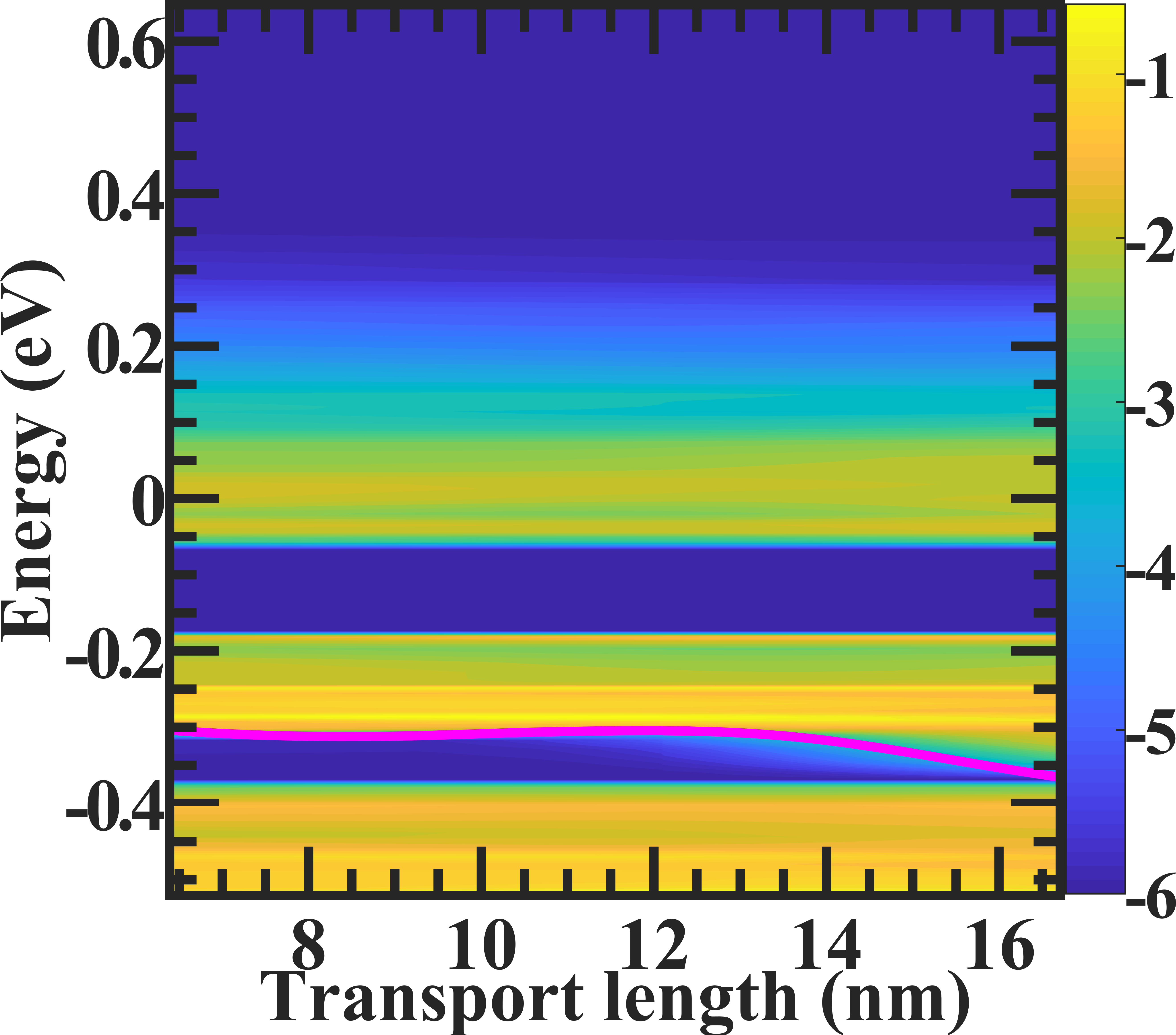}\subcaption{energy-resolved flux density profile 1}\label{fig-QTBM-2-p}\end{subfigure}
\begin{subfigure}{0.19\textwidth}\centering \includegraphics[width=\textwidth]{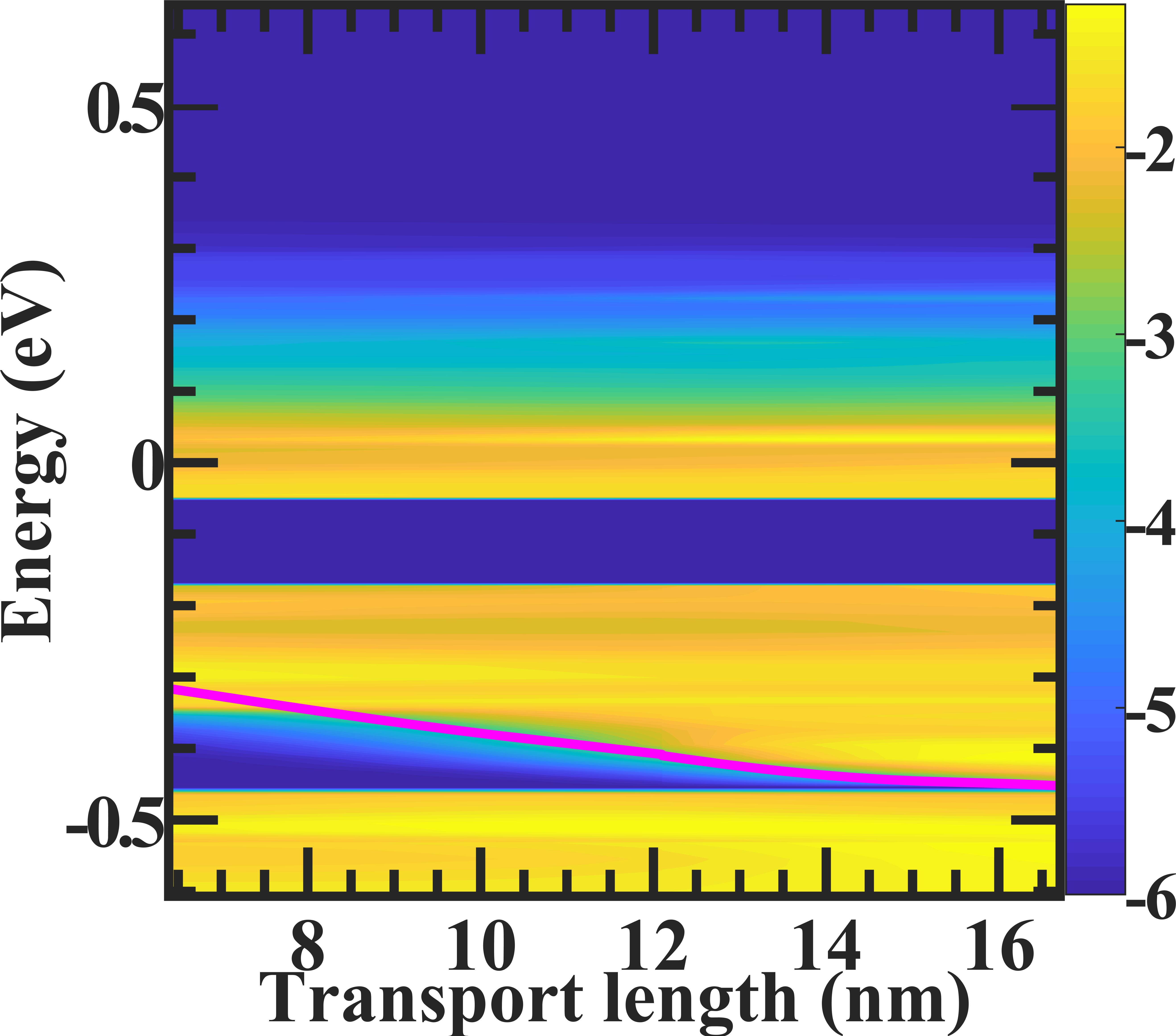}\subcaption{energy-resolved flux density profile 2}\label{fig-QTBM-2-t}\end{subfigure}
\begin{subfigure}{0.19\textwidth}\centering \includegraphics[width=\textwidth]{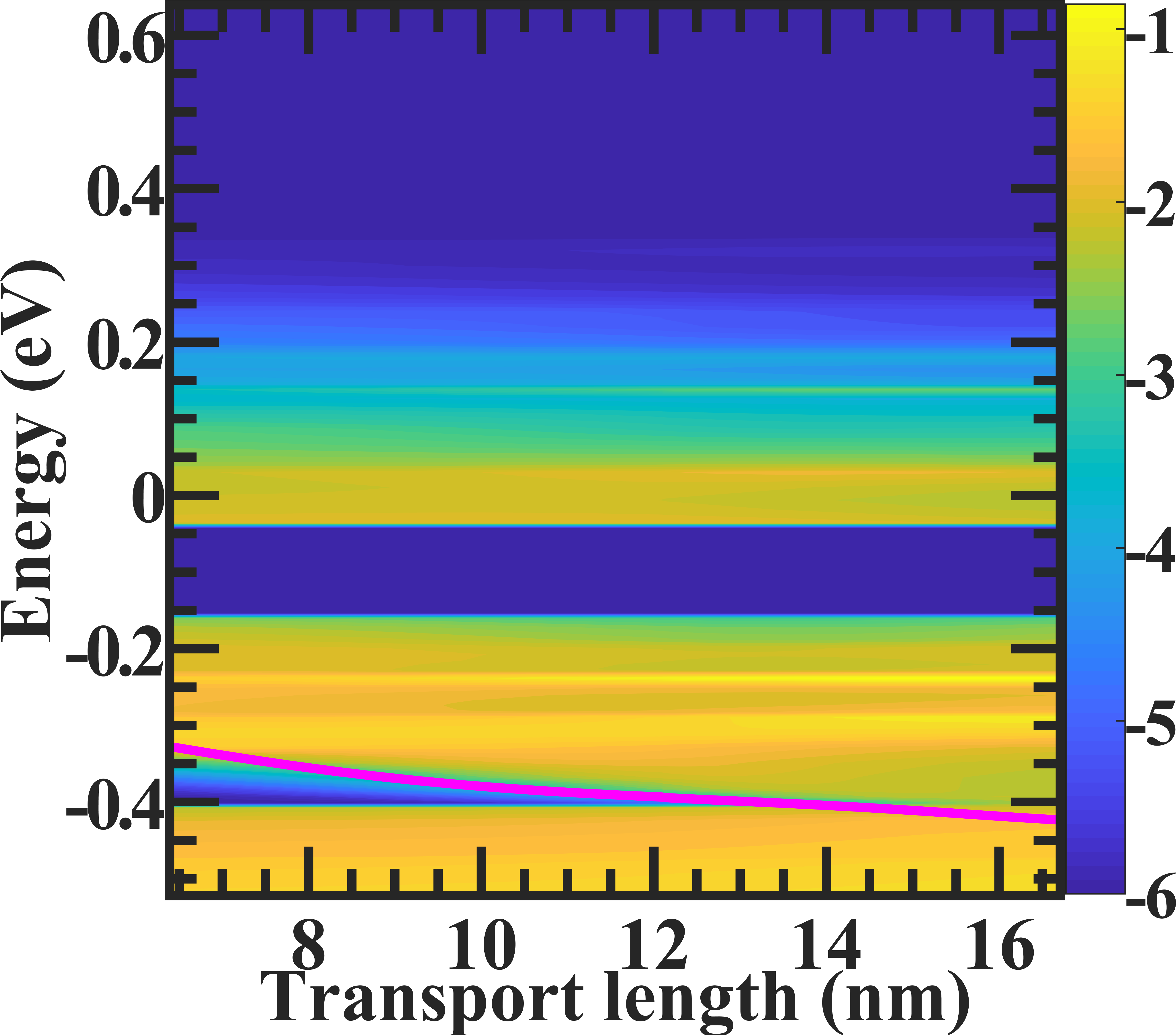}\subcaption{energy-resolved flux density profile 3}\label{fig-QTBM-2-s}\end{subfigure}
\begin{subfigure}{0.19\textwidth}\centering \includegraphics[width=\textwidth]{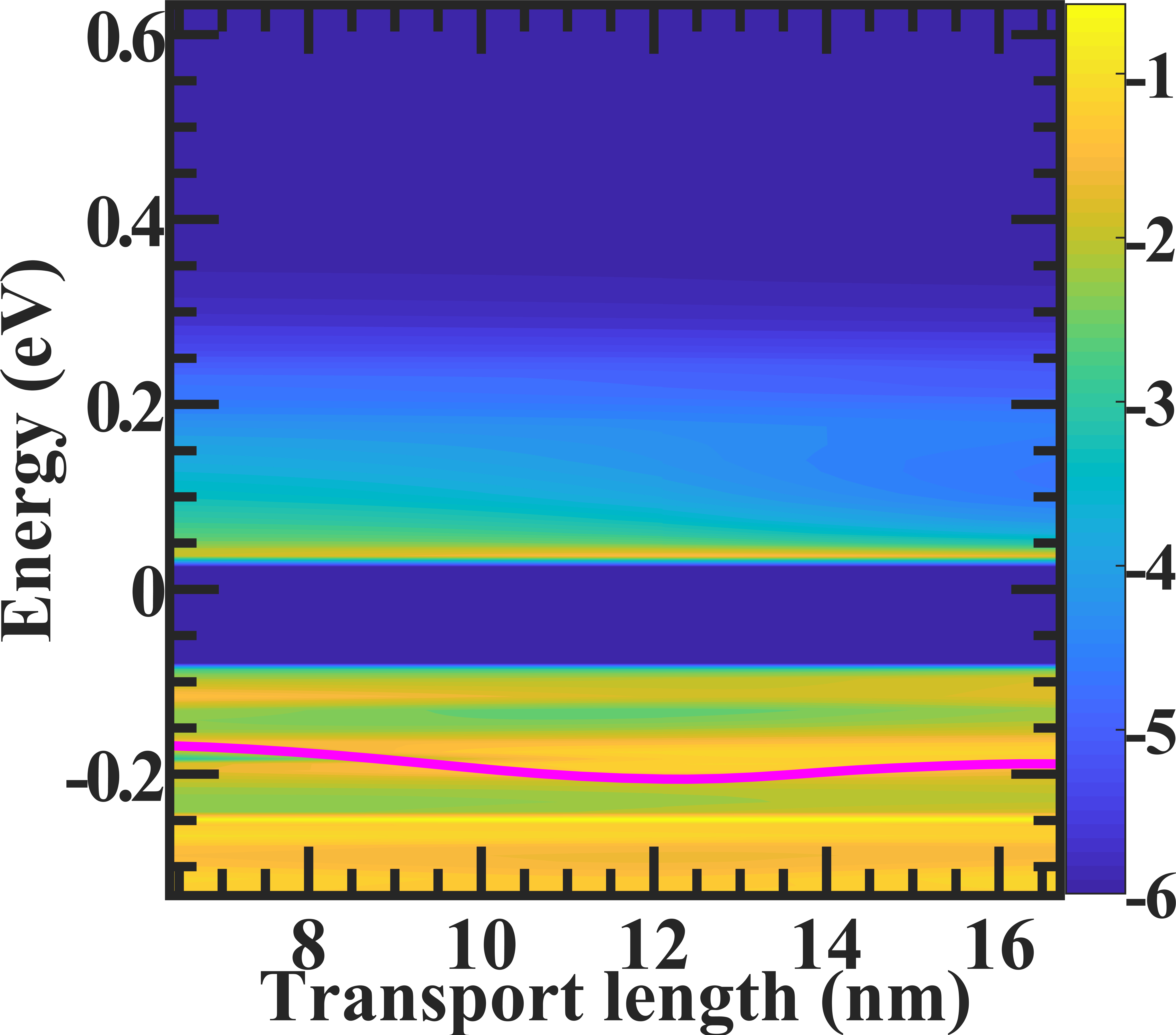}\subcaption{energy-resolved flux density profile 4}\label{fig-QTBM-2-q}\end{subfigure}
\begin{subfigure}{0.19\textwidth}\centering \includegraphics[width=\textwidth]{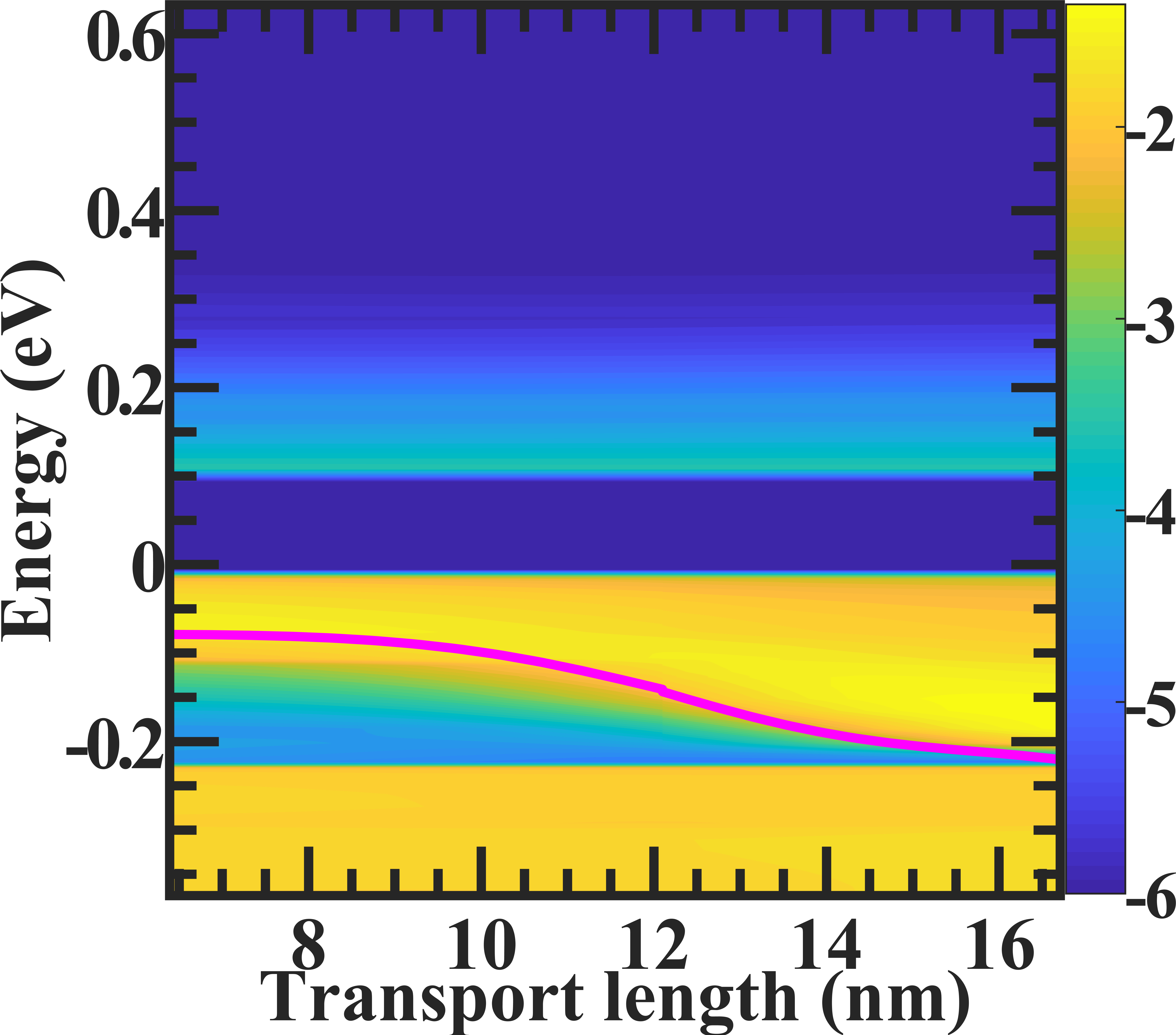}\subcaption{energy-resolved flux density profile 5}\label{fig-QTBM-2-r}\end{subfigure}
\begin{subfigure}{0.24\textwidth}\centering \includegraphics[width=\textwidth]{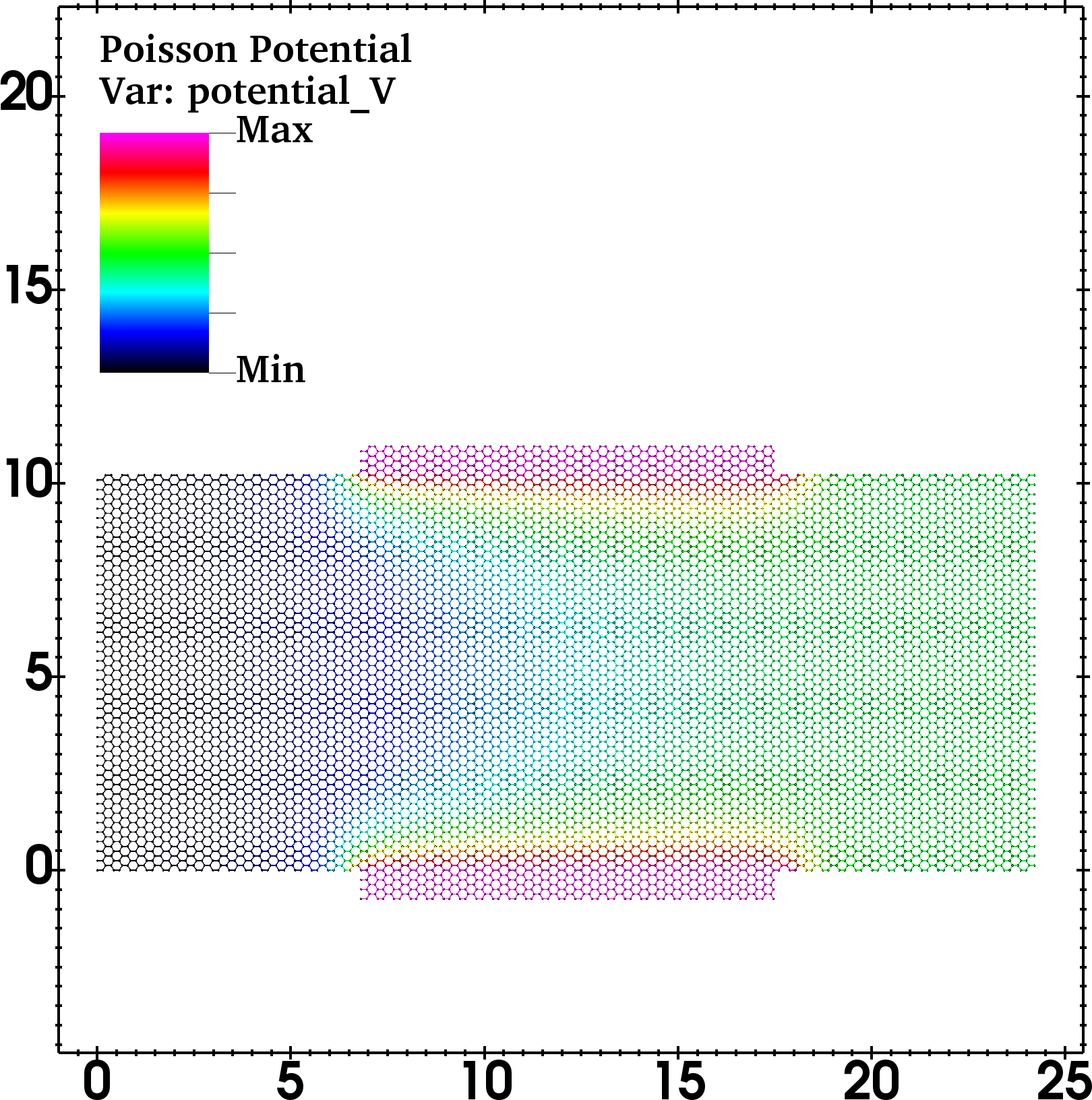}\subcaption{Poisson device at gate bias sweep of +2 Volt}\label{fig-QTBM-2-af}\end{subfigure}
\begin{subfigure}{0.24\textwidth}\centering \includegraphics[width=\textwidth]{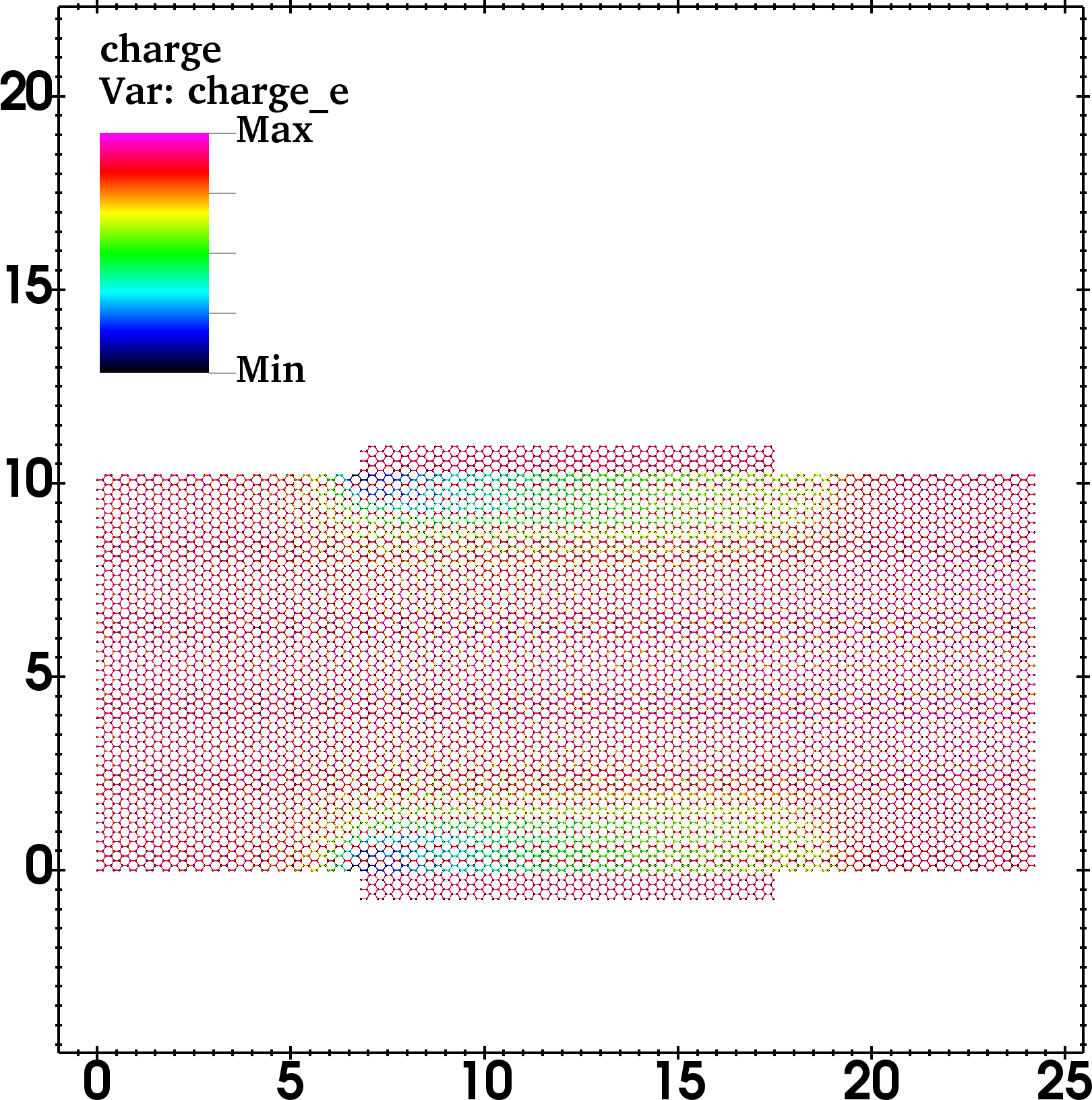}\subcaption{Charge distribution at gate bias sweep of +2 Volt}\label{fig-QTBM-2-ag}\end{subfigure}
\begin{subfigure}{0.24\textwidth}\centering \includegraphics[width=\textwidth]{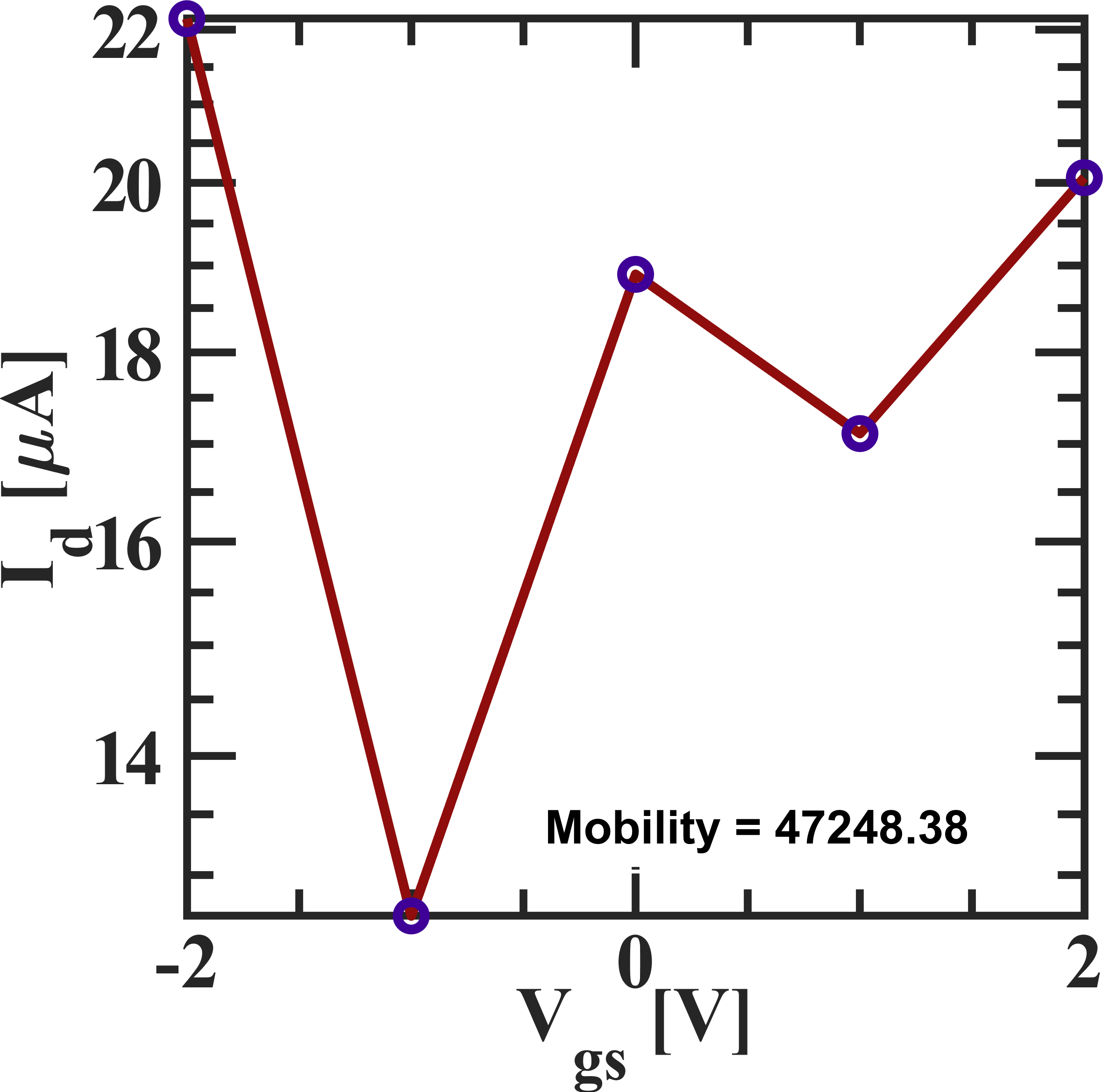}\subcaption{Current-voltage characteristics} \label{fig-QTBM-2-ae}\end{subfigure}
\begin{subfigure}{0.24\textwidth}\centering \includegraphics[width=\textwidth]{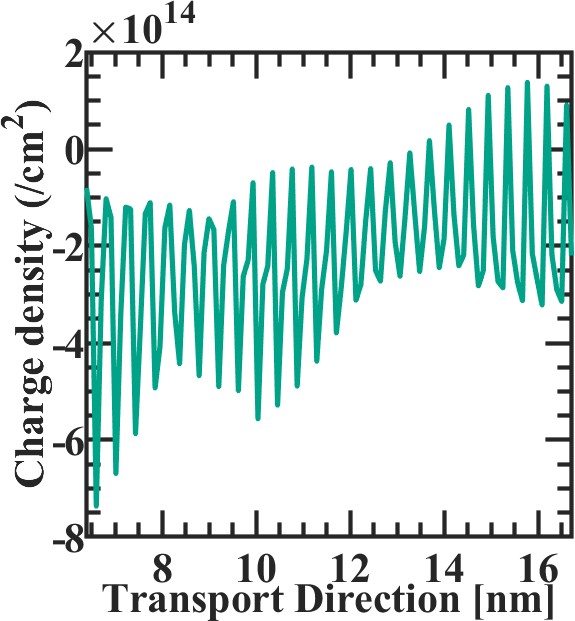}\subcaption{Charge variation redistribution at gate-field action} \label{fig-QTBM-2-z}\end{subfigure}
\caption{Transmission profile \cref{fig-QTBM-2-k} to \cref{fig-QTBM-2-n} with varying gate-field bias points, Corresponding energy-resolved flux density from \cref{fig-QTBM-2-p} to \cref{fig-QTBM-2-r} with varying gate-field bias points, Poisson potential profile of ballistic device \cref{fig-QTBM-2-af} at gate-field bias sweep of +2 Volt, Corresponding charge profile from \cref{fig-QTBM-2-ag} at gate-field bias sweep of +2 Volt, Current-voltage characteristics \cref{fig-QTBM-2-ae} of a ballistic device, Charge variation redistribution \cref{fig-QTBM-2-z} in the conducting device due to gate-field action at one bias point condition}
\label{fig-QTBM-2}
\end{figure}

The quantum transmitting boundary method (QTBM) is a purely ballistic charge transport algorithm in the space of quantum propagating lead modes. In this time-resolved transport method, wave-function propagates in time with a time-dependent potential to achieve a steady-state solution. The energy grid is homogeneous and contains multiple injection energies solved in parallel with the time propagator solver algorithm. \cite{ting_multiband_1992} In this approach, calculations numerical load is less than the ballistic non-equilibrium Green's function formalism or Recursive Green's Function (RGF) algorithm where all the propagating modes consider for transport. Furthermore, a small rank of original tight-binding achieves by incorporating incomplete spectral transformations of non-equilibrium Green's function equations into the Hilbert space. \cite{zeng_low_2013} 
We simulated the ballistic device with a gate bias sweep of -2 Volt to +2 Volt in step 1 Volt with a transport oxide barrier in the lateral direction, and a maximum source-drain bias of 0.8 Volt was applied. For the first bias point, the semi-classical initial guess uses, and after full step size, the Newton-Raphson method operates to achieve convergence. The second bias point uses the initial value from the previous solution loop. The extrapolation method based upon the last two solutions uses for the third and subsequent following bias points projection. The convergence scheme takes a maximum of 30 total iterations and five bias points simulated. In the complete step size method, the divergence of the solution protects by the initial semi-classical guess's quality. We need an efficient energy grid, a well-thought modest initial guess, and a self-consistent algorithm to solve the nonlinear Poisson self-consistent loop. Deploying an inhomogeneous energy grid resolves sharp features but increases overhead computations. We have used the adaptive energy grid, which gives an optimal solution. A modest initial guess, close to the final solution, will significantly reduce the convergence cycle. For the device calculations, the jellium model of background doping assumed a hundred percent ionization efficiency, and we have not incorporated the substitutional doping. The \cref{fig-QTBM-2} correspond to a simulated graphene device, and its electrical characteristics with channel length $\mathrm{L_G}$ of 10.3189 nm, source length $\mathrm{L_S}$ and drain length $\mathrm{L_D}$ of 6.391 nm with typical uniform doping density of $ 50  \times 10^{13} $ per cm$^2$ observe in the CVD/PECVD roll-to-roll batch sample. In the roll-to-roll CVD/PECVD graphne production, impurity doping density varies from ultra-clean and pure batch of $ 1  \times 10^{12} $ per cm$^2$ to extremely dirty and rough sample of $ 50 \times 10^{14} $ per cm$^2$. The side gate oxide thickness $\mathrm{t_{ox}}$ is $ 0.76 $ nm on the each side of device and gate dielectric constant $ \epsilon_{r} $ is $ 3.9 $. We have chosen the transport dimension of the test device, keeping sanity that a momentum relaxation length in a typical low-dimensional material is $ 10's $ of nanometer for the doping density used in the structure. Similarly, the test device width is broad enough to test the hall bar field-effect device action with minimal confinement, inducing a nanoribbon-type broader bandgap opening with the edge effect reduced to a minimum limit. The gate size keeps small as no transport happens in the lateral direction and is used only to provide an electrostatic gate-field action. Due to the effect of the induced charges by the parallel plate classical capacitance action, this gate-field will redistribute the charge in the graphene device by depleting or accumulating action based on the polarity of the gate-field. This gate-field will also alter the intrinsic doping density of graphene sheets by gate-induced carrier concentration and hence intrinsic conductivity or resistivity shift from the original value in an electrically operating device contacted with the external reservoir or battery terminal. However, the gate size can be expanded perpendicular to the transport direction but will add more atoms to the test device and penalize the simulation with computational overhead. The primitive unit cell has four atoms per cell and 10426 atoms simulated by a finite element mesh of 41,704 point domain size in the simulated device. The $ \mathrm {P-D} $ tight-binding model contains three orbitals, namely carbon $\mathrm{P_{z}}$, and carbon-hydrogen passivated $\mathrm{D_{yz}}$, $\mathrm{D_{xz}}$ orbitals. Therefore total degree of freedom in hamiltonian is 31,278 variable-sized. The gate oxide is treated as an impenetrable potential barrier in the Poisson equation in the device simulation due to the high graphene to oxide barrier height in the band alignment. \cite{robertson_band_2006} Nevertheless, a trivial penetration of the wave function into the thin gate oxide or surrounding dielectric encapsulation layer can significantly alter the effective electrical characteristic length scale. Therefore, proper care should be taken to adjust the desired bound state energies. We have analyzed the device operation to design a more suitable electrostatic control with optimal current-voltage characteristics and mobility. We have investigated the device operation at five different gate field-bias point sets and plotted the transmission profile in \cref{fig-QTBM-2-k} to \cref{fig-QTBM-2-n}; we have calculated the transmission $ T (\boldsymbol{k}_{\mathrm{t}};E) $ from \cref{eq-B52} and plotted the transmission profile with the varying applied gate-field points. In \cref{fig-QTBM-2-p} to \cref{fig-QTBM-2-r}, we have plotted the energy-resolved flux density profile superimposed color contour plots for the various gate bias point respectively for comparison. The propagating transmission mode occupation profile superimposes on top of the plot's local density of states (LDOS). In \cref{fig-QTBM-2-af}, \cref{fig-QTBM-2-ag}, we plotted the Poisson potential profile of the simulated device and its charge distribution at a gate bias sweep of +2 Volt. In \cref{fig-QTBM-2-ae}, we have plotted $\mathrm{I_{ON}}/\mathrm{I_{OFF}}$ current ratio distributions in the device via the $\mathrm{I_{D}}-\mathrm{V_{GS}}$ curve. In \cref{fig-QTBM-2-z}, we have plotted the corresponding charge variation distribution in the device due to gate-field action at one bias point condition.

We have made many observations from the device engineering point of view, which we discuss. First, we can observe that the contacted transmission is plotted on the Fermi-level resolved energy scale as clearly defined in the big thermalized reservoir bath. Furthermore, the propagating energy-resolved flux density is plotted on the energy scale as it is a localized property of the channel. Its quasi-Fermi-level varies across the transport direction to match source and drain contact Fermi-level. Also, the reservoir bath contacted electrons from all energy levels to try to kick in to influence the channel operation. Second, due to varying gate potential, the Fermi-level shift in the step of $\sim 0.05$ eV in the channel for the step of different gate potential points can be observed from these transmission profile plots from \cref{fig-QTBM-2-k} to \cref{fig-QTBM-2-n}. Third, graphene test device width dimensionality and acting gate potentials create a mini quasi-bandgap of size $\sim 0.125$ eV seen from transmission profile \cref{fig-QTBM-2-k}, \cref{fig-QTBM-2-o}, \cref{fig-QTBM-2-l} in the \cref{fig-QTBM-2-m} and \cref{fig-QTBM-2-n} configuration this mini bandgap tries to broden and split. The transmission falls to zero in this mini quasi-forbidden gap, as seen in all the transmission plots. Fourth, with the applied gate potential action, this dark-blue mini quasi-bandgap also shifts up and down with a quasi-Fermi-level shift in the channel in the step of $\sim 0.05$ eV, as seen in the propagating energy-resolved flux density plots with a dark blue region shift from \cref{fig-QTBM-2-p} to \cref{fig-QTBM-2-r}. The sharpness of its edge color transiting from dark blue to golden yellow depends upon the corresponding transmission plots for that bias point. In the case of transmission plot \cref{fig-QTBM-2-l} and \cref{fig-QTBM-2-m} have clearly defined mode step-plateau in the energy window, and modes not able to interact in momenta and energy and in \cref{fig-QTBM-2-r} and \cref{fig-QTBM-2-q} transition from gap state to conducting state is slow. It physically signifies that electrons propagate in their corresponding mode with the corresponding constant velocity on which they are injected from the contact. However, in \cref{fig-QTBM-2-k} and \cref{fig-QTBM-2-o}, transmission mode does not plateau and becomes linearized by mixing energy and in corresponding \cref{fig-QTBM-2-t} to \cref{fig-QTBM-2-p} energy-resolved flux density transition is quick from dark blue to golden yellow. Furthermore, in \cref{fig-QTBM-2-n} transmission modes fluctuate with energy spikes, and the corresponding \cref{fig-QTBM-2-s} energy-resolved flux density also fluctuates with the irregular mix of the blue and yellow color region. Fifth, in the energy-resolved flux density plots, another fictitious quasi-gap that emerged towards the source end was distinguished by the pink color potential line at lower energy. However, it disappeared towards the drain end as drain potential increased along the transport length and gate potential is comparable to it; instead, a gate-source potential difference at the source end and propagation mode mixed up with phase exchange. Hence, the electron can spill into this fictitious-looking quasi-gap at the drain end. As discussed earlier, it can be inferred as the effective gate potential playing in this device dimension and fluctuating and trying to alter a more robust confinement-induced dark-blue mini quasi-bandgap. Sixth, Another piece of information is that as the energy increases though contacted transmission mode is available for the electrons to propagate, as seen in transmission plots up to $\sim 0.6$ eV. However, reservoir baths do not supply high-energy electrons to join those high-energy transmission modes in this test device simulation. Therefore, the energy-resolved propagating flux density quickly turns dark blue with a minimum value. However, this transition for mini quasi-bandgap upper edge and above in energy and mini quasi-bandgap lower edge and below in energy have entirely different transitions profile. The reason is that as the reservoir baths still provide the low energy electron, the energy-resolved propagating flux density color resolution is stark golden yellow as most transport happens with valence electron in graphene. Also, energy-resolved propagating flux density in \cref{fig-QTBM-2-r} to \cref{fig-QTBM-2-t}, the stark golden yellow curve shows a successive wave pattern on the drain edge as the wavefunction reflects at contact. It shows the quantum wave nature of simulation; this description is not available in the preview of the device's semi-classical or drift-diffusion bound treatment. The effect of contact was further investigated in the self-consistent Born transport study. In this regard, for reference fluctuating charge profile in the device operation is also plotted in the figure \cref{fig-QTBM-2-z} for one bias-gate voltage point in device operation and conduction and valance electron redistributed from the initial uniform charge density profile. In the \cref{fig-QTBM-2-af}, we have calculated and visualized the potential profile over the device cross-sections. There is a depletion of charges below the gate circumference. The charge distribution depends on the device's electrostatic design, the relative density of states, and the effective mass of different materials at the interface regions. Moreover, wave-function can penetrate the device's surrounding oxide layer or passivation layer at the interface. This leaking off wave function will fluctuate the electronic density in the device. The strong infusion of the wave function can increase the possibility of trap-assisted tunneling current in the gate electrode and divert the source to drain the current path. The presence of irregular defects can also influence the source, drain injection current, and overall current path in the channel. Also, high doping in the source and drain can cause a high energetic charge injection from the band trail and increase the thermal effects. In the current-voltage characteristics plot \cref{fig-QTBM-2-ae} in the electronic branch, there is a dip in current value from 19 $\mu A$ to 17 $\mu A$ at one voltage point. We have investigated it, and it arises due to a nonlinear Newton-Raphson initial guess and delayed convergence and does not have any insightful physical phenomena. We have another test device simulation with a reasonably smooth curve also. However, we deliberately include this plot to bring readers' attention to spurious behavior due to the high-performance scientific, mathematical library used to solve the non-equilibrium Green's function physical framework. The Drude mobility extract from the current-voltage characteristics and drive current is high as the device is highly doped and transport is ballistic. Next, we will simulate the low-doped channel device with a phase-coherent regime. 

\subsection*{Phase-coherent Recursive Green's Function Transport}

\begin{figure}[!htbp]
\centering
\begin{subfigure}{0.19\textwidth}\centering \includegraphics[width=\textwidth]{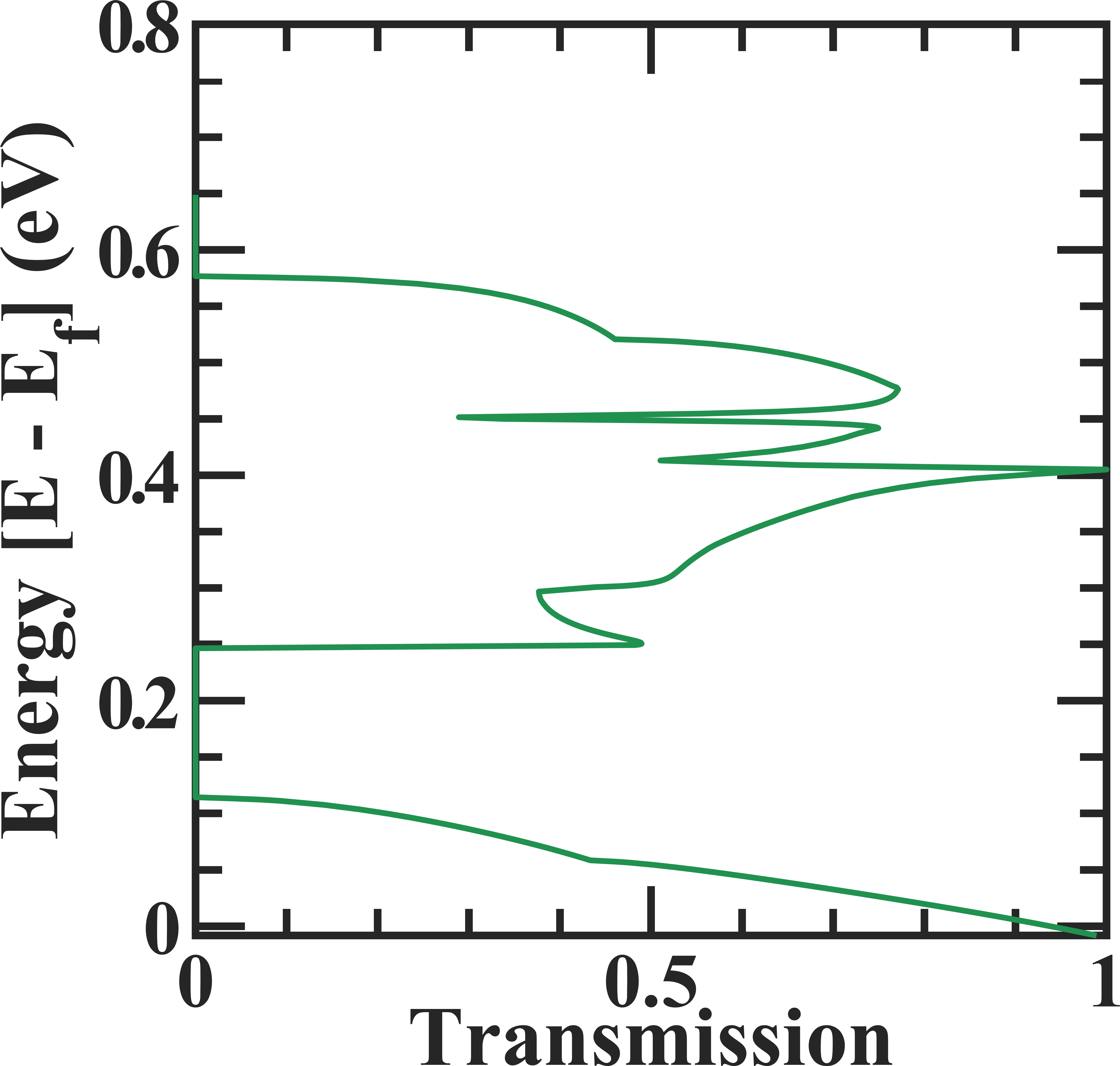}\subcaption{Transmission profile }\label{fig-RGF-1-h}\end{subfigure}
\begin{subfigure}{0.19\textwidth}\centering  \includegraphics[width=\textwidth]{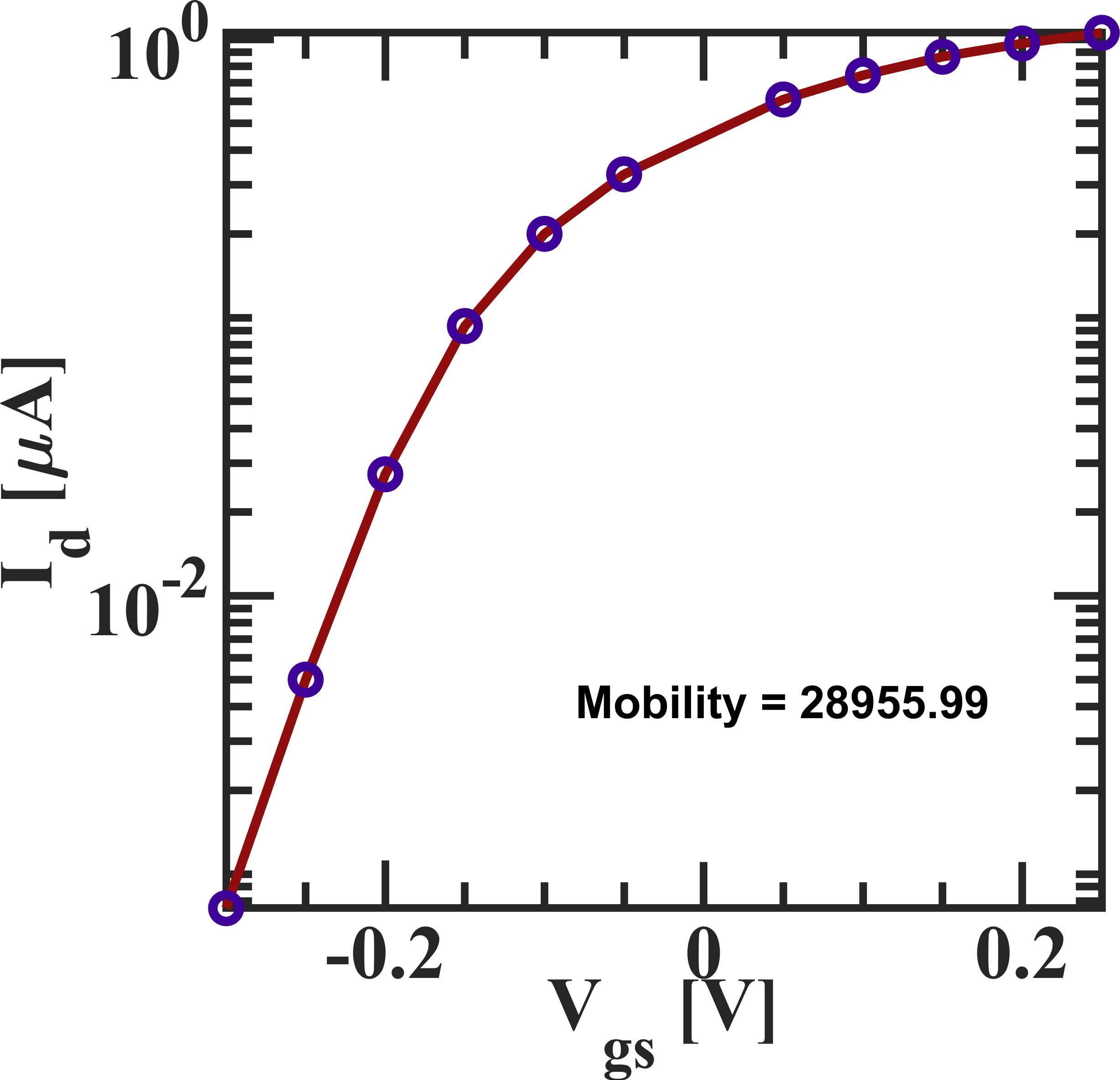}\subcaption{Current-voltage characteristics}\label{fig-RGF-1-g}\end{subfigure}
\begin{subfigure}{0.19\textwidth}\centering  \includegraphics[width=\textwidth]{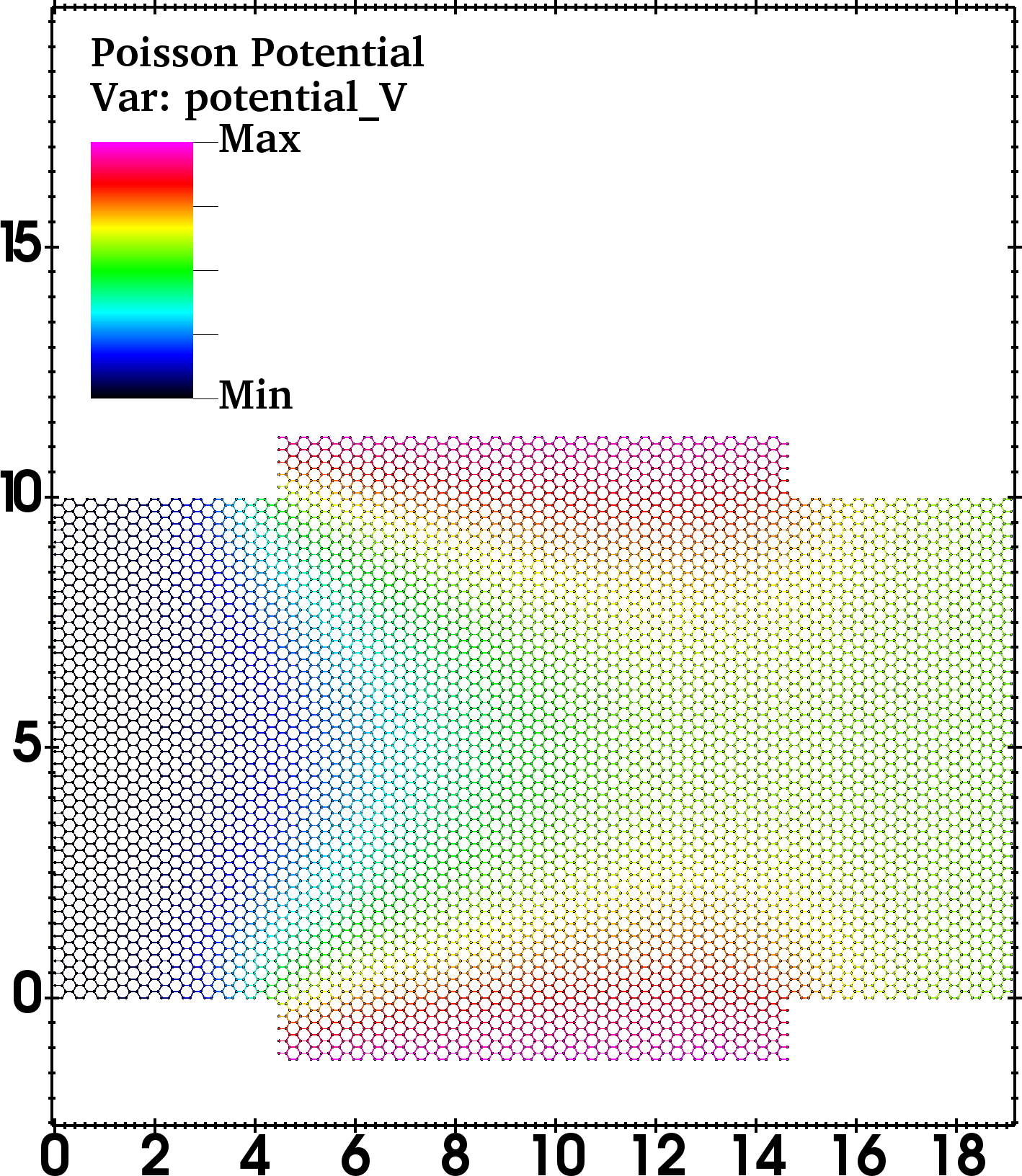}\subcaption{Device Poisson potential profile $\mathrm{V_{GS}}$ 0.25V}\label{fig-RGF-1-a}\end{subfigure}
\begin{subfigure}{0.19\textwidth}\centering  \includegraphics[width=\textwidth]{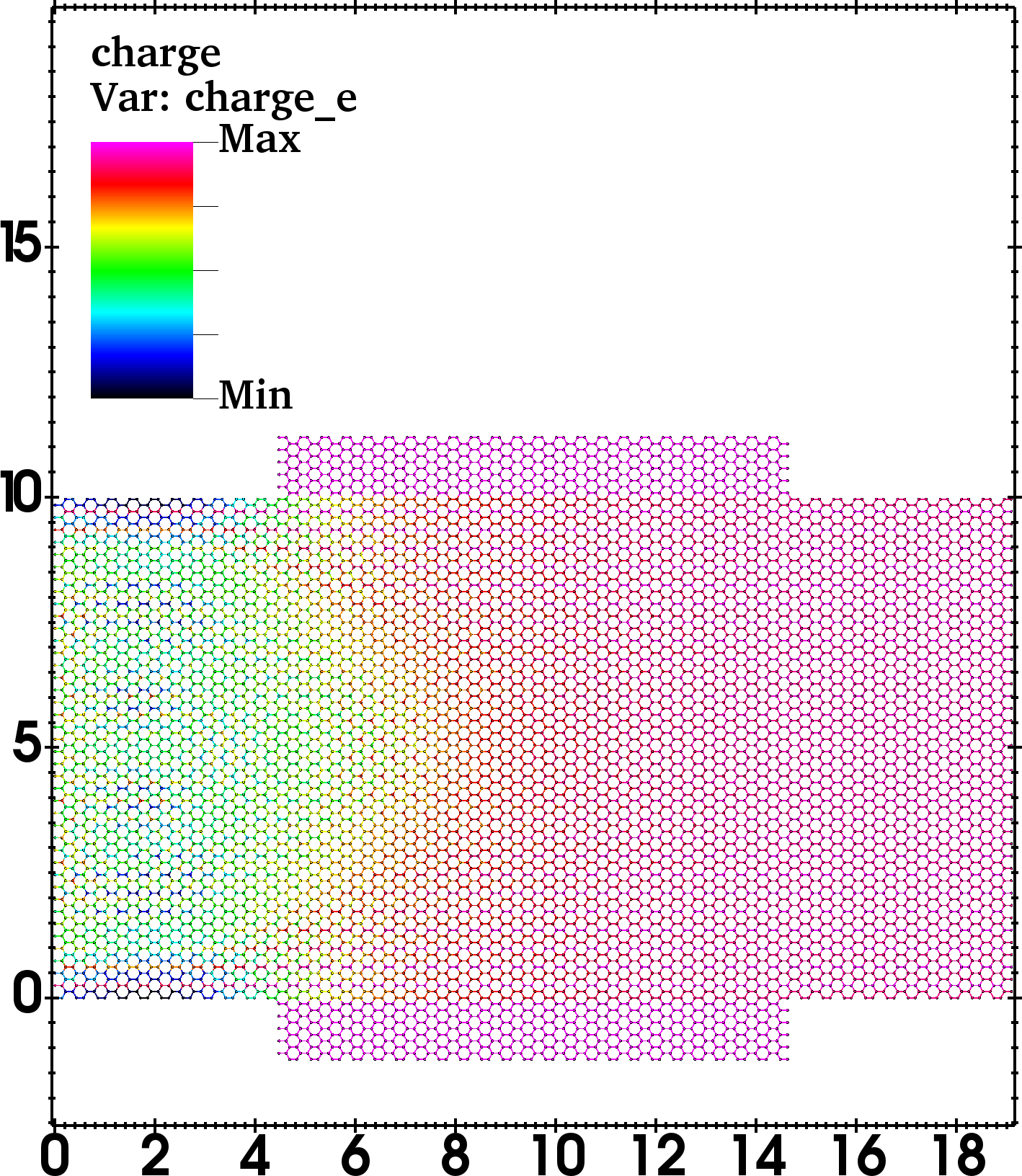} \subcaption{Device Free charge profile $\mathrm{V_{GS}}$ 0.25V}\label{fig-RGF-1-b}\end{subfigure}
\begin{subfigure}{0.19\textwidth}\centering \includegraphics[width=\textwidth]{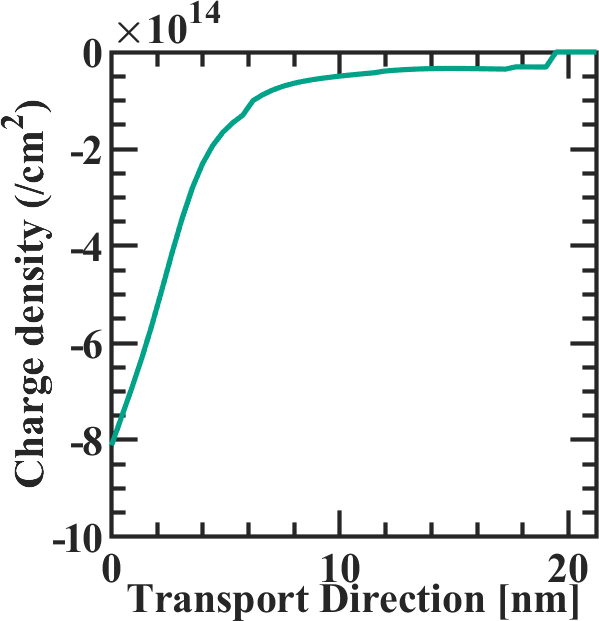}\subcaption{Charge variation redistribution at gate-field action} \label{fig-RGF-1-i}\end{subfigure}
\caption{Total transmission profile of simulated device in the phase-coherent transport regime \cref{fig-RGF-1-h}, Current-voltage characteristics \cref{fig-RGF-1-g}, Device Poisson potential \cref{fig-RGF-1-a} and charge profile \cref{fig-RGF-1-b} at $\mathrm{V_{GS}}$ +0.25V, Charge variation redistribution \cref{fig-RGF-1-i} in the conducting device due to gate-field action at one bias point condition}
\label{fig-RGF-1}
\end{figure}

In this section, we have calculated the transmission and current in the phase-coherent regime, \cite{buttiker_generalized_1985, meir_landauer_1992} from \cref{eq-B52} transmission $ T (\boldsymbol{k}_{\mathrm{t}};E) $ is given as $ \big[\Gamma_{ll_{1}}^{L}G_{l_{1}l_{2}}^{R}\Gamma_{l_{2}l_{3}}^{R}G_{l_{3}l}^{A}\big](\boldsymbol{k}_{\mathrm{t}};E) $. The carrier wave function is scattered back and forth by the interface potential barriers or the disorder potential and creates the standing waves in the transport direction. The tunneling between localized states governs the carrier's transport and the average transmission probability. Transport is in the localization regime, and with an increase in device length, transmission probability decreases exponentially. \cite{anderson_new_1980} The \cref{fig-RGF-1} correspond to a simulated graphene device in the phase-coherent transport regime, and its electrical characteristics with channel length $\mathrm{L_G}$ of 6 nm, source length $\mathrm{L_S}$ and drain length $\mathrm{L_D}$ of 6 nm with drain, with typical channel doping density of $ 5  \times 10^{13} $ per cm$^2$ observe in the CVD/PECVD roll-to-roll batch sample, source doping density of $ 50\times10^{13} $ per cm$^2$. The side gate oxide thickness $\mathrm{t_{ox}}$ is 1.3 nm on the each side of device and gate dielectric constant $ \epsilon_{r} $ is 3.9. We simulated the phase-coherent device with a gate bias sweep of -0.3 Volt to +0.25 Volt in step 0.05 Volt with a transport oxide barrier in the lateral direction, and a maximum source-drain bias of 0.2 Volt was applied. In the simulated device, the primitive unit cell has four atoms per cell and 8556 atoms simulated by a finite element mesh of 34,224 point domain size. The $ \mathrm {P-D} $ tight-binding model contains three orbitals, namely carbon $\mathrm{P_{z}}$, and carbon-hydrogen passivated $\mathrm{D_{yz}}$, $\mathrm{D_{xz}}$ orbitals. Therefore total degree of freedom in hamiltonian is 25,668 variable-sized. In the \cref{fig-RGF-1-h} we have plotted total transmission profile of simulated device in the phase-coherent transport regime. In \cref{fig-RGF-1-g}, we have plotted $\mathrm{I_{ON}}/\mathrm{I_{OFF}}$ current ratio distributions in the device via the $\mathrm{I_{D}}-\mathrm{V_{GS}}$ current-voltage characteristics curve. In the \cref{fig-RGF-1-a} and \cref{fig-RGF-1-b} we have plotted Poisson potential and free charge profile at $\mathrm{V_{GS}}$ 0.25V for the phase-coherent transport regime. In the \cref{fig-RGF-1-i} charge variation redistribution in the conducting device due to gate-field action at one bias point condition. 

\subsection*{Self-consistent Born Transport}

\begin{figure}[!htbp]
\centering 
\begin{subfigure}{0.24\textwidth}\centering  \includegraphics[width=\textwidth]{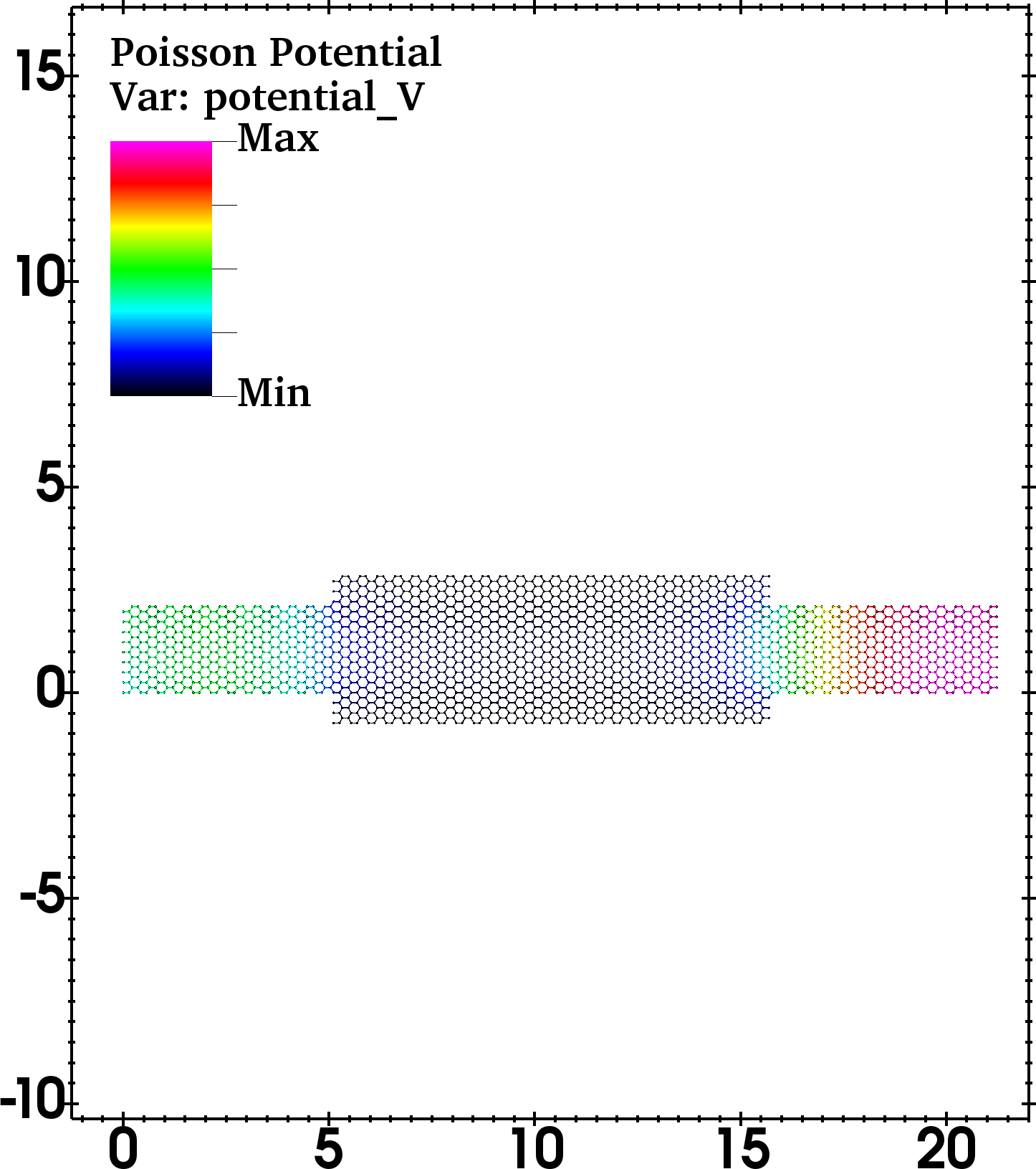}\subcaption{Device Poisson potential profile $\mathrm{V_{GS}}$0.25V}\label{fig-SCBA-1-a}\end{subfigure}
\begin{subfigure}{0.24\textwidth}\centering  \includegraphics[width=\textwidth]{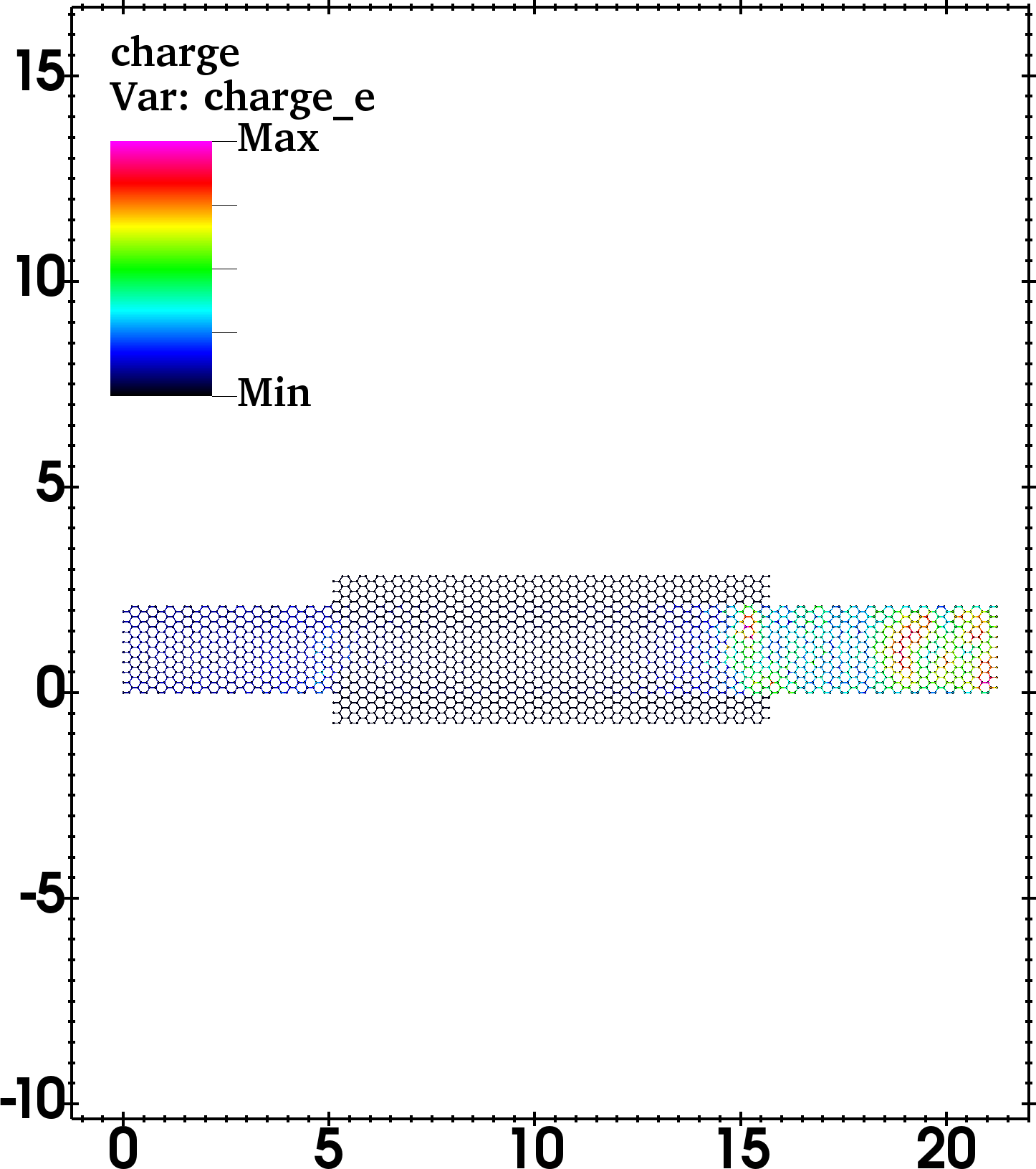}\subcaption{Device free charge profile $\mathrm{V_{GS}}$ 0.25V}\label{fig-SCBA-1-b}\end{subfigure}
\begin{subfigure}{0.24\textwidth}\centering  \includegraphics[width=\textwidth]{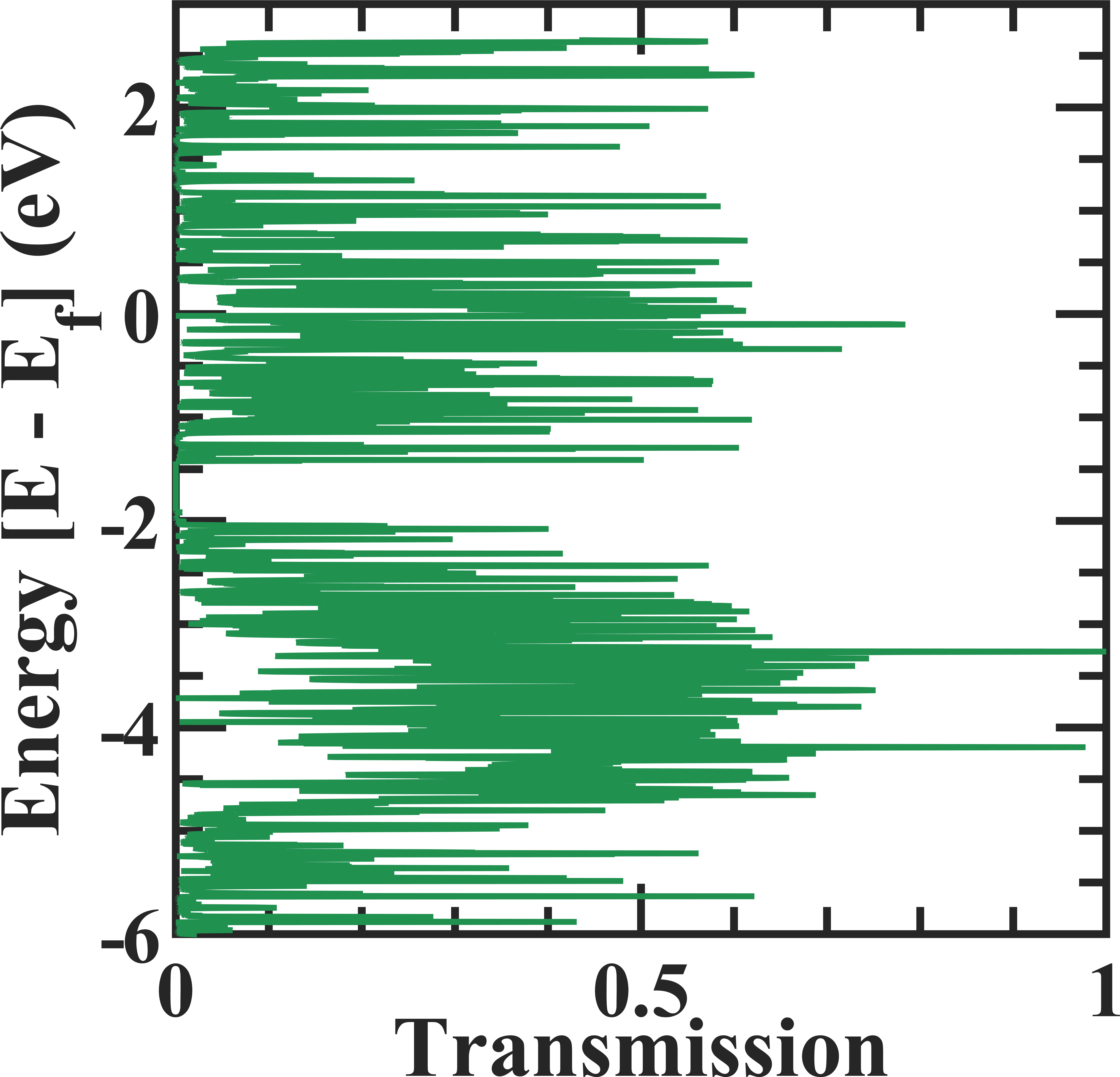}\subcaption{Device Transmission profile $\mathrm{V_{GS}}$ 0.25V}\label{fig-SCBA-1-c}\end{subfigure}
\begin{subfigure}{0.24\textwidth}\centering  \includegraphics[width=\textwidth]{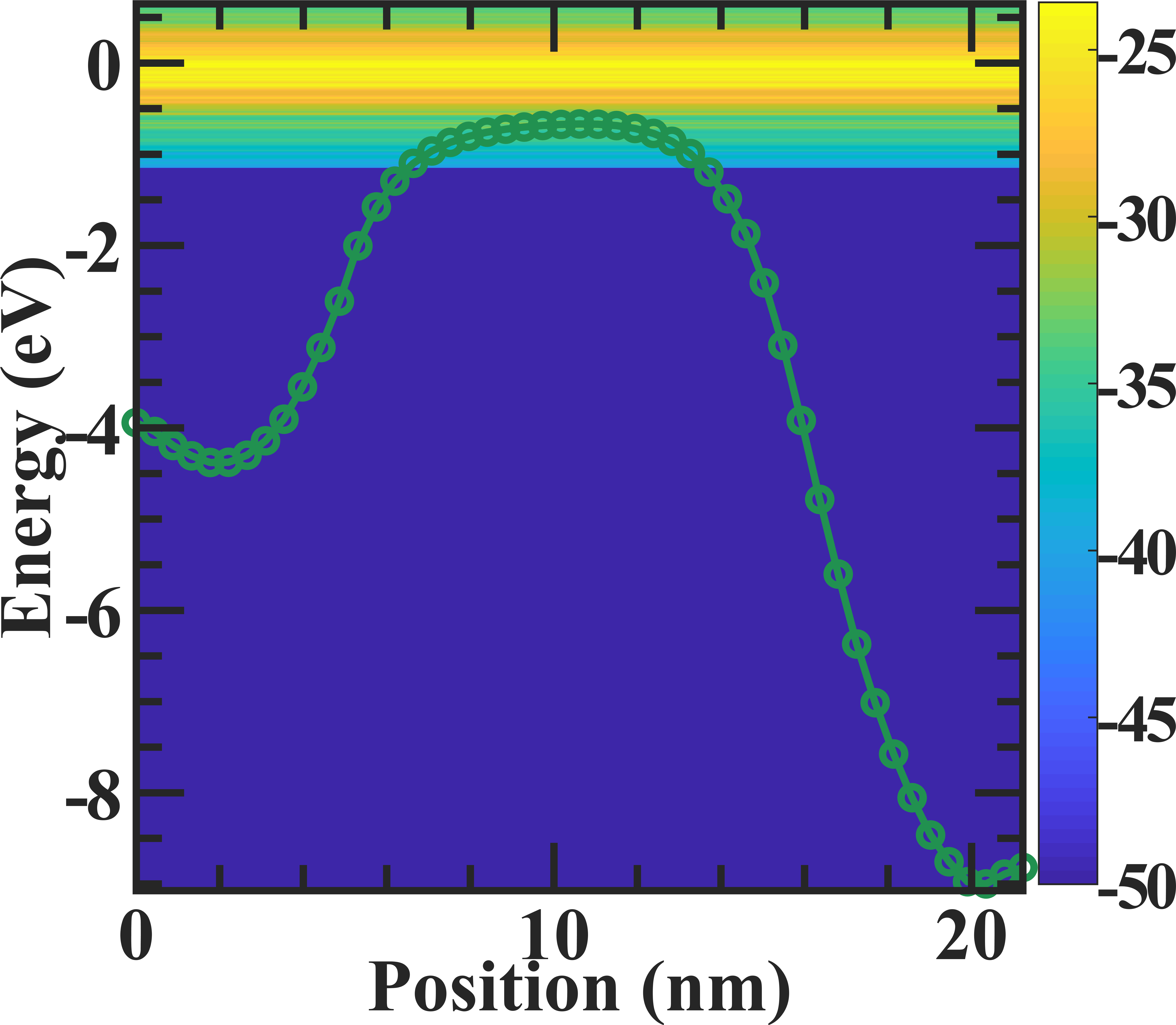}\subcaption{Device energy-resolved flux density $\mathrm{V_{GS}}$ 0.25V}\label{fig-SCBA-1-d}\end{subfigure}
\begin{subfigure}{0.24\textwidth}\centering  \includegraphics[width=\textwidth]{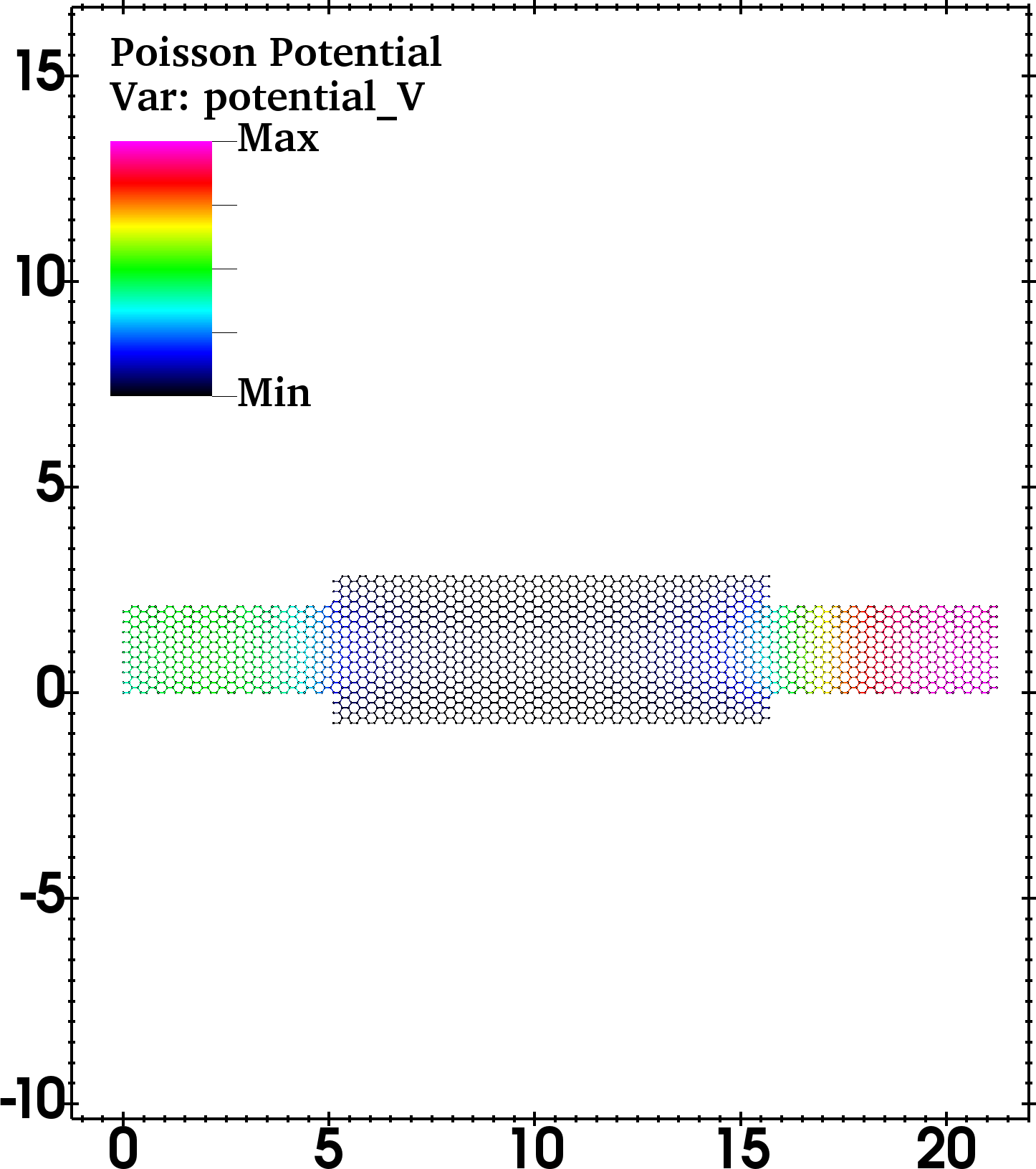}\subcaption{Device Poisson potential profile $\mathrm{V_{GS}}$ 0V}\label{fig-SCBA-1-e}\end{subfigure}
\begin{subfigure}{0.24\textwidth}\centering  \includegraphics[width=\textwidth]{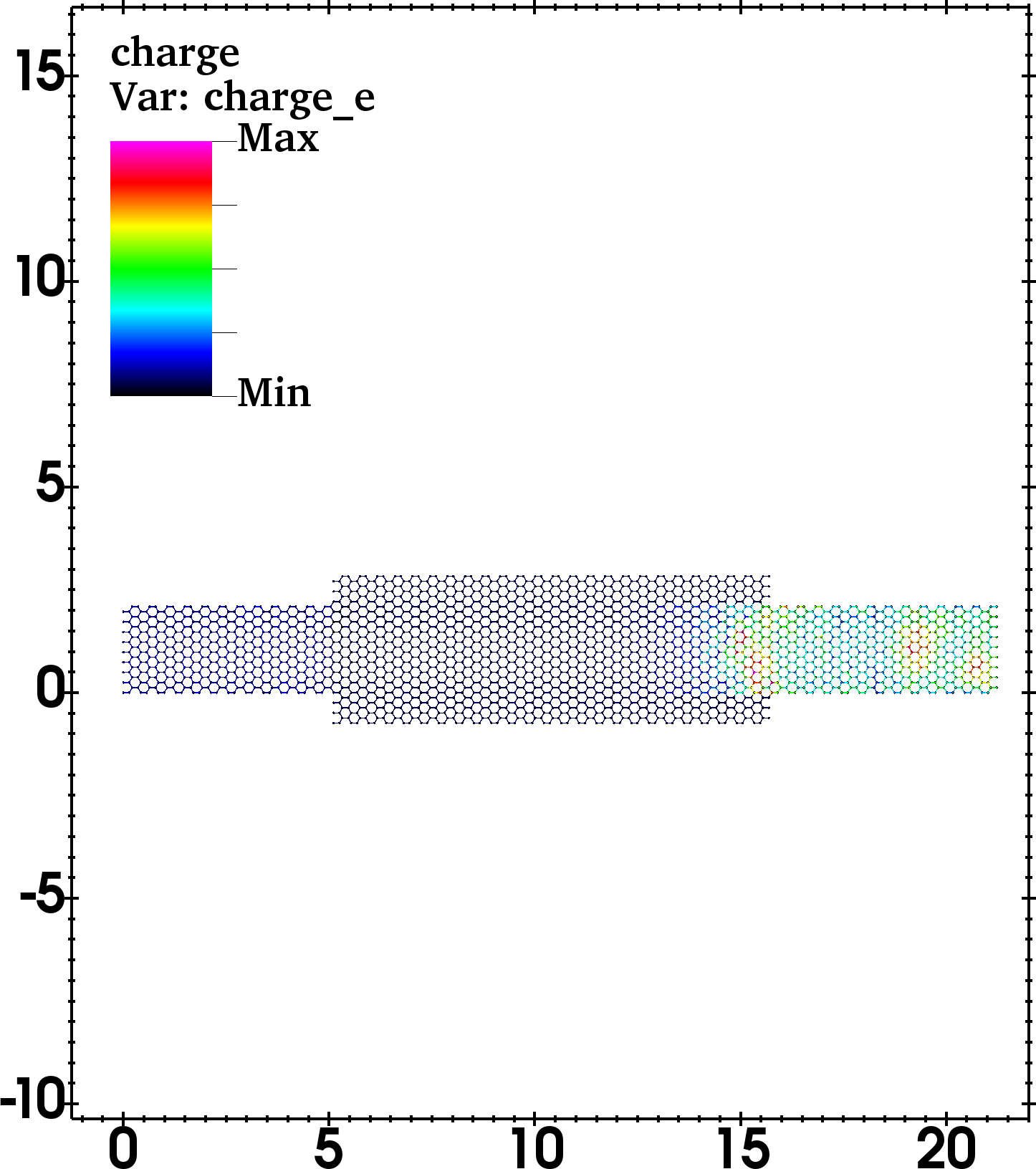}\subcaption{Device free charge profile $\mathrm{V_{GS}}$ 0V}\label{fig-SCBA-1-f}\end{subfigure}
\begin{subfigure}{0.24\textwidth}\centering  \includegraphics[width=\textwidth]{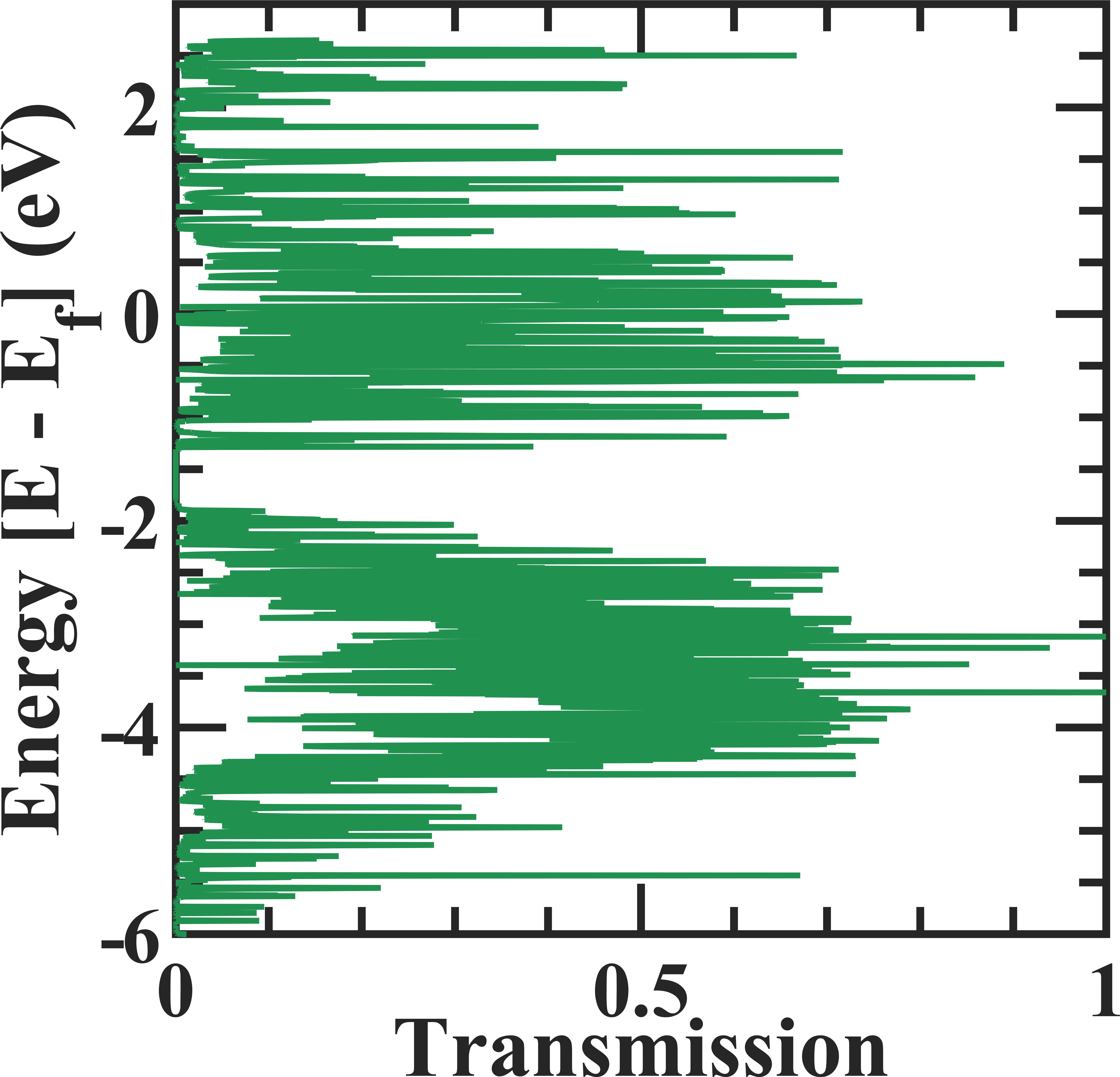}\subcaption{Device Transmission profile $\mathrm{V_{GS}}$ 0V}\label{fig-SCBA-1-g}\end{subfigure}
\begin{subfigure}{0.24\textwidth}\centering  \includegraphics[width=\textwidth]{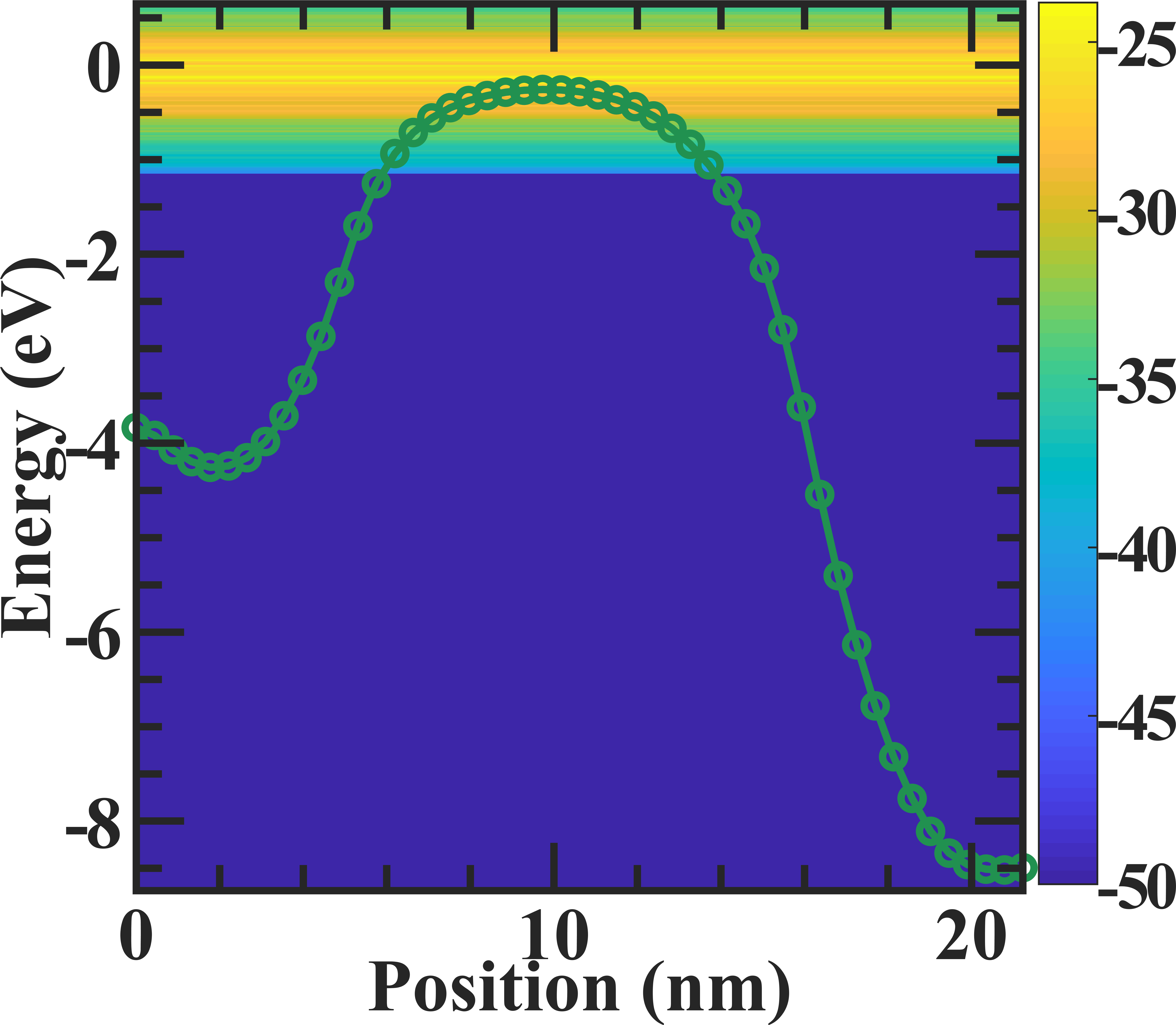}\subcaption{Device energy-resolved flux density $\mathrm{V_{GS}}$ 0V}\label{fig-SCBA-1-h}\end{subfigure}
\begin{subfigure}{0.24\textwidth}\centering  \includegraphics[width=\textwidth]{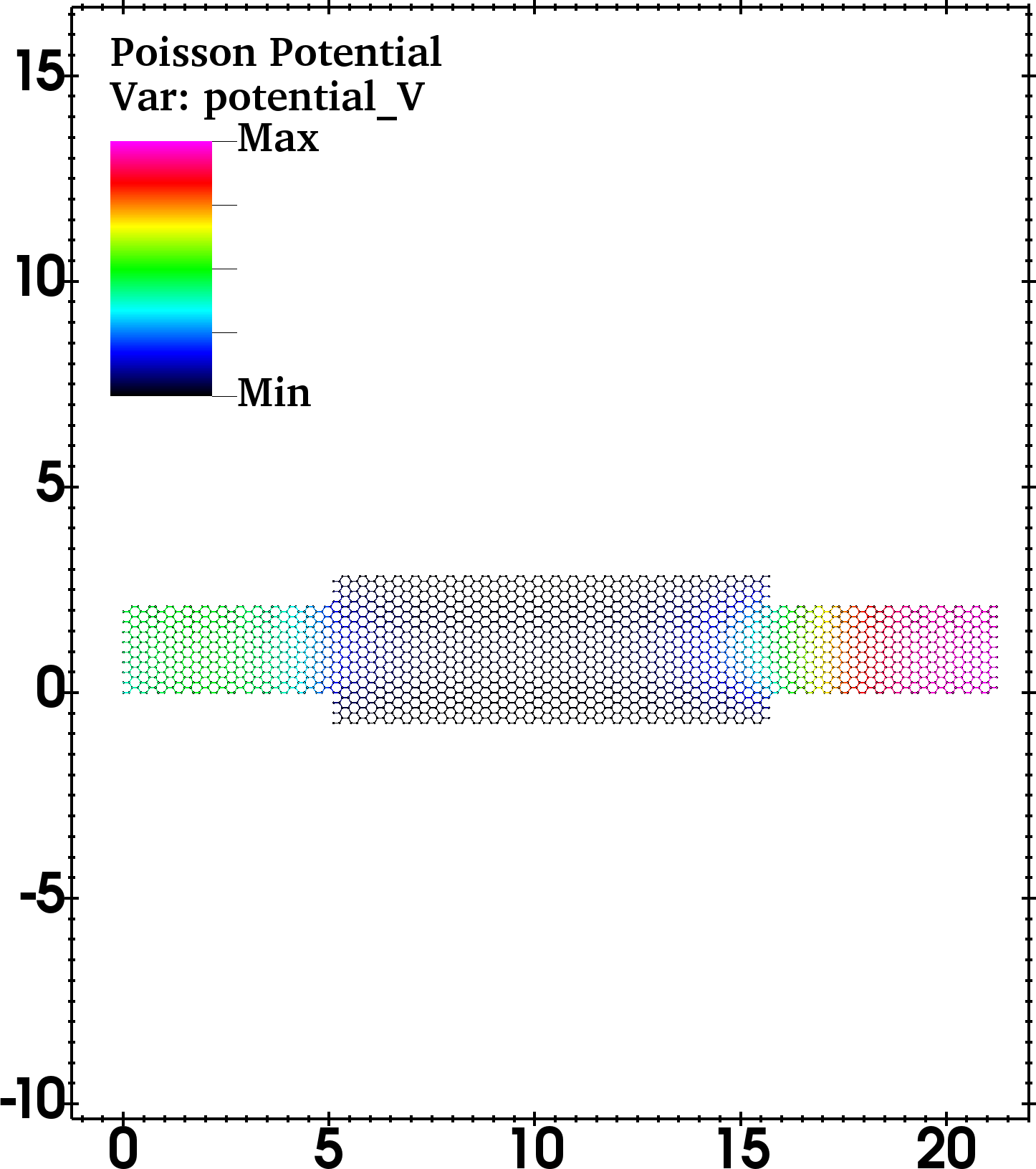}\subcaption{Device Poisson potential profile $\mathrm{V_{GS}}$ -0.25V}\label{fig-SCBA-1-i}\end{subfigure}
\begin{subfigure}{0.24\textwidth}\centering  \includegraphics[width=\textwidth]{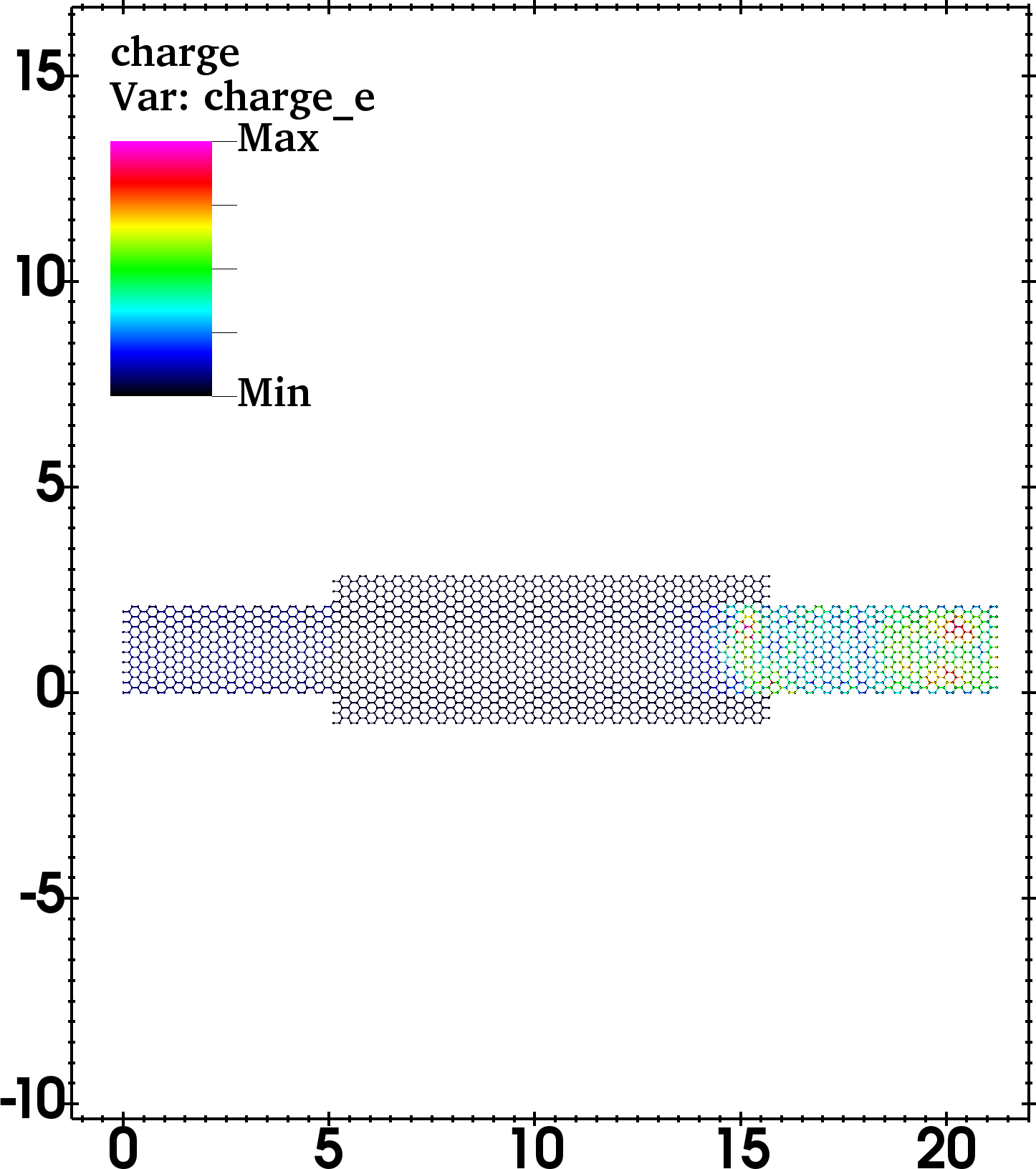}\subcaption{Device free charge profile $\mathrm{V_{GS}}$ -0.25V}\label{fig-SCBA-1-j}\end{subfigure}
\begin{subfigure}{0.24\textwidth}\centering  \includegraphics[width=\textwidth]{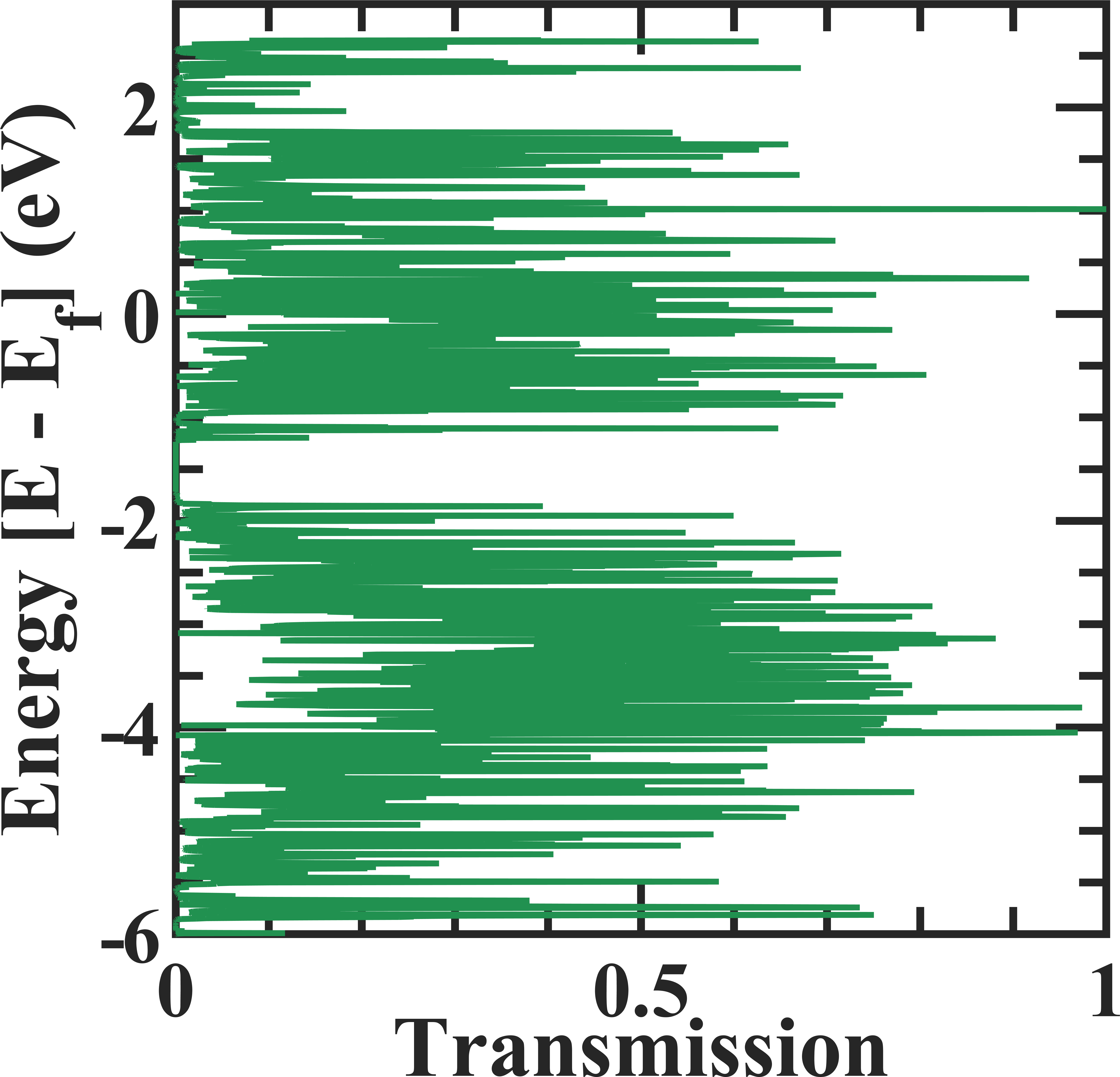}\subcaption{Device Transmission profile $\mathrm{V_{GS}}$ -0.25V}\label{fig-SCBA-1-k} \end{subfigure}
\begin{subfigure}{0.24\textwidth}\centering  \includegraphics[width=\textwidth]{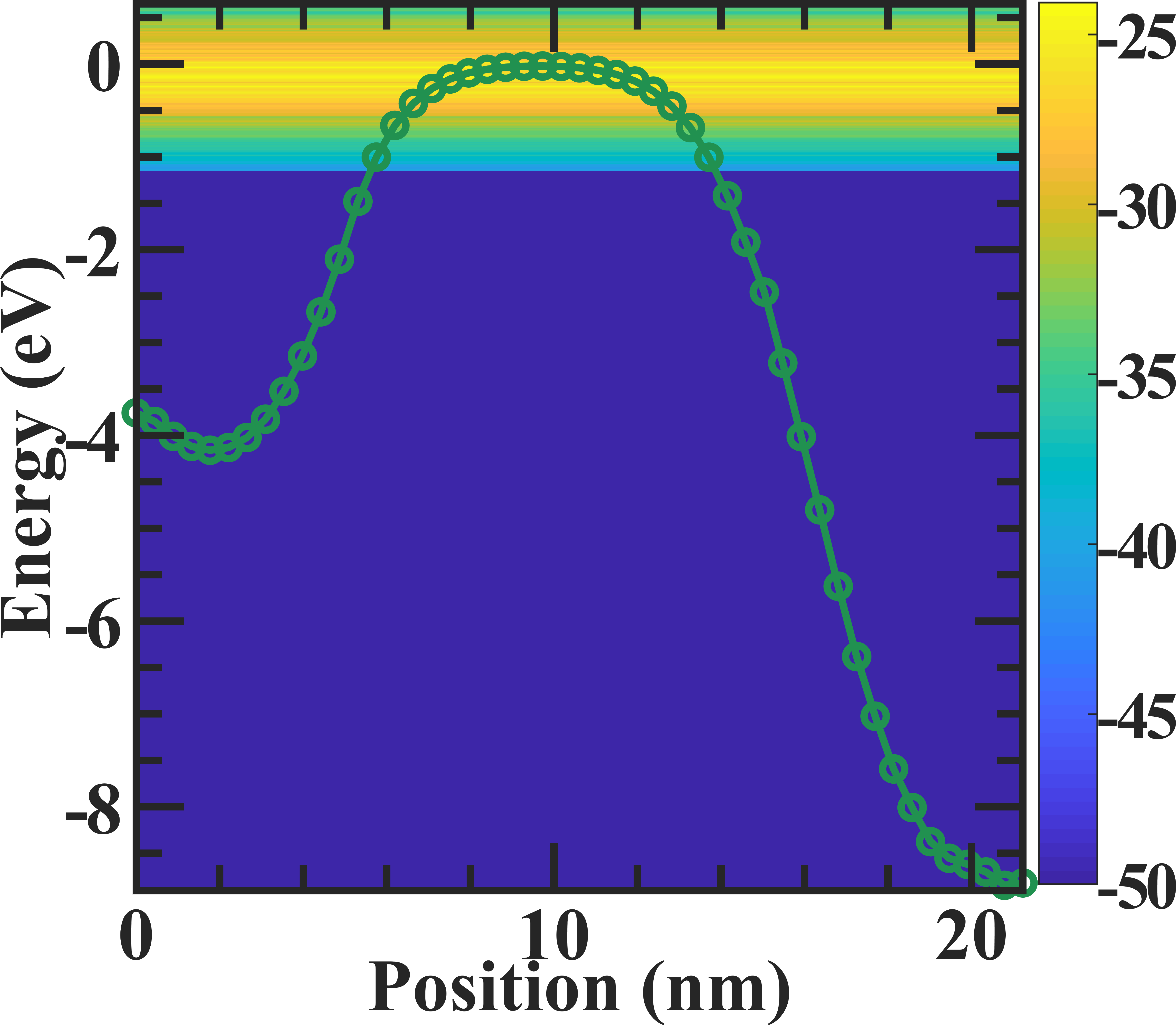}\subcaption{Device energy-resolved flux density $\mathrm{V_{GS}}$ -0.25V}\label{fig-SCBA-1-l}\end{subfigure}
\begin{subfigure}{0.24\textwidth}\centering  \includegraphics[width=\textwidth]{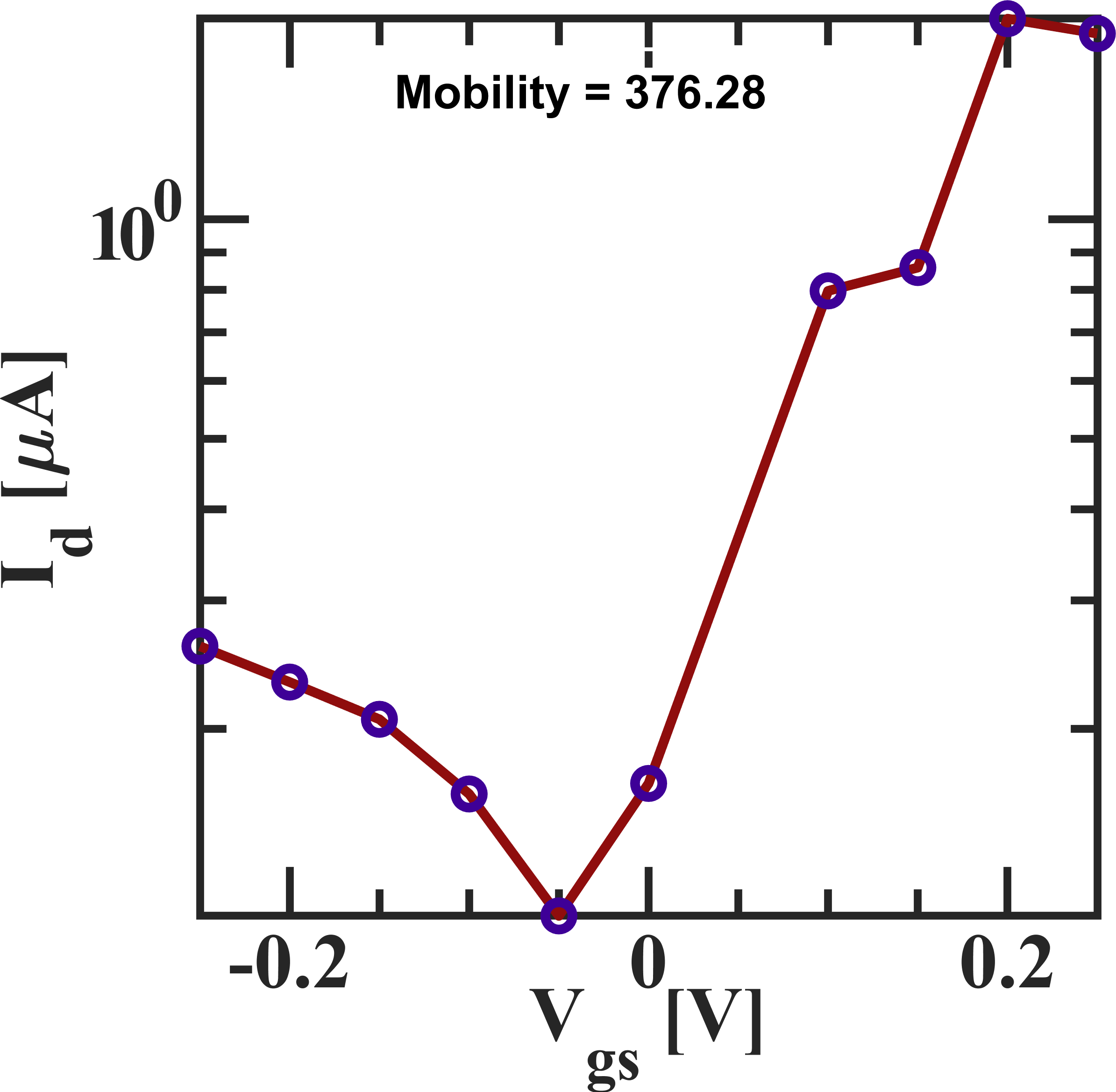}\subcaption{Current-voltage characteristics}\label{fig-SCBA-1-m}\end{subfigure}
\begin{subfigure}{0.24\textwidth}\centering  \includegraphics[width=\textwidth]{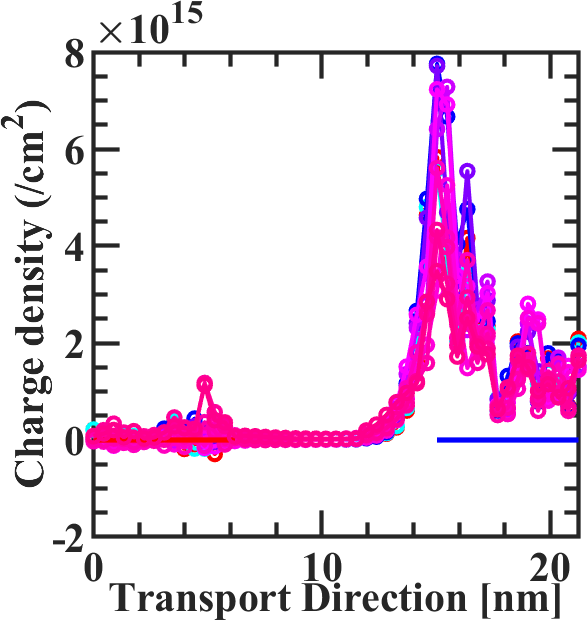}\subcaption{Charge density variation}\label{fig-SCBA-1-n}\end{subfigure}
\begin{subfigure}{0.24\textwidth}\centering  \includegraphics[width=\textwidth]{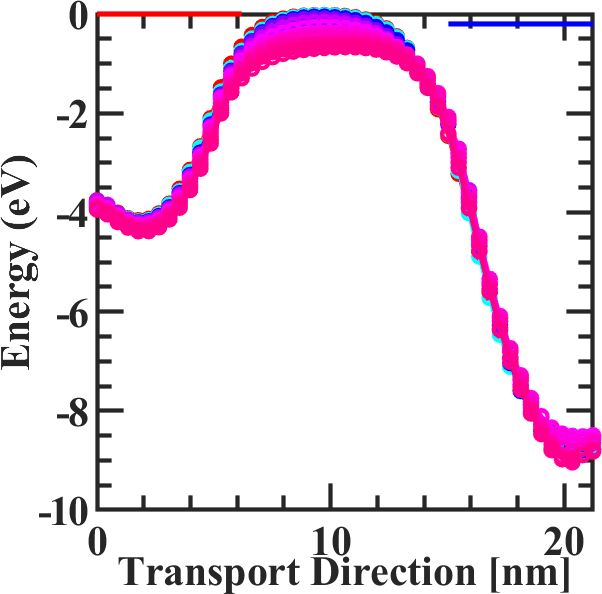}\subcaption{Potential variation}\label{fig-SCBA-1-p}\end{subfigure}
\caption{Device structure graphene Poisson potential \cref{fig-SCBA-1-a}, free charge \cref{fig-SCBA-1-b},Transmission \cref{fig-SCBA-1-c}, and energy-resolved flux density \cref{fig-SCBA-1-d} profile at 0.25V, Poisson potential \cref{fig-SCBA-1-e}, free charge \cref{fig-SCBA-1-f},Transmission \cref{fig-SCBA-1-g}, and energy-resolved flux density \cref{fig-SCBA-1-h} profile at 0V, Poisson potential \cref{fig-SCBA-1-i}, free charge \cref{fig-SCBA-1-j},Transmission \cref{fig-SCBA-1-k}, and energy-resolved flux density \cref{fig-SCBA-1-l} profile at -0.25V for the three bias point, Current-voltage characteristics \cref{fig-SCBA-1-m}, Charge density variation in the conducting device due to scattering and gate-field action at different bias point condition  \cref{fig-SCBA-1-n}, and potential variation in the conducting device due to gate-field action at different bias point condition \cref{fig-SCBA-1-p} in the conducting device in the incoherent transport regime}
\label{fig-SCBA-1}
\end{figure}

We have solved transport calculations in the self-consistent Born algorithm with Poisson iterations through the Recursive Green's Function (RGF) algorithm. The simulation begins with solving the hamiltonian matrix for different geometric sections, and by using Schr\"{o}dinger solvers, the band structure in the source and drain are derived. After that, the initial potential profile determines by semi-classical consideration through a Poisson solver. Next, the density solver uses this initial potential profile to self-consistently calculate the first electronic density through \cref{eq-B43}. Next, the density solver implements the scattered Recursive Green's Function calculations and combines them with the nonlinear Poisson solver to construct the final electronic density in the following loop. Ultimately, with the final electronic density and Poisson potential from \cref{eq-B44}, the current is calculated through the \cref{eq-B54} and \cref{eq-B55}. We have incorporated 50 Sancho Rubio-based Recursive Green's Function calculations, two self-consistent Poisson iterations, and one scattering iteration in the device. In addition, we have used deformation potential type scattering self-energy from \cref{eq-B30} and \cref{eq-B32} for the acoustic phonon and \cref{eq-B33} and \cref{eq-B34} for the optical phonon to calculate the current in the device. By solving the \cref{eq-B54} and \cref{eq-B55} output of the self-consistent computation are plotted in the \cref{fig-SCBA-1} correspond to a simulated graphene device of $ 21 $ nm with the $\mathrm{P-D}$ tight-binding basis in the incoherent transport regime. The device is separated into three geometry sections, source, channel, and drain, and its electrical characteristics with channel length $\mathrm{L_G}$ of 10.3189 nm, source length $\mathrm{L_S}$ of 4.8879 nm, and drain length $\mathrm{L_D}$ of 5.431 nm with drain, and $ 2.15 $nm width with typical channel doping density of $ 50 \times 10^{13} $ per cm$^2$ observe in the CVD/PECVD roll-to-roll batch sample. The side gate oxide thickness $\mathrm{t_{ox}}$ is 0.76 nm on the each side and gate dielectric constant $ \epsilon_{r} $ is 3.9. We simulated the incoherent transport device with a gate bias sweep of -0.25 Volt to +0.25 Volt in step 0.05 Volt with the presence of a transport  barrier oxide, and a maximum source-drain bias of 0.2 Volt was applied. In the simulated device, the primitive unit cell has four atoms per cell and 2620 atoms simulated by a finite element mesh of 10,480 point domain size. The $ \mathrm {P-D} $ tight-binding model contains three orbitals, namely carbon $\mathrm{P_{z}}$, and carbon-hydrogen passivated $\mathrm{D_{yz}}$, $\mathrm{D_{xz}}$ orbitals. Therefore total degree of freedom in hamiltonian is 7,860 variable-sized. We have made the thin body field-effect device by reducing the lateral dimension by one-third of the previously simulated ballistic study. It will help to reduce the number of atoms in the device, hence propagating Green's function's matrix inversion bottleneck. Dyson's scattering loop is involved in this device with a self-consistent Poisson's-non-equilibrium Greens function loop, and therefore computation overhead scales up by order of three. By reducing the lateral dimension by making a thin body, we also introduce a nanoribbon confinement-induced bandgap in the range of $ \sim 0.633 $ eV in this device, which will help to throttle the change transport on the applying gate field action and to deduce field-effect mobility. We have plotted the device Poisson potential in \cref{fig-SCBA-1-a}, \cref{fig-SCBA-1-e}, and \cref{fig-SCBA-1-i} for the three gate bias sweep of 0.25V, 0V, and -0.25V. Correspondingly \cref{fig-SCBA-1-b}, \cref{fig-SCBA-1-f}, and \cref{fig-SCBA-1-j} charge distribution, \cref{fig-SCBA-1-c}, \cref{fig-SCBA-1-g}, and \cref{fig-SCBA-1-k} transmission, and \cref{fig-SCBA-1-d}, \cref{fig-SCBA-1-h}, and \cref{fig-SCBA-1-l} energy-resolved flux density profile plotted for the three gate bias sweep of 0.25V, 0V, and -0.25V. In \cref{fig-SCBA-1-m} current-voltage characteristics of corresponding device in the incoherent transport regime are plotted. In the \cref{fig-SCBA-1-n} charge density variation in the conducting device due to scattering and gate-field action at different bias point conditions and \cref{fig-SCBA-1-p} Potential variation in the conducting device due to gate-field action at different bias point condition in the conducting incoherent transport regime plotted.																																																																								

The scattering interaction increases the coupling between the individual current mode by mixing the carrier's energy. Now, more path is available for the carrier to scatter through the available density of states. The scattering interaction usually broadens the coupling and resonance mode. The peak resonance in the transmissions usually reduces, but increasing broadening will increase the total transmission flux and current output. In the current spectra, an additional peak appears due to the inclusion of scattering. The increasing scattering rate becomes prominent, and scattering leakage dominates the total current flow. We have calculated the field-effect mobility value from the slope of the conductivity curve, which is a more exact treatment than the effective mass formulation for graphene. Furthermore, it matches a variety of device scenarios. An explicit treatment of strong and close to thermalizing scattering interaction with a self-consistent Born algorithm loop is a computationally intensive task. \cite{klimeck_elastic_1994,lake_single_1997,bowen_quantitative_1997} The method is also two to three orders of expansive in comparison to ballistic simulations. \cite{luisier_atomistic_2009} An additional self-consistent Born loop requires between Green's function and the self-energies. Nevertheless, a thermalizing scattering is critical to quantitative device modeling. In the graphene literature, various device studies have shown excessively high mobility value for the ultra-clean sample device. However, our numerical work examines to benchmark nominal roll-to-roll graphene sheet-based production device mobility degradation, non-ideality, and variability. The critical input from this bottom-up multi-scale study scrutinizes the device engineering aspect in the low-dimensional material industry. It provides feedback on the process loop optimization to achieve improved mobility yield in the bulk CVD/PECVD fabrication process. To investigate and mitigate these underlying microscopic effects, we tried to incorporate as much variability and non-ideality physics as possible in the quantum transport framework to solve at a supercomputer cluster with a limited CPU time to benchmark and commercialize CVD graphene-based device products.
																																												
\subsection*{Generalized Contact Self-energy Transport}

The source and drain contact in the electronic device usually has a higher doping density, and a ballistic quantum transport-dominated regime governs the channel in the generalized contact self-energy transport regime. A large density of states dominated reservoir size to thermalize the carrier in the output path. It models as locally defined quasi-Fermi levels, electron scattering, and decoherence-dominated regime, modeled by drift-diffusion transport. The device separated into strong scattering and non-scattering regions. This assumption is reasonably valid in graphene and two-dimensional material, where the channel has an order of magnitude less scattering rate. Moreover, scattering in the contact increases the electron transmission in the channel region and leaks into the channel bandgap's forbidden gap. These effects model as generalized contact in one-dimensional, two-dimensional, and three-dimensional geometries. The embedded single scattering rate incorporated in this method can represent the effect of multiple scattering mechanisms in the transport direction. \cite{klimeck_quantum_1995,klimeck_quantum_2003} The method directly deduced experimentally relevant parameters such as momentum relaxation time and mean free path. The diagonal elements of the thermalized contact's Green function are computed using the Recursive Green's Function (RGF) algorithm. This modeling approach limits scattering to diagonal elements and scattering limits at subatomic resolutions. On the other hand, the self-consistent Born loop does more physical scattering, incorporates non-diagonal elements, and has a more accurate subatomic field. The computation overhead in this generalized contact method where we treat the scattering effect phenomenological is around five times higher than the ballistic quantum transmitting boundary method, where we can not incorporate complex scattering in the contact. \cite{luisier_atomistic_2006} The source and drain have higher carrier densities and are in local thermal equilibrium, and channels have coherent transport and are in a non-equilibrium state. This approach will help reduce the spending on computational resources to obtain non-equilibrium Green's function calculation. An empirical scattering rate describes the thermalized equilibrium reservoirs, equivalent to the realistic physics-based scattering rate, depending on the electron's momentum, energy, and spatial distribution. The phenomenological empirical scattering rate deduces some energy constant scattering potential above the transverse momentum-dependent band edge. It exponentially decays below the band edge from the experimentally observed scattering rates. The above energy and momentum dependent imaginary potential $ i \eta $ to account for the scattering-induced broadening added to the on-site elements of the reservoir Hamiltonian in the phenomenological scattering model as, from the equation \cref{eq-B21} and \cref{eq-B28} as,

\begin{equation} \label{eq76} 
\lbrack G_{mn}^{R}(\boldsymbol{k}_{\boldsymbol{t}};E)\rbrack^{-1} = E-H_{mn}(\boldsymbol{k_t},E)-qV_{mn} + i\eta_{mn}{(\boldsymbol{k_t},E)}
\end{equation}

\begin{equation*}
\text{Where}\;\; \eta_{mn}{(\boldsymbol{k_t},E)}=
\begin{cases}
\eta_0, & \text{if}\ E>E^{c}_{mn}(\boldsymbol{k_t}) \\
\eta_{0}\exp{\frac{E-E^{c}_{mn}(\boldsymbol{k_t})}{\eta_0}}, &\text{if}\ E\leq E^{c}_{mn}(\boldsymbol{k_t})
\end{cases}
\end{equation*}

Where $ E^{c} $ is the conduction band edge, $ {\lbrack G_{mn}^R\rbrack}^{-1} $ represents the on-site elements of the inverse of the Green's function at position $ mn $, $ V_{mn} $ is the potential at position $ mn $, $ E $ is the energy, $ i\eta $ is a small imaginary potential related to the scattering rate, $ \boldsymbol{k_t} $ is the transverse wave vector, and $ H_{mn} $ is the diagonal element of the Hamiltonian. The Recursive Green's Function (RGF)  algorithm obtains the inverted matrix elements. \cite{stovneng_multiband_1994} from \cref{eq-B43} the equilibrium reservoirs electrons are in the local quasi-Fermi levels as,

\begin{equation}\label{eq77}
n_{mn}^{res}=\frac{1}{A}\sum_{\boldsymbol{k}_{\mathrm{t}}}\sum_{nm}\int\frac{dE}{2\operatorname\pi}A_{mn}(\boldsymbol{k}_{\mathrm{t}};E)\:f_{FD}{(E-E_{Fmn})}
\end{equation}

Where $ A_{mn} $ is $ \boldsymbol{k_t} $ and $ E $ dependent  reservoirs spectral function, $ E_{Fmn} $ is local quasi-Fermi level, and Fermi-Dirac distribution function $ f_{FD}$,

\begin{equation}\label{eq78}
A_{mn}(\boldsymbol{k}_{\mathrm{t}};E) = i\big\lbrack G_{mn}^{R}(\boldsymbol{k}_{\mathrm{t}};E)-G_{mn}^{A}(\boldsymbol{k}_{\mathrm{t}};E)\big\rbrack
\end{equation}

The solution of the continuity equation in the reservoirs will determine the quasi-Fermi levels $ E_{Fmn} $ as,

\begin{equation}\label{eq79}
\nabla\cdot J_{mn}=0
\end{equation}

Where in the reservoirs the total drift-diffusion current density is $ J_{{mn}} = \mu n_{{mn}} \nabla E_{Fmn}$ and carrier density $ n_{{mn}} $ from \cref{eq77} and $ \mu  $ is carrier mobility. Here the drift-diffusion equation contains complex local charge densities defined by quantum mechanical treatment. To solve the drift-diffusion equation we have employed $ E_{Fmn} = E_{FD} (E_{FS})  $ drain (source) boundary condition at the right (left) edge of device with corresponding contact drain (source) Fermi level of $ E_{FS} (E_{FD}) $. From \cref{eq-B52}, we will now enforce current continuity to hold throughout the device by matching the current value at the central region and current value at the reservoir-channel interfaces as, \cite{klimeck_quantum_2003}

\begin{equation}\label{eq80}
J_{L-R} =\frac{e}{\hbar A}\sum_{ll_{1}}\sum_{l_{2}l_{3}}\sum_{\boldsymbol{k_t}}\int\frac{\mathrm{d}E}{2\pi}\big[\mathcal{T}{(\boldsymbol{k_t},E)}\big]\big[f_{FD}(E-E_{FL})-f_{FD}(E-E_{FR})\big] 
\end{equation}

Where $ \mathcal{T} $ transmission coefficient is $ \boldsymbol{k_t} $ and $ E $ dependent and it is calculated at the channel and thermalized reservoir boundary, and the quasi-Fermi levels $ E-E_{FR} $ and $ E-E_{FL} $ are right and left boundary condition. The Recursive Green's Function (RGF) algorithm marched through the entire device and incorporated channel and reservoir contact. We have used Newton's method to solve the nonlinear \cref{eq79}. After solving for quasi-Fermi levels from \cref{eq77}, contact electronic density distribution is obtained. To solve for the channel density distribution by,

\begin{equation}\label{eq81}
n_{mn}^{ch}=\frac{1}{A}\sum_{\boldsymbol{k_t}}\sum_{nm}\int\frac{dE}{2\operatorname\pi}{\bigg[A_{mn}^{L}(\boldsymbol{k}_{\boldsymbol{t}};E)f_{FD}{(E-E_{FL})}+A_{mn}^{R}(\boldsymbol{k}_{\boldsymbol{t}};E)f_{FD}{(E-E_{FR})}\bigg]}
\end{equation}

\begin{figure}[!htbp]
\centering
\begin{subfigure}{0.24\textwidth}\centering  \includegraphics[width=\textwidth]{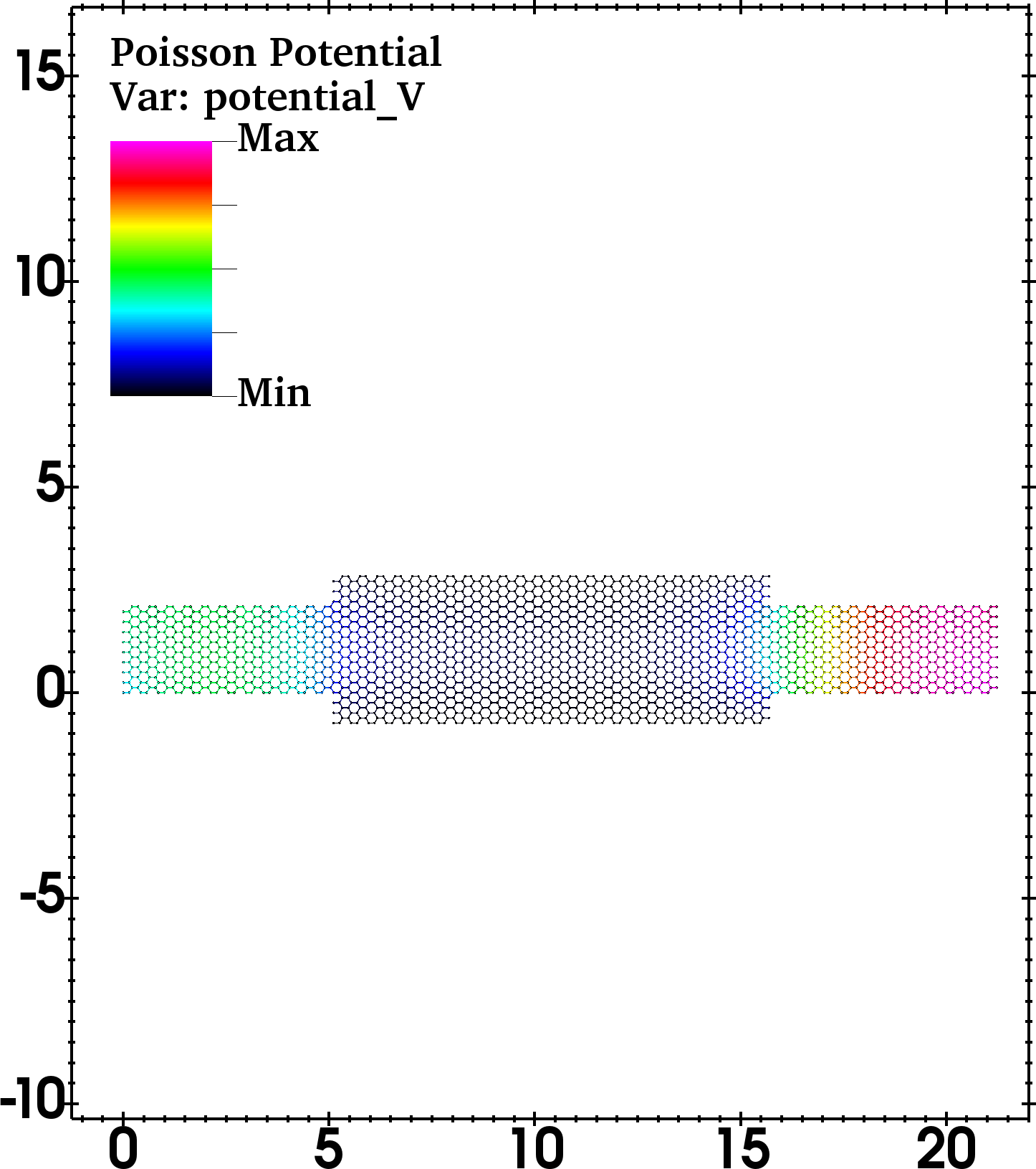}\subcaption{Device Poisson potential profile $\mathrm{V_{GS}}$ 0.3V} \label{fig-LamdaG-1-a}\end{subfigure}
\begin{subfigure}{0.24\textwidth}\centering  \includegraphics[width=\textwidth]{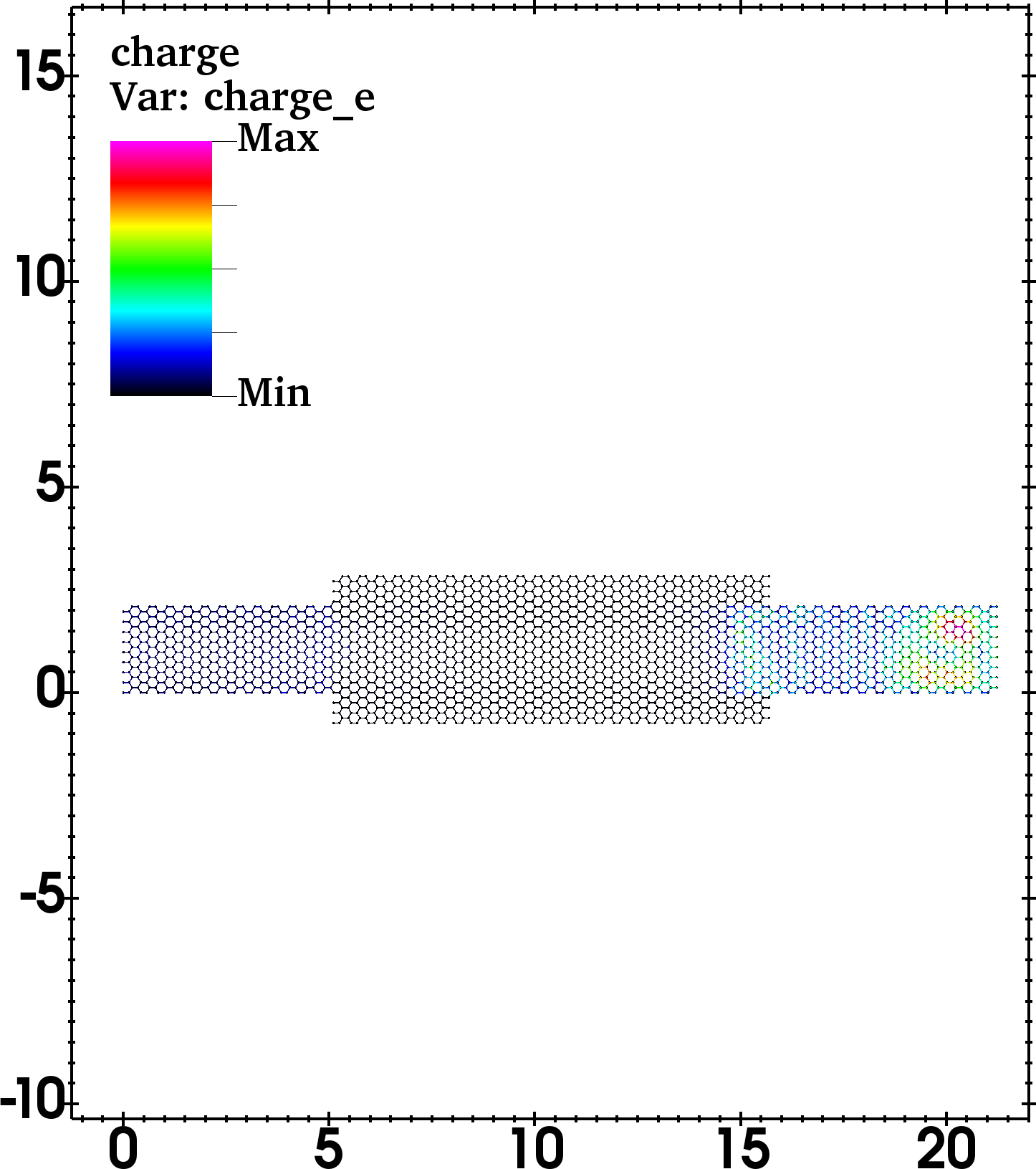}\subcaption{Device free charge profile $\mathrm{V_{GS}}$ 0.3V}\label{fig-LamdaG-1-b}\end{subfigure}
\begin{subfigure}{0.24\textwidth}\centering  \includegraphics[width=\textwidth]{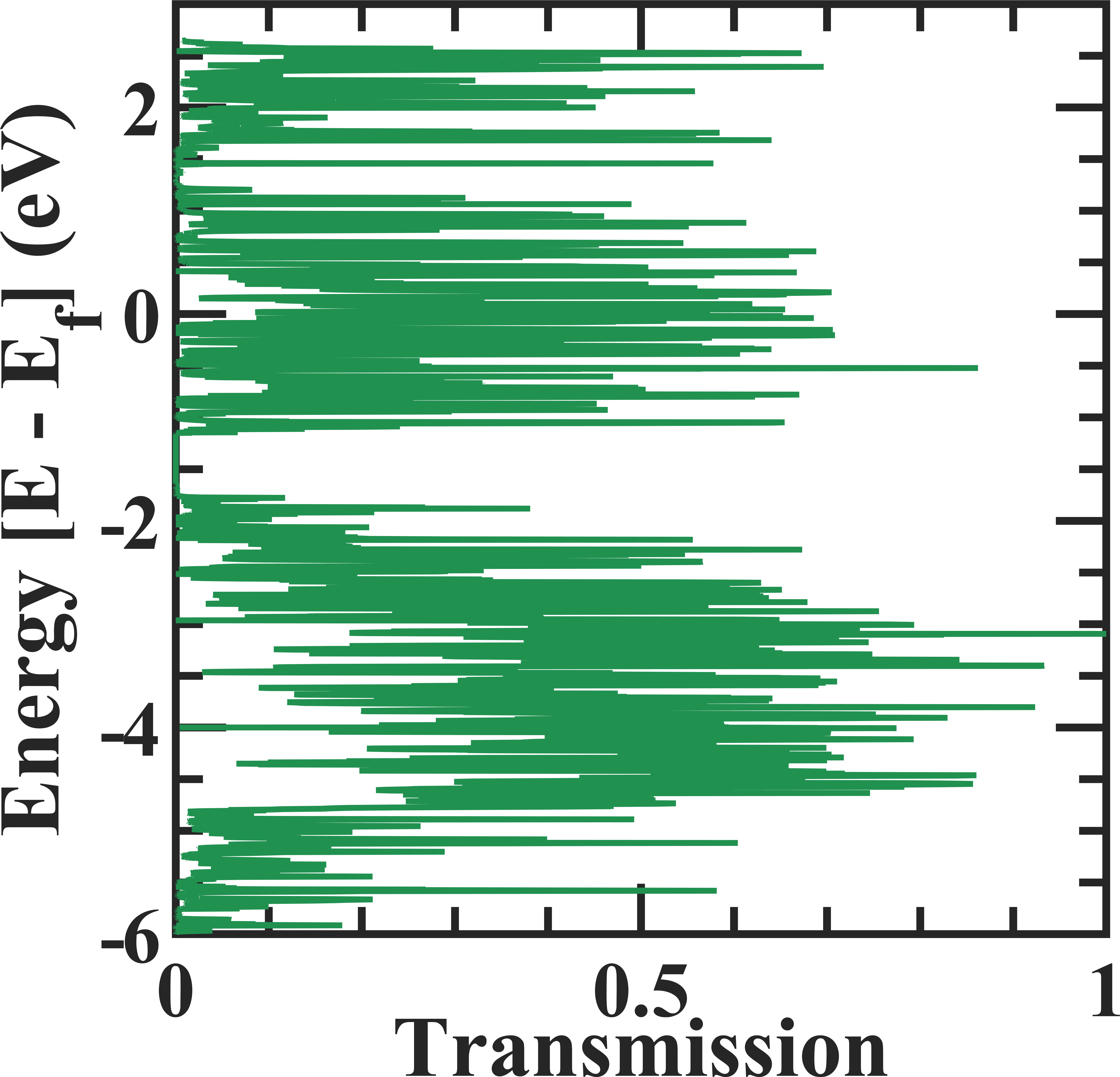}\subcaption{Device Transmission profile $\mathrm{V_{GS}}$ 0.3V}\label{fig-LamdaG-1-c}\end{subfigure}
\begin{subfigure}{0.24\textwidth}\centering  \includegraphics[width=\textwidth]{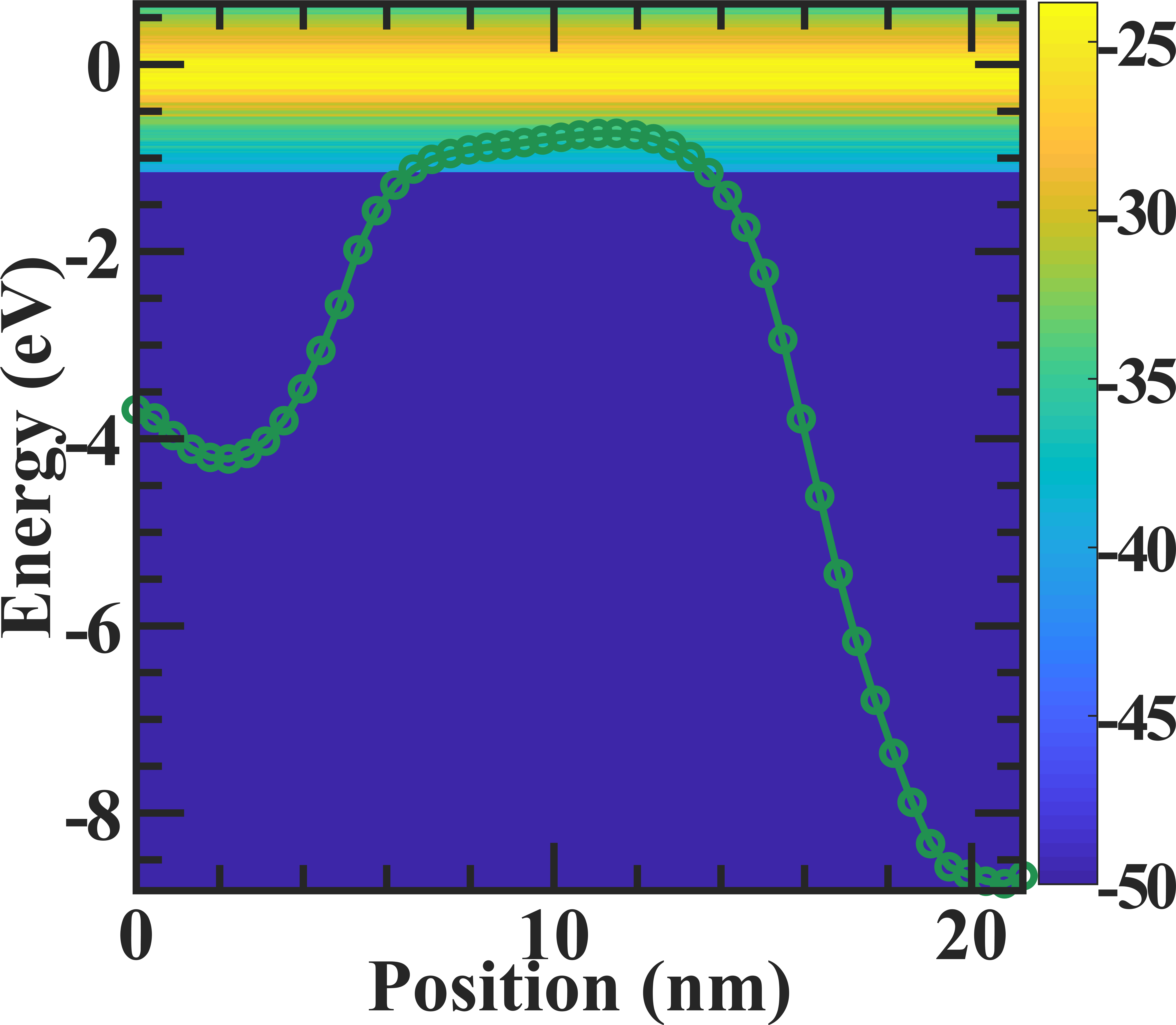}\subcaption{Device energy-resolved flux density $\mathrm{V_{GS}}$ 0.3V}\label{fig-LamdaG-1-d}\end{subfigure}
\begin{subfigure}{0.24\textwidth}\centering  \includegraphics[width=\textwidth]{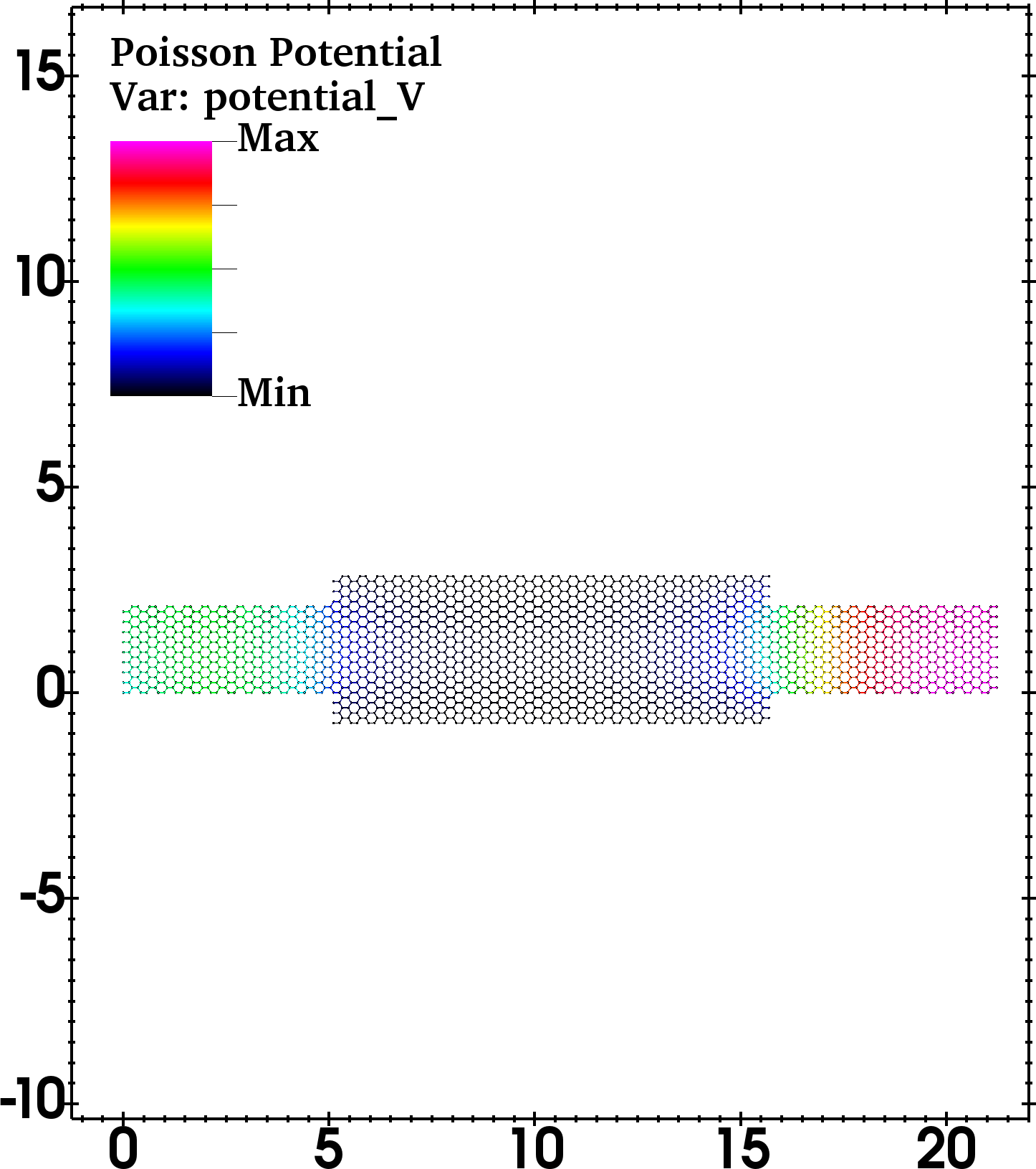}\subcaption{Device Poisson potential profile $\mathrm{V_{GS}}$ 0V}\label{fig-LamdaG-1-e}\end{subfigure}
\begin{subfigure}{0.24\textwidth}\centering  \includegraphics[width=\textwidth]{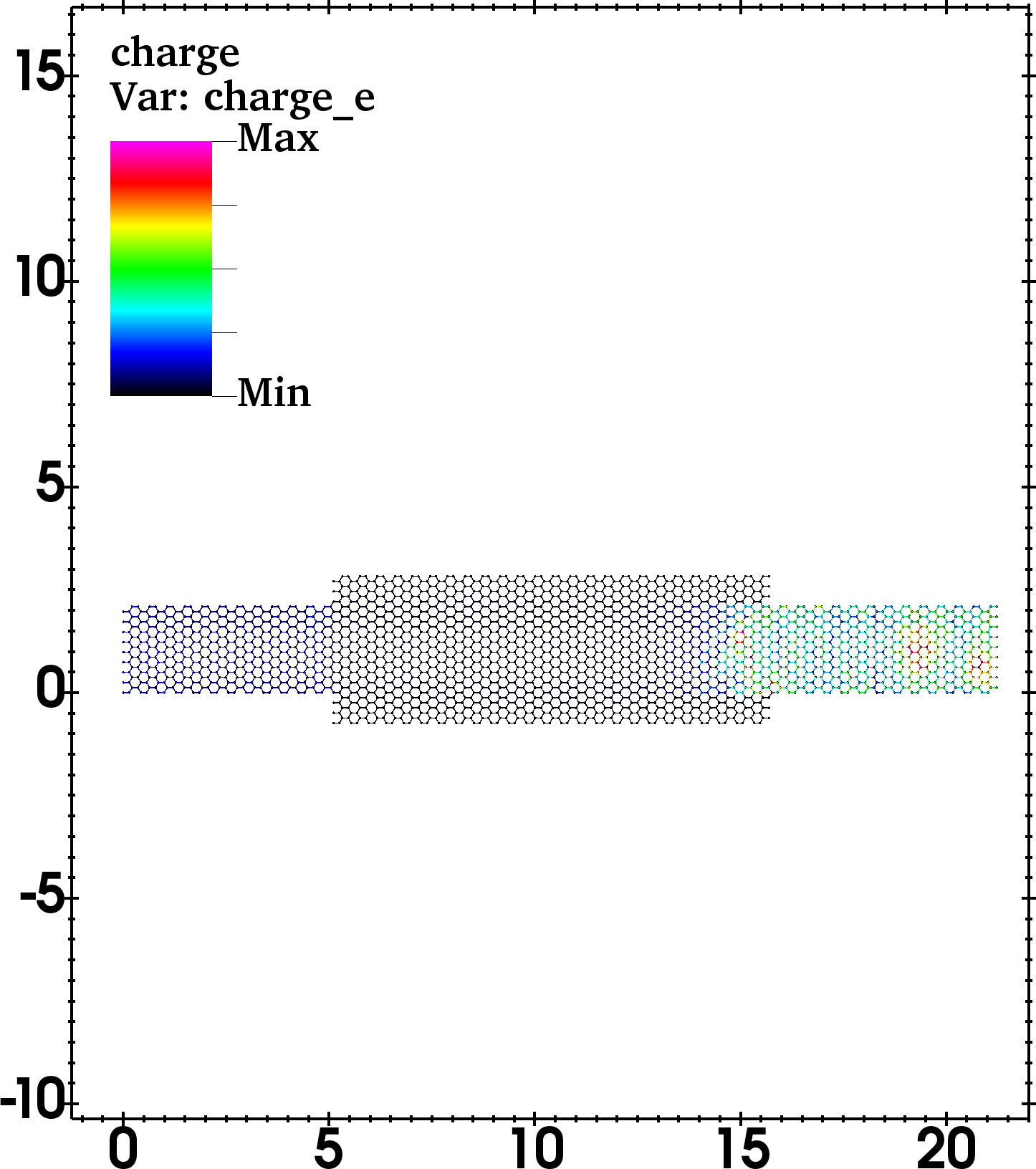}\subcaption{Device free charge profile $\mathrm{V_{GS}}$ 0V}\label{fig-LamdaG-1-f}\end{subfigure}
\begin{subfigure}{0.24\textwidth}\centering  \includegraphics[width=\textwidth]{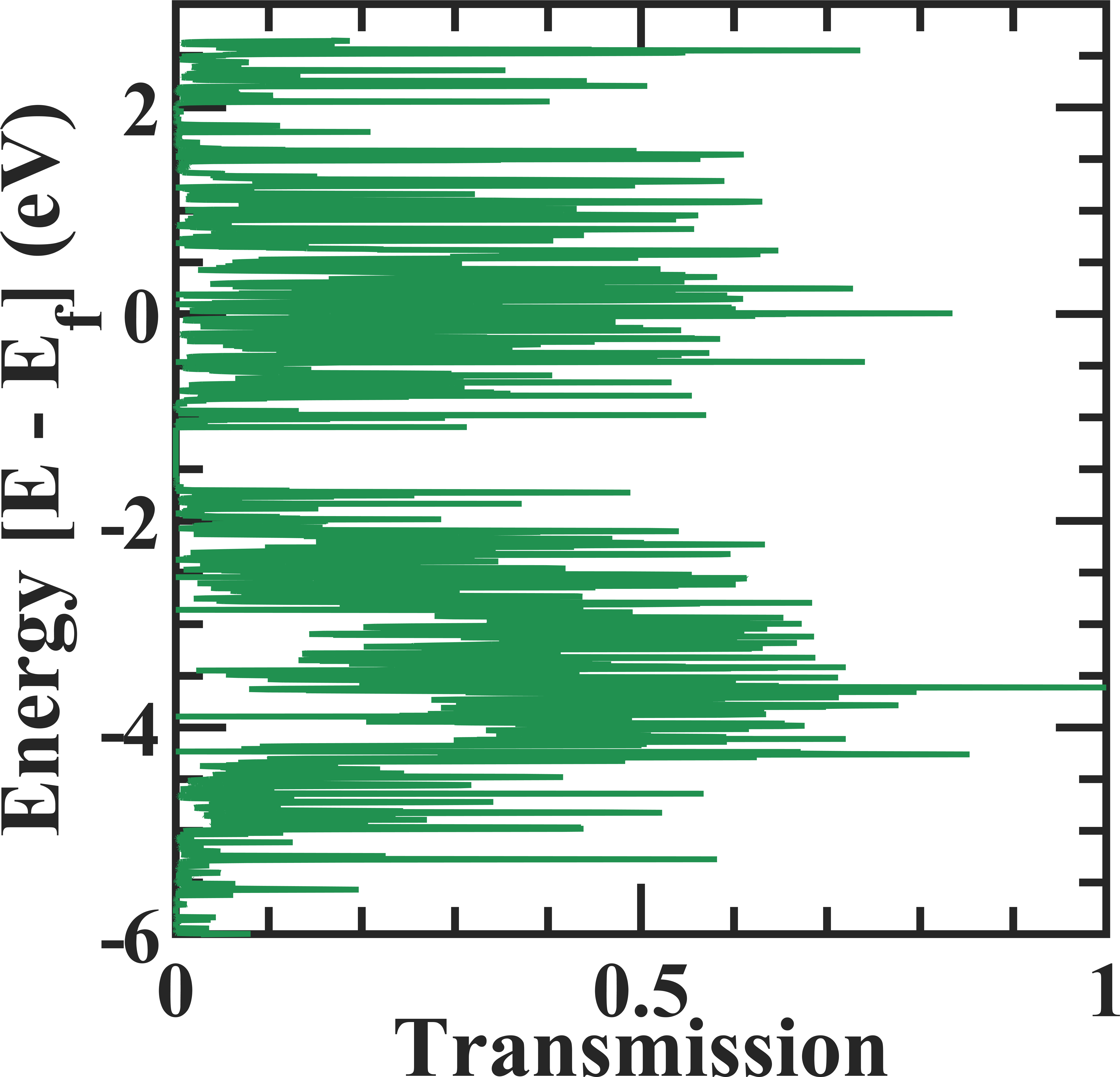}\subcaption{Device Transmission profile $\mathrm{V_{GS}}$ 0V} \label{fig-LamdaG-1-g}\end{subfigure}
\begin{subfigure}{0.24\textwidth}\centering  \includegraphics[width=\textwidth]{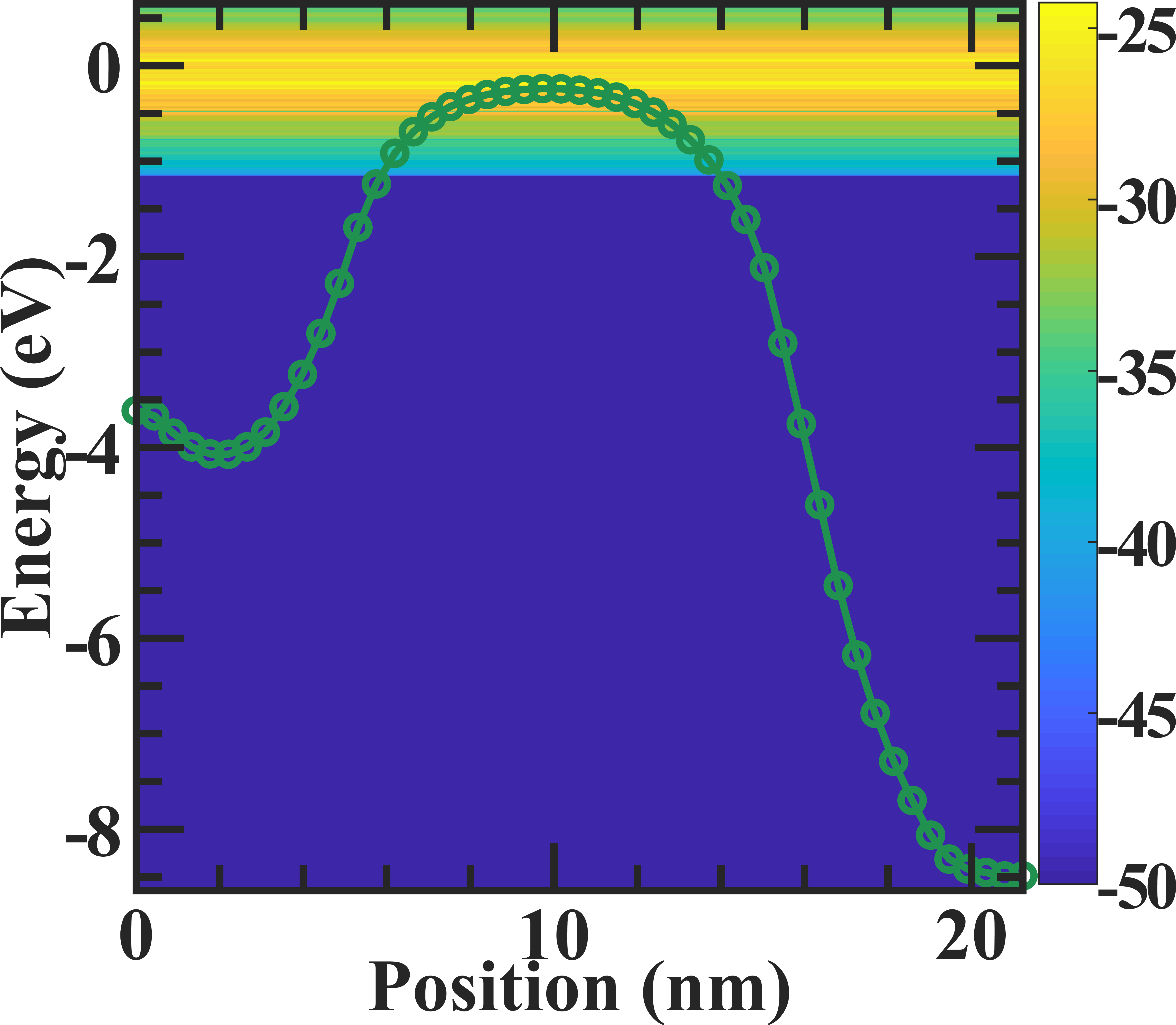}\subcaption{Device energy-resolved flux density $\mathrm{V_{GS}}$ 0V}\label{fig-LamdaG-1-h}\end{subfigure}
\begin{subfigure}{0.24\textwidth}\centering  \includegraphics[width=\textwidth]{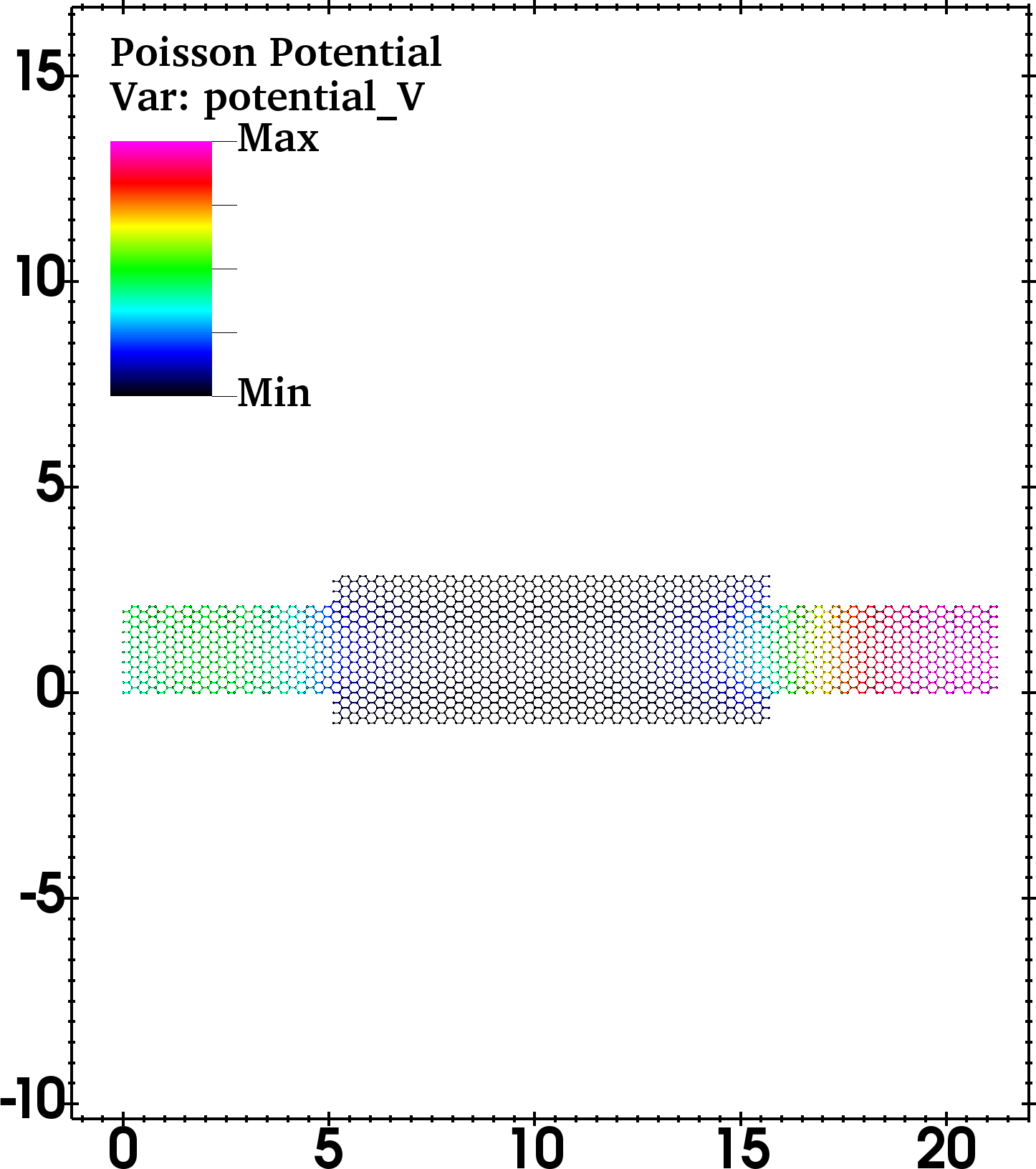}\subcaption{Device Poisson potential profile $\mathrm{V_{GS}}$ -0.3V}\label{fig-LamdaG-1-i}\end{subfigure}
\begin{subfigure}{0.24\textwidth}\centering  \includegraphics[width=\textwidth]{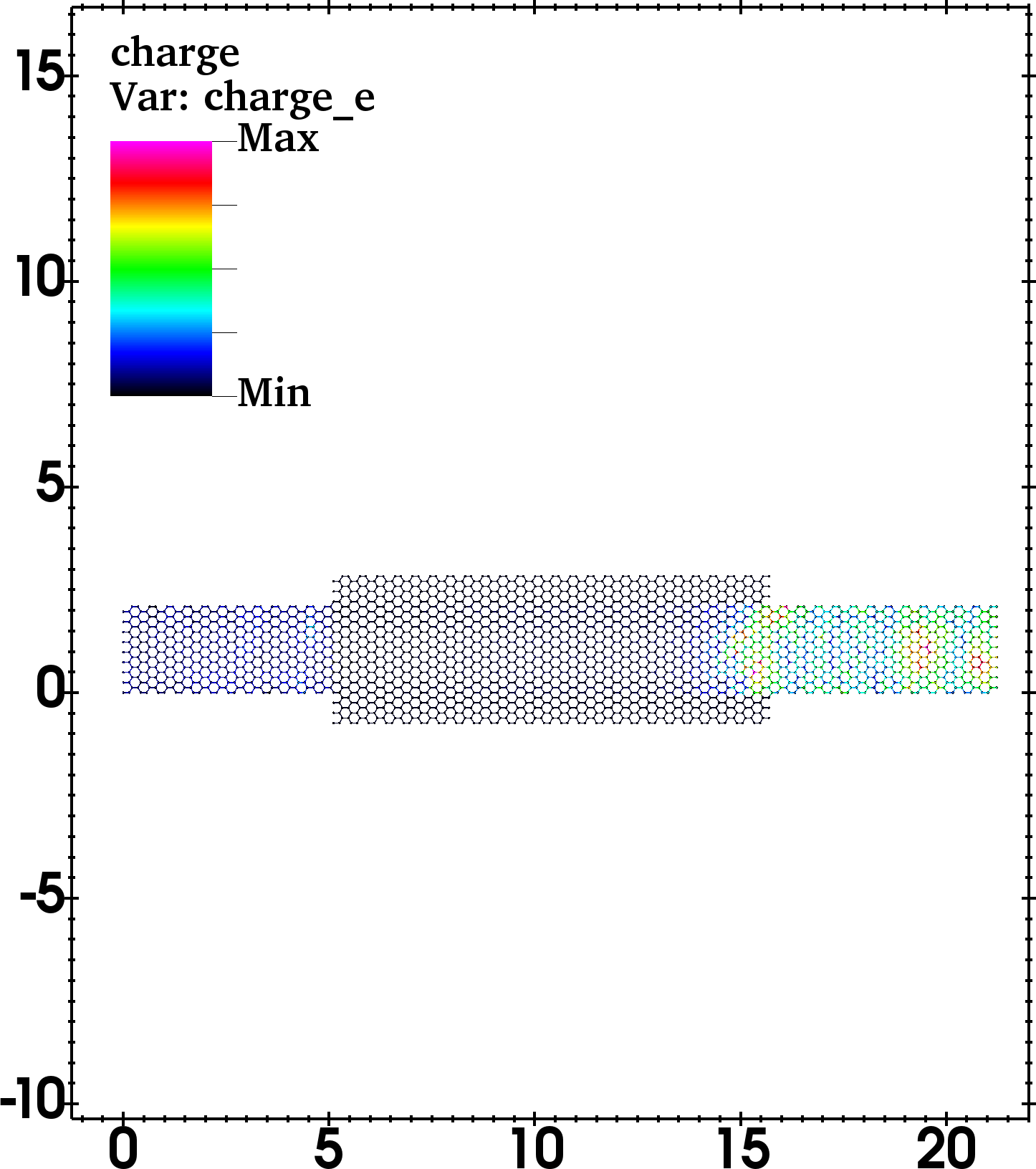}\subcaption{Device free charge profile $\mathrm{V_{GS}}$ -0.3V}\label{fig-LamdaG-1-j}\end{subfigure}
\begin{subfigure}{0.24\textwidth}\centering  \includegraphics[width=\textwidth]{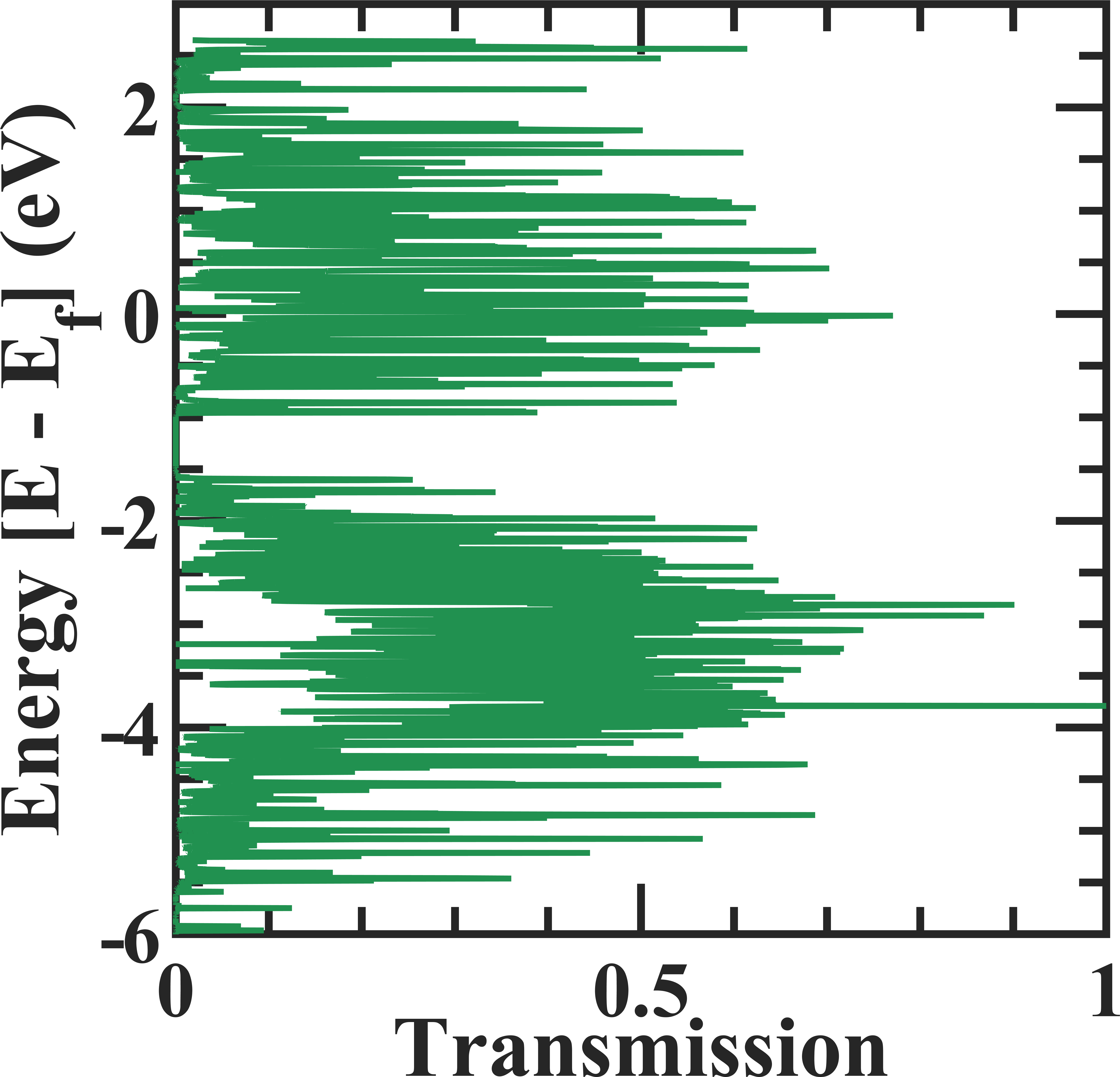}\subcaption{Device Transmission profile $\mathrm{V_{GS}}$ -0.3V}\label{fig-LamdaG-1-k}\end{subfigure}
\begin{subfigure}{0.24\textwidth}\centering  \includegraphics[width=\textwidth]{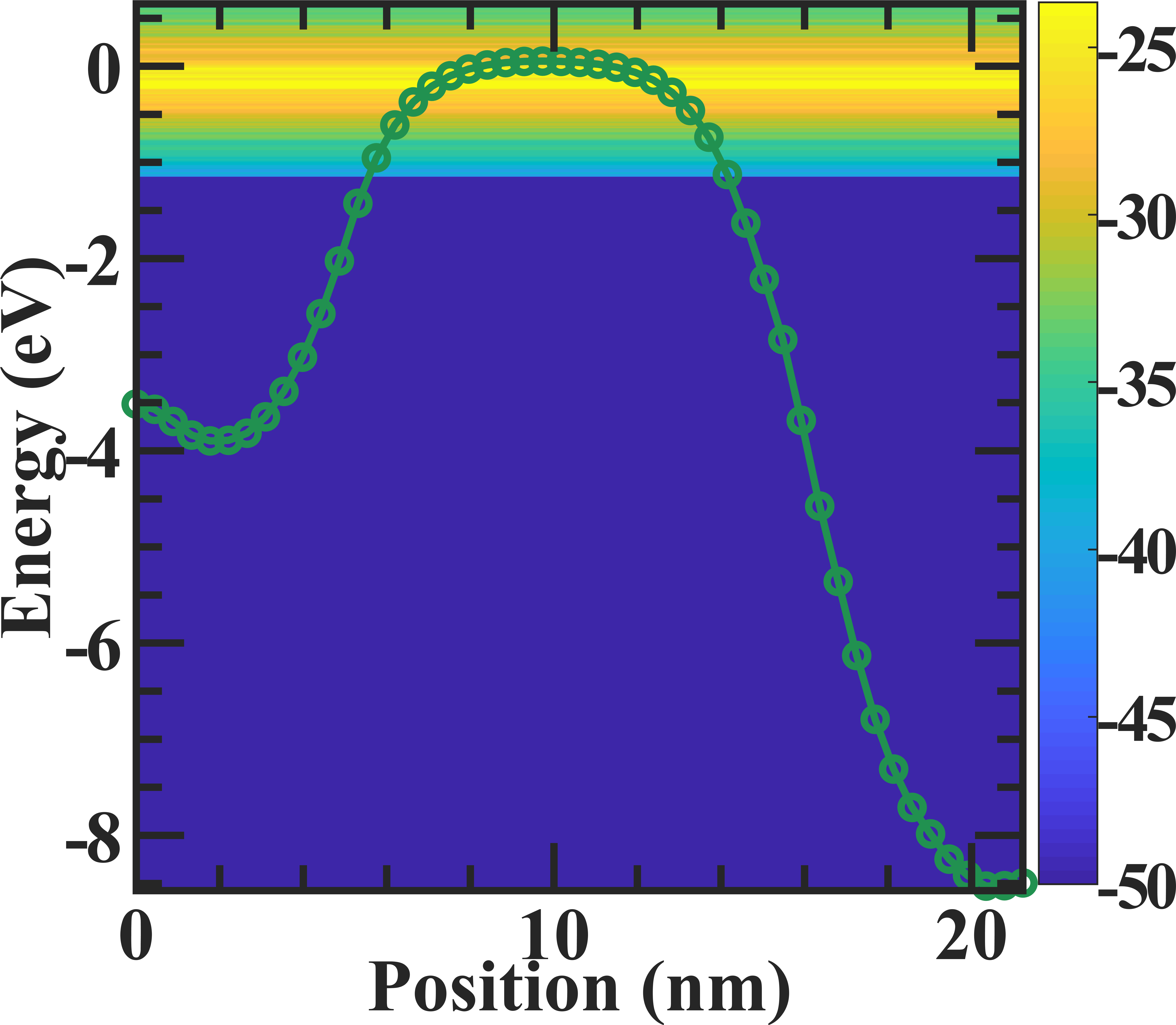}\subcaption{Device energy-resolved flux density $\mathrm{V_{GS}}$ -0.3V}\label{fig-LamdaG-1-l}\end{subfigure}
\begin{subfigure}{0.24\textwidth}\centering  \includegraphics[width=\textwidth]{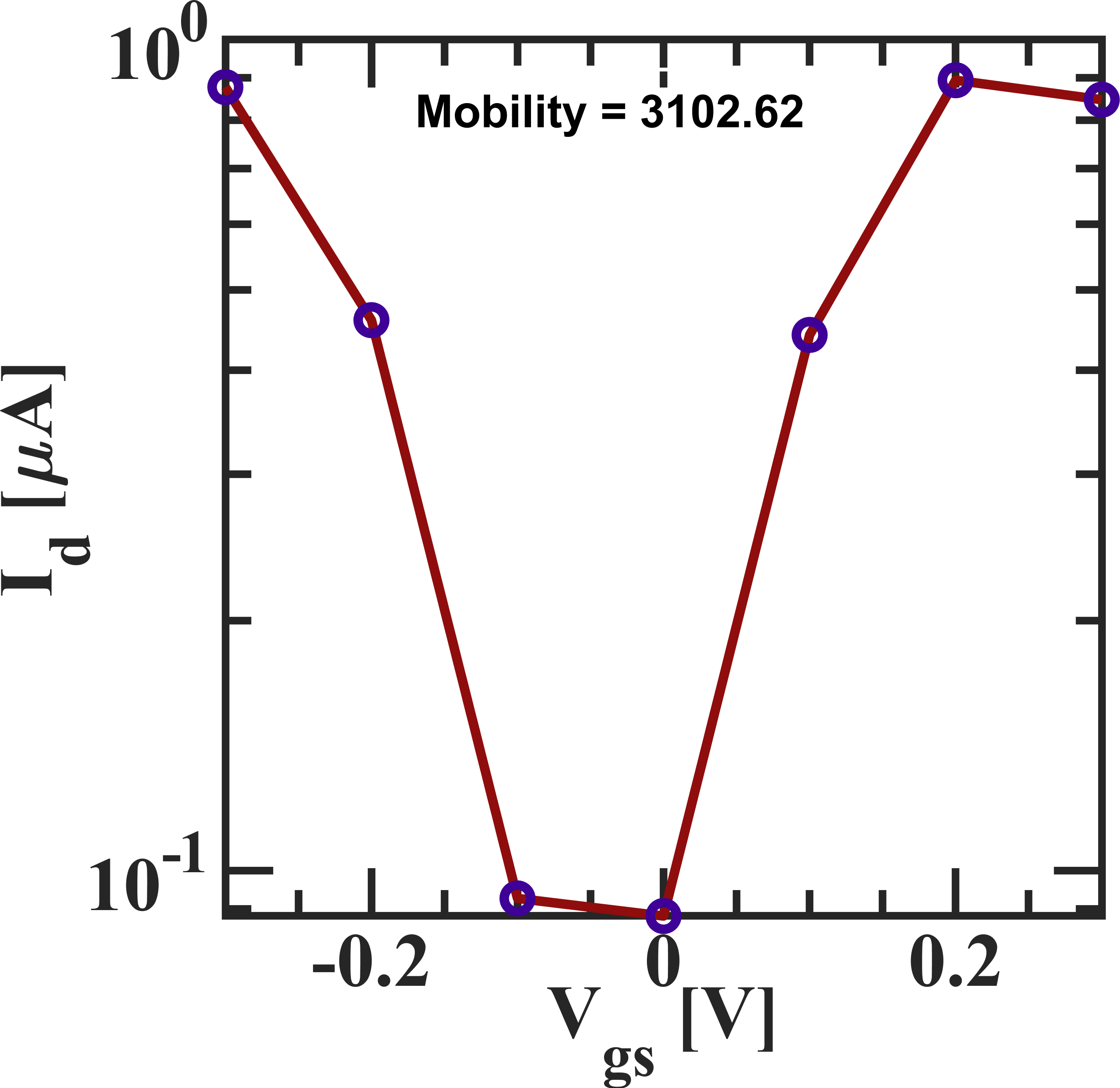}\subcaption{Current-voltage characteristics}\label{fig-LamdaG-1-m}\end{subfigure}
\begin{subfigure}{0.24\textwidth}\centering  \includegraphics[width=\textwidth]{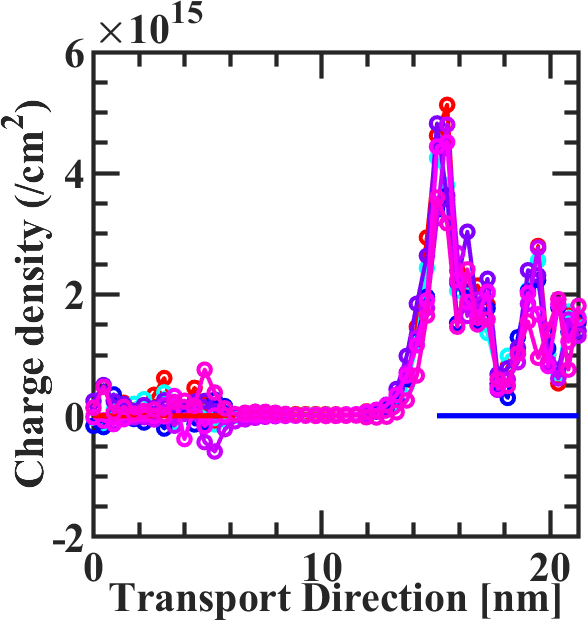}\subcaption{Charge density variation}\label{fig-LamdaG-1-n}\end{subfigure}
\begin{subfigure}{0.24\textwidth}\centering  \includegraphics[width=\textwidth]{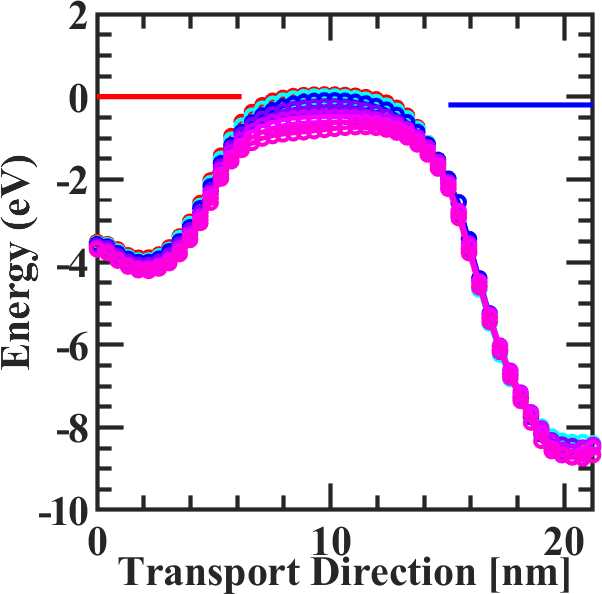}\subcaption{Potential variation}\label{fig-LamdaG-1-p}\end{subfigure}
\caption{Device structure graphene Poisson potential \cref{fig-LamdaG-1-a}, free charge \cref{fig-LamdaG-1-b},Transmission \cref{fig-LamdaG-1-c}, and energy-resolved flux density \cref{fig-LamdaG-1-d} profile at 0.3V, Poisson potential \cref{fig-LamdaG-1-e}, free charge \cref{fig-LamdaG-1-f},Transmission \cref{fig-LamdaG-1-g}, and energy-resolved flux density \cref{fig-LamdaG-1-h} profile at 0V, Poisson potential \cref{fig-LamdaG-1-i}, free charge \cref{fig-LamdaG-1-j},Transmission \cref{fig-LamdaG-1-k}, and energy-resolved flux density \cref{fig-LamdaG-1-l} profile at -0.3V for the three bias point, and Current-voltage characteristics \cref{fig-LamdaG-1-m}, Charge density variation in the conducting device due to scattering and gate-field action at different bias point condition \cref{fig-LamdaG-1-n}, and potential variation in the conducting device due to gate-field action at different bias point condition  \cref{fig-LamdaG-1-p} in the generalized contact self-energy limited transport regime}
\label{fig-LamdaG-1}
\end{figure}

Where right/left-connected spectral functions $ A^R $/$ A^L $ in the channel. \cite{datta_quantum_2005} Furthermore, due to high carrier concentrations in the lead, scattering in the leads is stronger than the electron-phonon assisted scattering in the channel. Hence the generalized scattering contact captures a fair amount of underline physics. We have used $ \eta =5 \mathrm{meV}$ scattering potential related to the scattering rate through the generalized contact to calculate the current in the device. The output of the self-consistent computation \cref{eq80} and \cref{eq81} plotted in the \cref{fig-LamdaG-1} corresponds to a simulated graphene device in the generalized contact self-energy limited transport regime. Its electrical characteristics with channel length $\mathrm{L_G}$ of 10.3189 nm, source length $\mathrm{L_S}$ of 4.8879 nm and drain length $\mathrm{L_D}$ of 5.431 nm, with typical channel doping density of $ 50  \times 10^{13} $ per cm$^2$ observe in the CVD/PECVD roll-to-roll batch sample. The side gate oxide thickness $\mathrm{t_{ox}}$ is 0.76 nm on the each side and gate dielectric constant $ \epsilon_{r} $ is 3.9.
We simulated the incoherent transport device with a gate bias sweep of -0.3 Volt to +0.3 Volt in step 0.1 Volt with the presence of a transport oxide barrier, and a maximum source-drain bias of 0.2 Volt was applied. In the simulated device, the primitive unit cell has four atoms per cell and 2620 atoms simulated by a finite element mesh of 10,480 point domain size. The $ \mathrm {P-D} $ tight-binding model contains three orbitals, namely carbon $\mathrm{P_{z}}$, and carbon-hydrogen passivated $\mathrm{D_{yz}}$, $\mathrm{D_{xz}}$ orbitals. Therefore total degree of freedom in hamiltonian is 7,860 variable-sized. We have made the thin body field-effect device by reducing the lateral dimension by one-third of the previously simulated ballistic study. It will help to reduce the number of atoms in the device, hence propagating Green's function's matrix inversion bottleneck. Dyson's scattering loop is involved in this device with a self-consistent Poisson's-non-equilibrium Greens function loop, and therefore computation overhead scales up by order of three. By reducing the lateral dimension by making a thin body, we also introduce a nanoribbon confinement-induced bandgap in the range of 0.611 eV in this device, which will help to throttle the change transport on the applying gate field action and deduce field-effect mobility. We have plotted the device Poisson potential in \cref{fig-LamdaG-1-a}, \cref{fig-LamdaG-1-e}, and \cref{fig-LamdaG-1-i} for the three gate bias sweep of 0.3V, 0V, and -0.3V. Correspondingly \cref{fig-LamdaG-1-b}, \cref{fig-LamdaG-1-f}, and \cref{fig-LamdaG-1-j} charge distribution, \cref{fig-LamdaG-1-c}, \cref{fig-LamdaG-1-g}, and \cref{fig-LamdaG-1-k} transmission, and \cref{fig-LamdaG-1-d}, \cref{fig-LamdaG-1-h}, and \cref{fig-LamdaG-1-l} energy-resolved flux density profile plotted for the three gate bias sweep of 0.3V, 0V, and -0.3V. In \cref{fig-LamdaG-1-m} current-voltage characteristics of corresponding device in generalized contact self-energy limited transport regime are plotted. In the \cref{fig-LamdaG-1-n} charge density variation in the conducting device due to scattering and gate-field action at different bias point conditions and \cref{fig-LamdaG-1-p} potential variation in the conducting device due to gate-field action at different bias point condition in generalized contact self-energy limited transport regime plotted.  

Typical scattering time value obtained in the chemical vapor deposited graphene is in the range of sub-one hundred of femtosecond $(\mathrm{fs})$, for the high-quality single-crystalline graphene is in the range of picosecond $(\mathrm{ps})$. The charge impurity scattering is the dominant scattering mechanism that limits the graphene's mobility value among various scattering mechanisms. Encapsulation of graphene devices remarkably reduces the residue charges density imposed by process gas and water molecules, polymer residual charge impurities, lithography damages, increases the device stability towards the environment, and reduces the electron-phonon scattering from the substrate. \cite{kim_synthesis_2012,chen_quantitative_2013,kim_synthesis_2015} While the charged impurity scatters and resonant scatters have a spatial variation on the graphene devices and give rise to lateral inhomogeneity in the electrical parameters. On the other hand, in the $(\mathrm{Si}{\mathrm{O}}_2)$ substrate, the background surface polar phonon scatters uniformly distributed over the entire substrate area. At the room temperature in atmospheric conditions on the $(\mathrm{SiO}_2)$ substrate, a typical value reported for the measured carrier mobility is in the range of 500-5000 $\mathrm{cm}^{2}\mathrm{v}^{-1}\mathrm{s}^{-1}$. In the high vacuum and for the optimized device design, it as high as 12000 $\mathrm{cm}^{2}\mathrm{v}^{-1}\mathrm{s}^{-1}$. \cite{chen_how_2014} In the graphene, the upper bound on the carrier mobility value imposed by polar optical phonon scattering on $(\mathrm{Si}{\mathrm{O}}_2)$ substrate is 40000 $\mathrm{cm}^{2}\mathrm{v}^{-1}\mathrm{s}^{-1}$, \cite{chen_intrinsic_2008} the upper limit by longitudinal acoustic phonon scattering in the room temperature is 120000 $\mathrm{cm}^{2}\mathrm{v}^{-1}\mathrm{s}^{-1}$. \cite{sarma_electronic_2011} Electron-electron scattering and optical phonon scattering are captured through $-i\eta$ modeling. Due to the resonant contact state's broadening, more states are connected to the channel and give rise to an additional current path. To simulate the device characteristic, charge filling from the thermalized equilibrium contact by \cref{eq77} using Fermi-Dirac statistics. By using the Heisenberg uncertainty for the energy relaxation time and approximating broadening $\eta$ through $\eta ~ \tau \approx \frac{\hbar}{2}$, \cite{wacker_semiconductor_2002} we have used the $ \eta $ scattering potentials of $ 3~\mathrm{meV} $ to $ 30~\mathrm{meV} $, which correspond to the scattering rate of $ 110 ~\mathrm{fs} $ and $ 10 ~\mathrm{fs} $ for the typical CVD graphene sample in the roll-to-roll production process. From semiclassical treatment if the thermal velocity $\mathrm{v_{th}}$ is assumed to be $ \sim 3 \times 10^7\mathrm{cm/s}$ at low carrier density for CVD graphene sample than we can deduce mean free path $ \Lambda = \mathrm{v_{th}}\tau$ for the $ \eta = 5\mathrm{meV} $ correspond to the scattering decoherence length $ \Lambda = 19.746\mathrm{nm} $. Furthermore, the empirical exploration of a simple mobility model as apparent mobility deduced $ \mu = q\tau /m^* $, while $ m^* $ assumes the Dirac electron effective mass value in the graphene. Vice versa, the Hall-bar experimental measurement conductivity curve result or the field-effect mobility $\mu$ measurement deduces the contemporary measurand to our bottom-up atomistic numerical approach in the incoherent transport regime. Further used in $\tau =\frac{m^* \mu}{q}$, where $m^*$ is the effective empirical mass for graphene devices to calculate energy relaxation time $\tau$. Furthermore, it gives the broadening $\eta$ from above mentioned Heisenberg uncertainty, \cite{klimeck_quantum_1995,sellier_nemo5_2012,ilatikhameneh_can_2016}

\begin{equation}\label{eq82}
\eta \approx \frac{q\hbar}{2m^{*}\mu}
\end{equation}
and the energy relaxation time is, 
\begin{equation}\label{eq83}
\tau \approx \frac{\hbar}{2\eta}
\end{equation}

From the optimal device design point of view, increasing the contact's doping density will decrease serial resistance and reduce the induced potential at the contact. However, it will come with a trade-off also and give rise to the quasi-Fermi level in the contact. The drop in potential will increase the transmission into the channel. However, an increase in quasi-fermi degeneracy reduces the Fermi conduction window and affects the device channel's total current flux. Moreover, with the increasing contact doping, Fermi's degeneracy in the contact increases the energy separation between the contact Fermi level and the conduction band edge of the channel. Furthermore, with the scattering enhanced Fermi conduction window, more states are coupled with the channel, enhancing the current throughput. However, through interaction with channel phonons or electrons, evanescent incident states of channel or contact can still couple with propagating states of contact or channel and give rise to the additional current flux at the room temperature operation. In the band misalignment overlap area of contact and channel, the state's ballistic local density is significantly less; hence, the electron spectral densities reduce. The slope of the spectral density increases with the increases in the scattering rate. By properly designing the electrostatic, phonon-assisted leakage reduces if the channel's propagating states keep at the order of optical phonon energy higher than the valence band edge. The energy separation of $ 3kT \sim  75~\mathrm{meV}$ is good design practice at room temperature for low-dimensional materials device design. Our detailed results will provide a helpful incentive for device design in graphene and related two-dimensional nanoelectronics device for further application in the device research industry. In future studies, we will try to incorporate the surface roughness scattering in the Dyson loop, which affect the two-dimensional materials device's performance. Graphene becomes intrinsically corrugated and rippled to compensate for the thermodynamic instabilities and stabilize the long-range crystal order. Also, substrate roughness, graphene negative thermal expansion coefficient on the substrate, and broken symmetry in the graphene with the substrate layer give rise to imposed roughness in the graphene devices; both intrinsic and imposed roughness bound the upper limit of the graphene mobility values. \cite{lui_ultraflat_2009} Furthermore, particular attention gives to the quantum confinement effect, which can alter energy bandgap and band splitting and, in the end, influence the carrier's injection and transmission in the two-dimensional material device design. Our objective is to comprehend the critical phenomena and device design principle to guide the new generation of two-dimensional material-based device modeling and emulate quantitative and qualitative results by employing a multi-scale, multi-physics modeling framework with efficient computational implementation. It will help build up and predict the new generations of the two-dimensional material-based device industry and a plethora of systems and architectures in the coming decade of quantum computing.

\subsection*{Multilayer Stacking Effect on Transport}

\begin{figure}[!htbp]
\centering
\begin{subfigure}{0.32\textwidth}\centering \includegraphics[width=\textwidth]{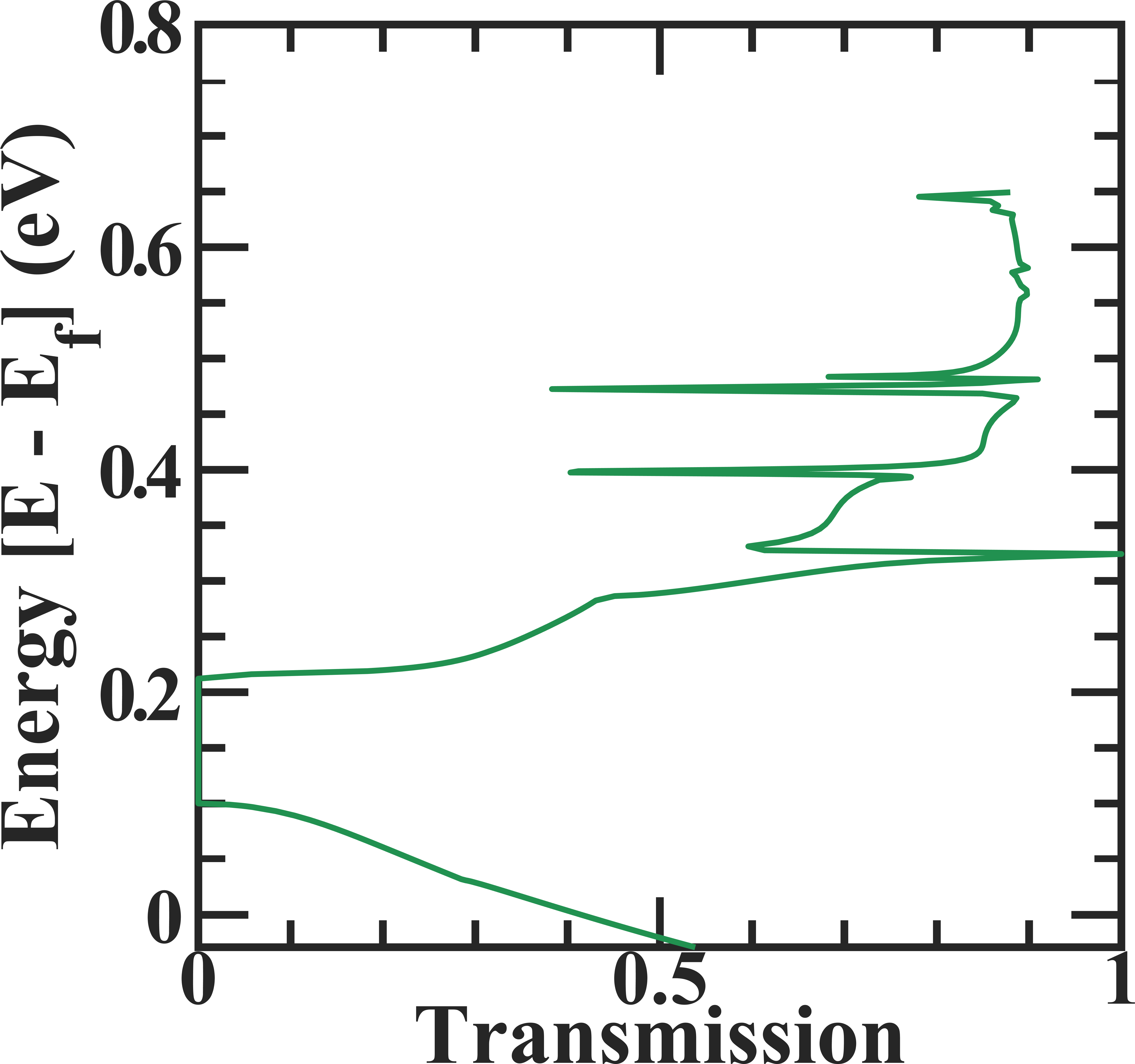}\subcaption{Transmission profile }\label{fig-MUL-1-h}\end{subfigure}
\begin{subfigure}{0.32\textwidth}\centering  \includegraphics[width=\textwidth]{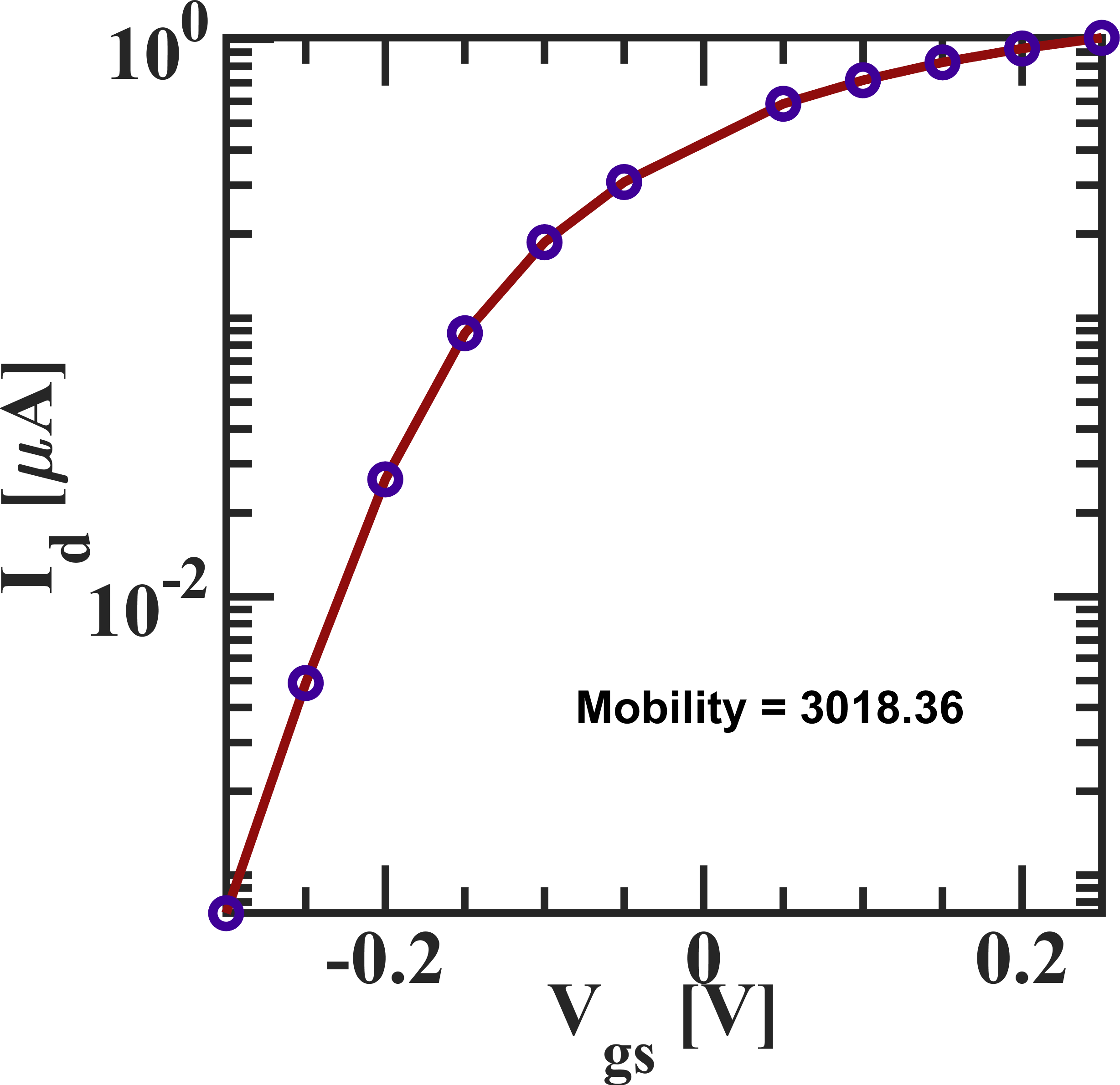}\subcaption{Current-voltage characteristics}\label{fig-MUL-1-g}\end{subfigure}
\begin{subfigure}{0.32\textwidth}\centering \includegraphics[width=\textwidth]{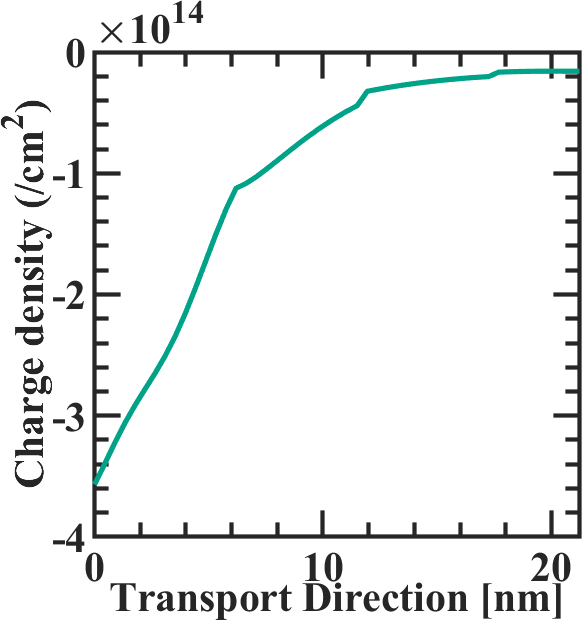}\subcaption{Charge variation redistribution at gate-field action} \label{fig-MUL-1-i}\end{subfigure}
\begin{subfigure}{0.32\textwidth}\centering  \includegraphics[width=\textwidth]{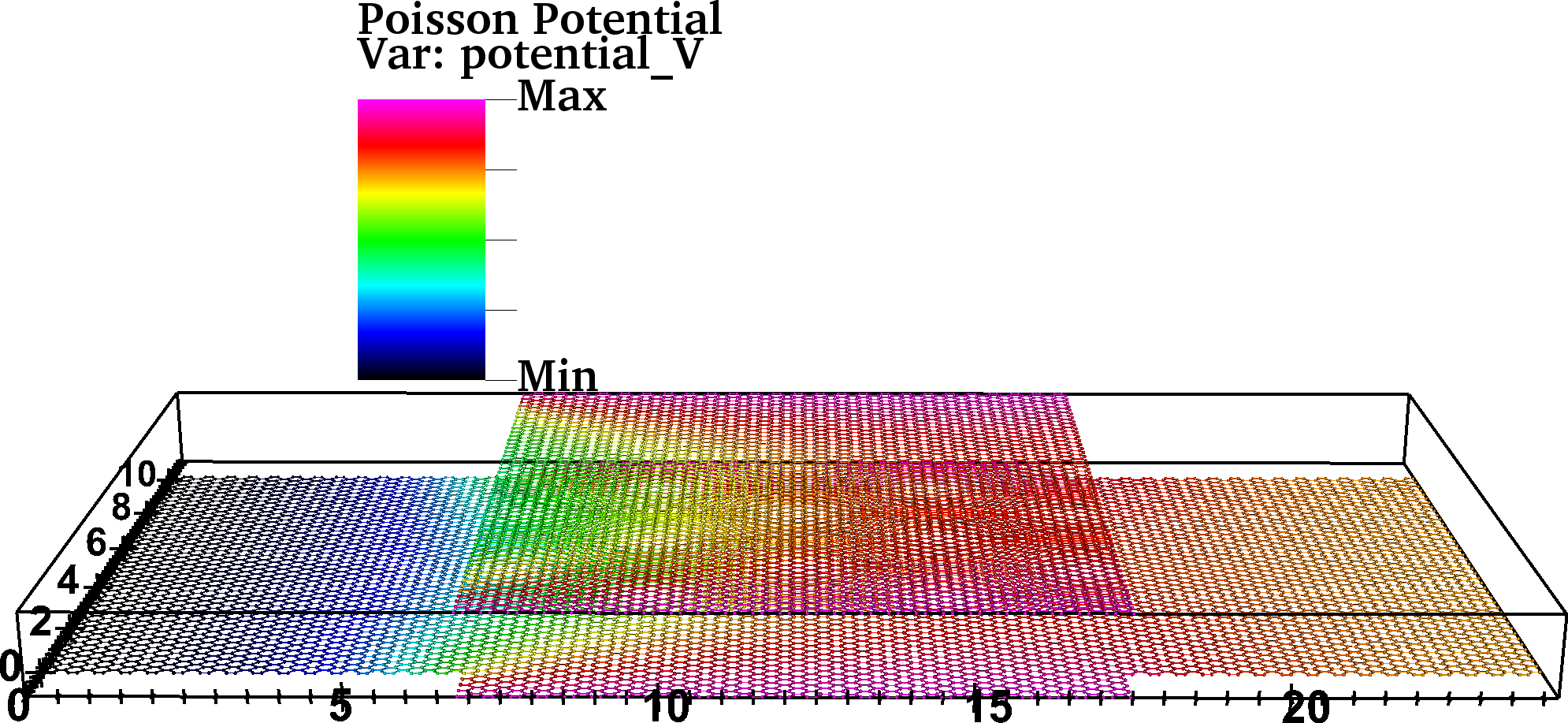}\subcaption{Device Poisson potential profile $\mathrm{V_{GS}}$ 0.25V} \label{fig-MUL-1-a}\end{subfigure}
\begin{subfigure}{0.32\textwidth}\centering  \includegraphics[width=\textwidth]{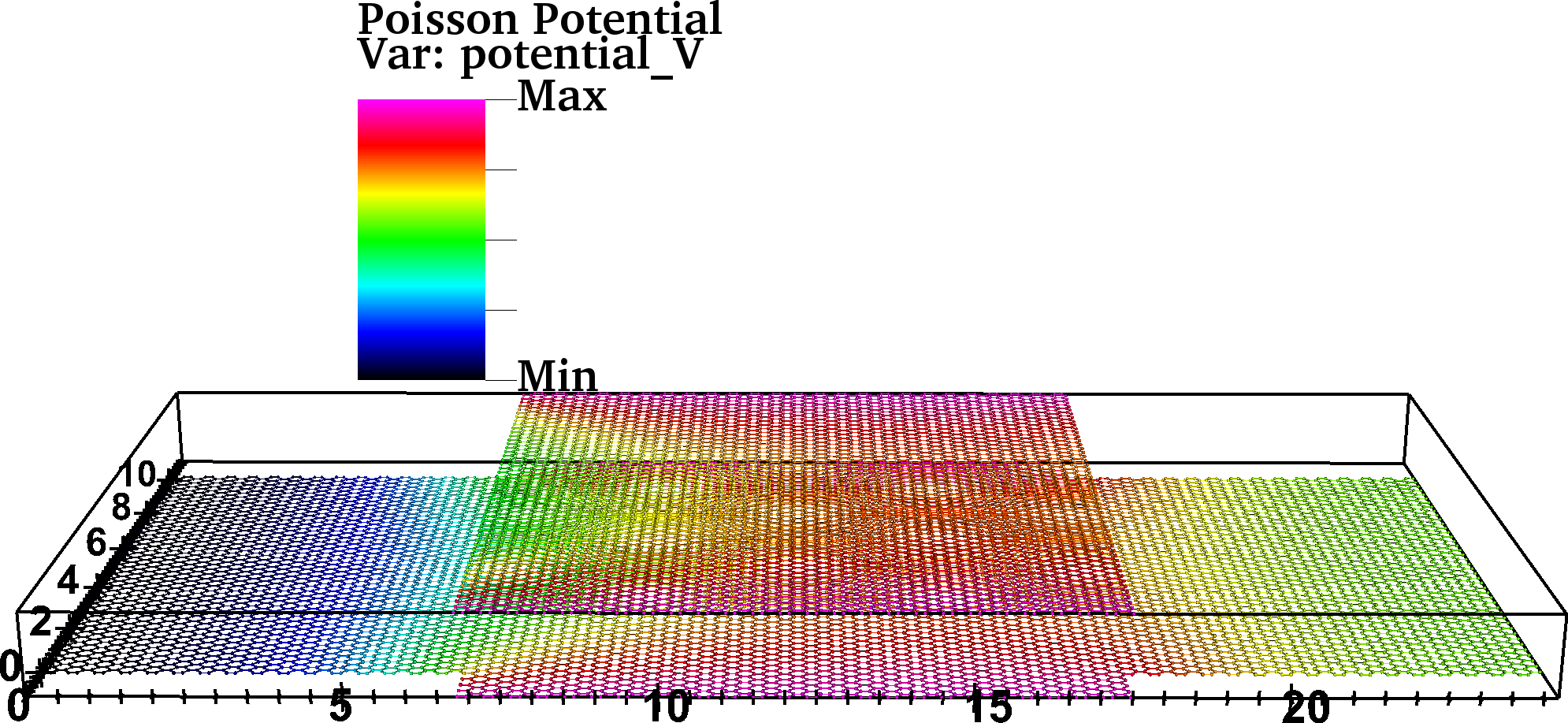}\subcaption{Device Poisson potential profile $\mathrm{V_{GS}}$ 0.05V} \label{fig-MUL-1-c}\end{subfigure}
\begin{subfigure}{0.32\textwidth}\centering  \includegraphics[width=\textwidth]{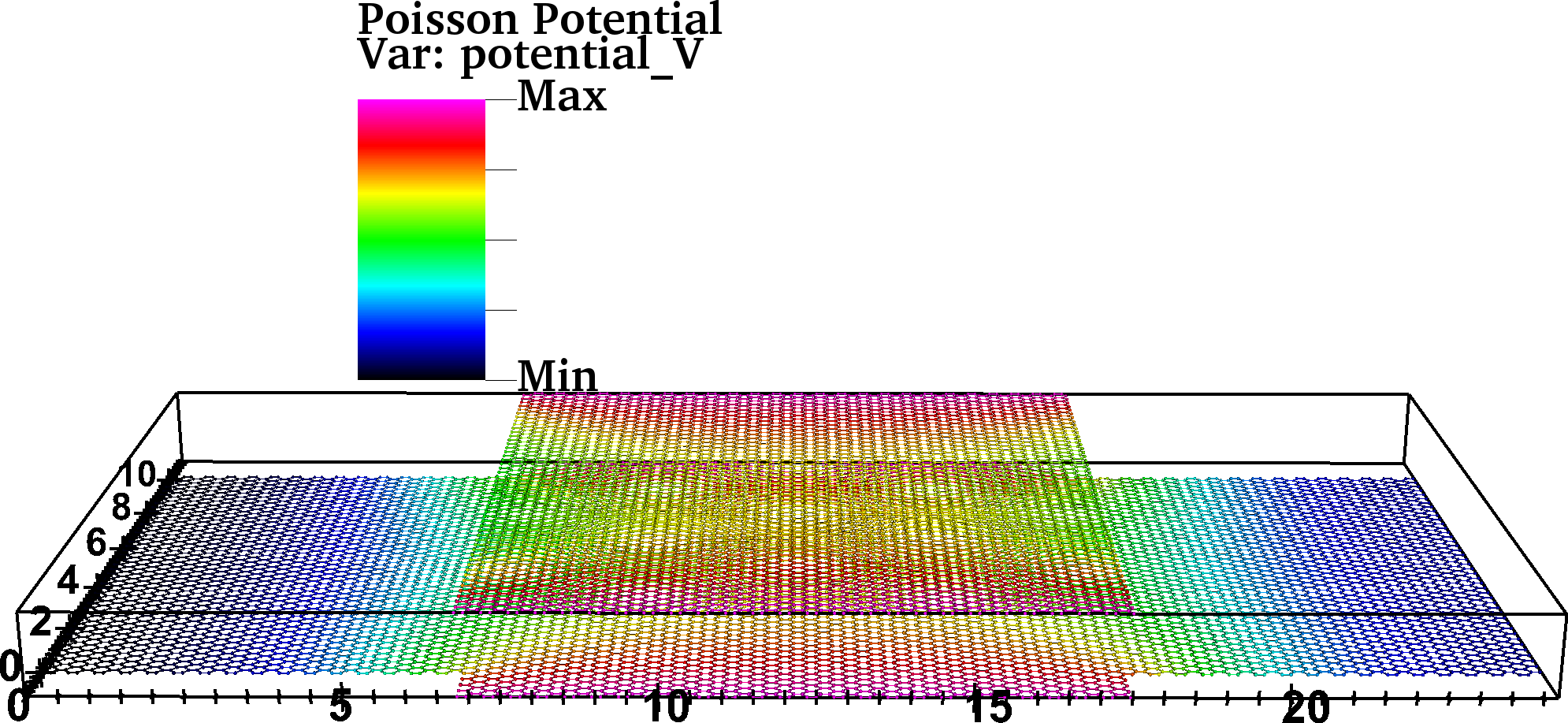}\subcaption{Device Poisson potential profile $\mathrm{V_{GS}}$ -0.25V}\label{fig-MUL-1-e}\end{subfigure}
\begin{subfigure}{0.32\textwidth}\centering  \includegraphics[width=\textwidth]{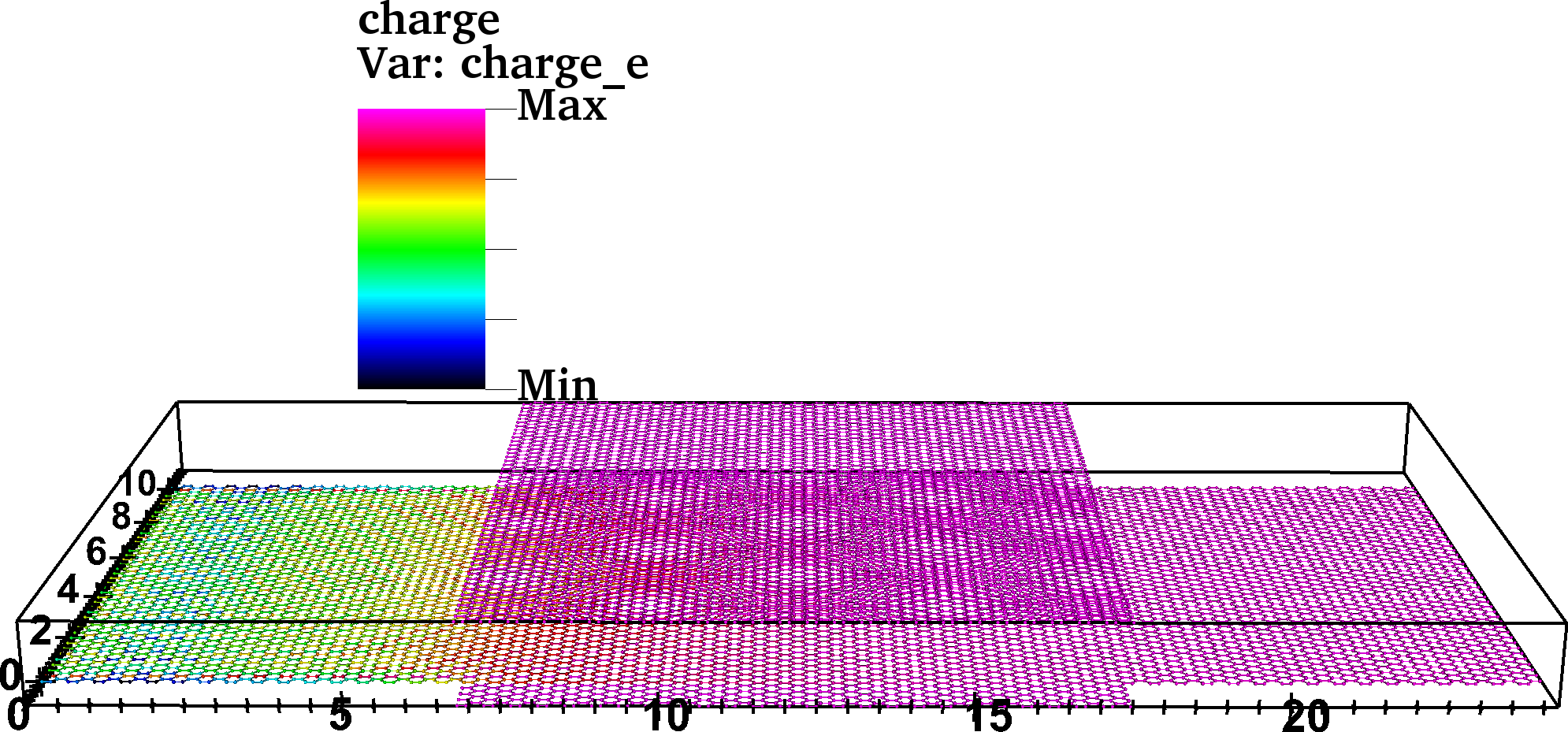}\subcaption{Device free charge profile $\mathrm{V_{GS}}$ 0.25V}\label{fig-MUL-1-b}\end{subfigure}
\begin{subfigure}{0.32\textwidth}\centering  \includegraphics[width=\textwidth]{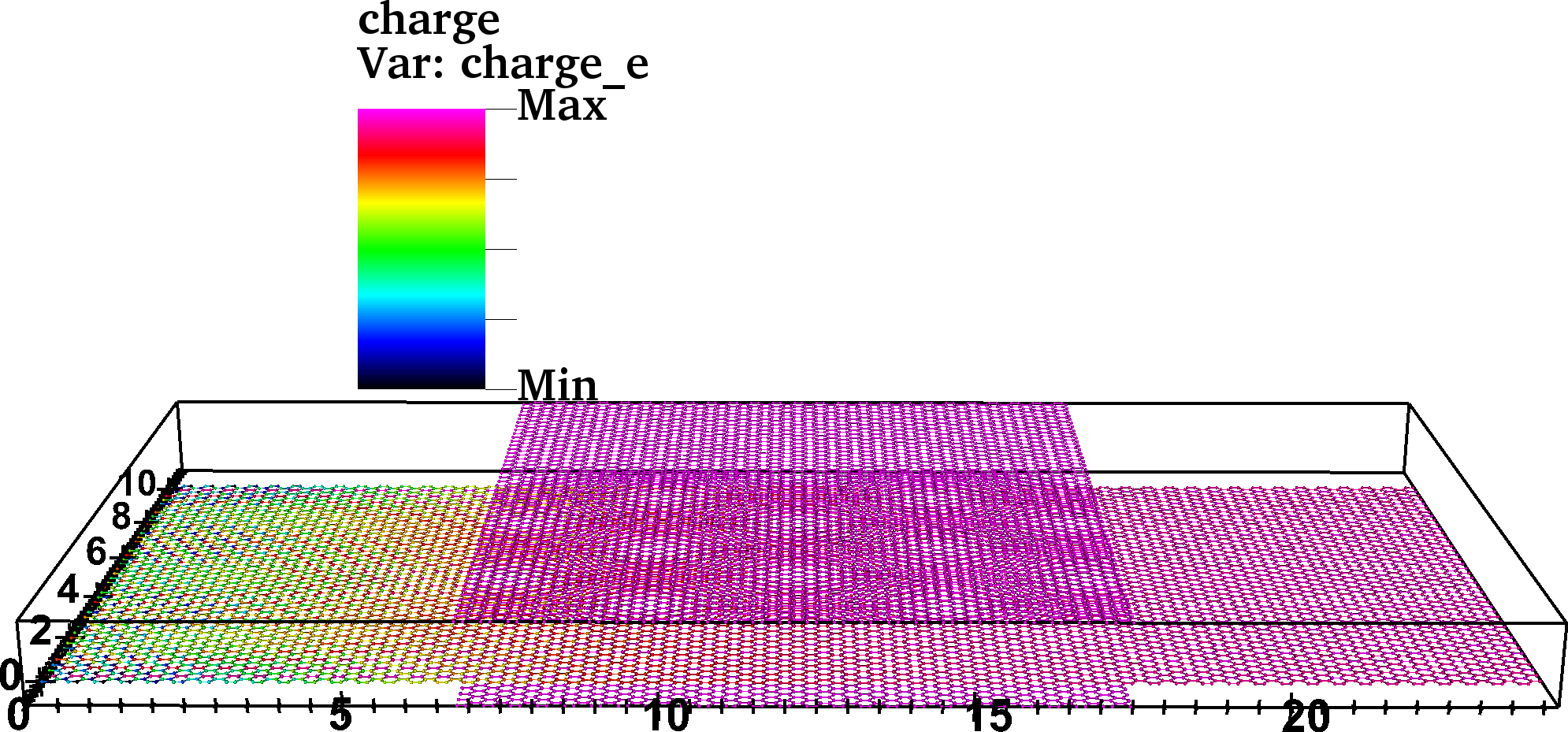}\subcaption{Device free charge profile $\mathrm{V_{GS}}$ 0.05V} \label{fig-MUL-1-d}\end{subfigure}
\begin{subfigure}{0.32\textwidth}\centering  \includegraphics[width=\textwidth]{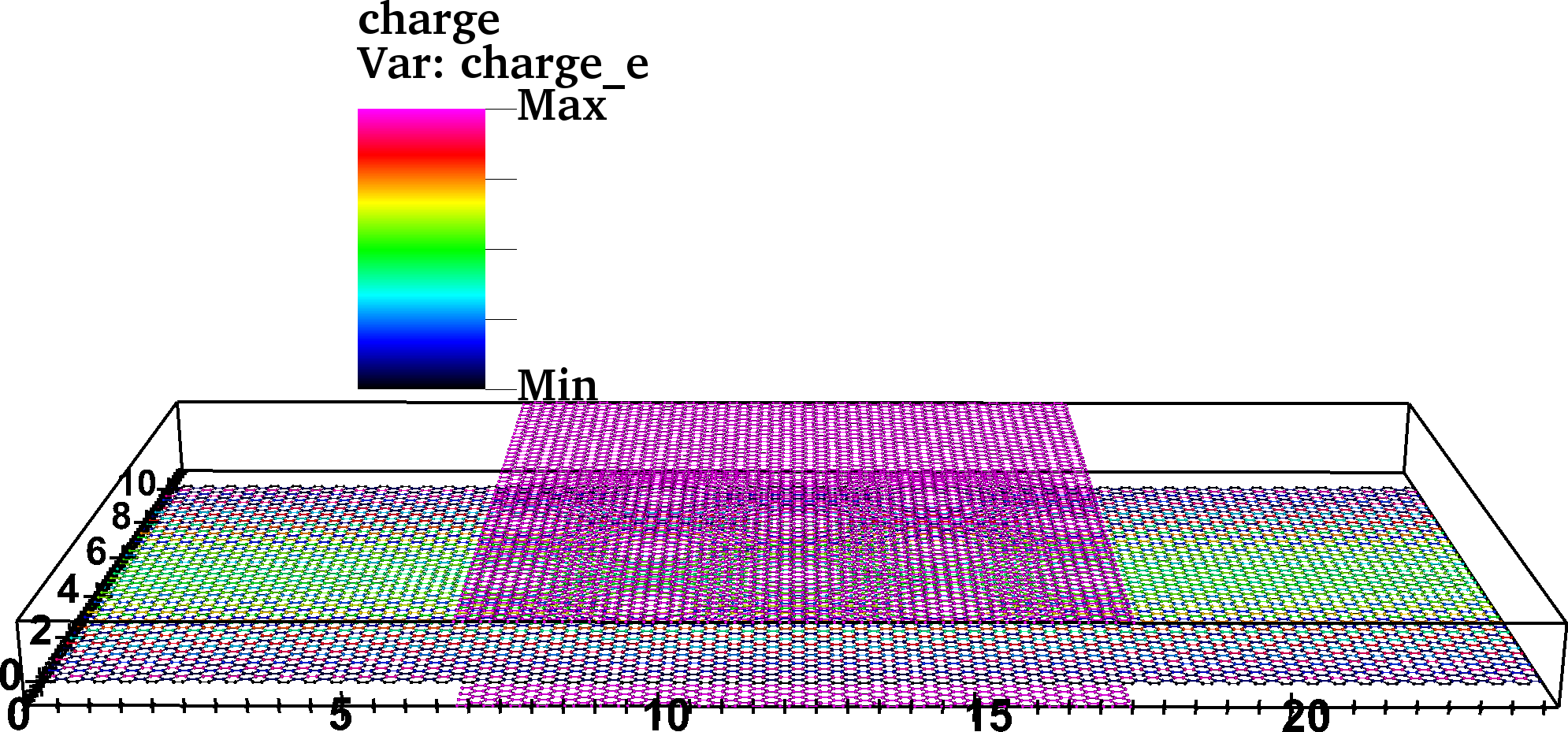}\subcaption{Device free charge profile $\mathrm{V_{GS}}$ -0.25V}\label{fig-MUL-1-f}\end{subfigure}
\caption{Total transmission profile of simulated device in the phase-coherent transport regime \cref{fig-MUL-1-h}, Current-voltage characteristics \cref{fig-MUL-1-g}, Charge variation redistribution \cref{fig-MUL-1-i} in the conducting device due to gate-field action at one bias point condition, Device structure graphene Poisson potential \cref{fig-MUL-1-a}, and free charge \cref{fig-MUL-1-b} profile at 0.25V, Poisson potential \cref{fig-MUL-1-c}, and free charge \cref{fig-MUL-1-d} profile at 0.05V, Poisson potential \cref{fig-MUL-1-e}, and free charge \cref{fig-MUL-1-f} profile at -0.25V for the three bias point}
\label{fig-MUL-1}
\end{figure}

Due to process variation, i.e., copper poly-crystal grains, two graphene layers are in arbitrary orientation stacked together in the active device during the growth step in the CVD roll-to-roll graphene sheet device. We have also investigated how the overlayered graphene influences the underlayer active graphene device transport. The under-layer graphene devices electrically contacts the source and drain contact. At the same time, the overlayer is electrically inactive but induces the image charge electrostatic effect on the underlayer channel by available $\pi$ orbitals. We investigated the induced charges, wave-function spilling/leaking effect, and the potential profile between the graphene layers. The \cref{fig-MUL-1} correspond to a simulated graphene device in the phase-coherent transport regime, and its electrical characteristics with channel length $\mathrm{L_G}$ of 6 nm, source length $\mathrm{L_S}$ and drain length $\mathrm{L_D}$ of 6 nm with drain, with channel doping density of $ 1  \times 10^{14} $ per cm$^2$ observe in the CVD/PECVD roll-to-roll batch sample. The side gate oxide thickness $\mathrm{t_{ox}}$ is 1.3 nm on the each side of device and gate dielectric constant $ \epsilon_{r} $ is 3.9. We simulated the phase-coherent device with a gate bias sweep of -0.3 Volt to +0.25 Volt in step 0.05 Volt with the presence of a transport oxide barrier, and a maximum source-drain bias of 0.2 Volt was applied. In the \cref{fig-MUL-1}, we have plotted \cref{fig-MUL-1-h} the complete transmission profile, and \cref{fig-MUL-1-g} we have plotted the current-voltage characteristics multilayer stacking transport regime. In the \cref{fig-MUL-1-i} charge variation redistribution in the conducting device due to gate-field action at one bias point condition plotted. We have plotted the Poisson potential \cref{fig-MUL-1-a}, and free charge \cref{fig-MUL-1-b} profile at 0.25V, Poisson potential \cref{fig-MUL-1-c}, and free charge \cref{fig-MUL-1-d} profile at 0.05V, Poisson potential \cref{fig-MUL-1-e}, and free charge \cref{fig-MUL-1-f} profile at -0.25V for the three bias point the in the multi-layer device's configuration. 

We have observed that overlayered graphene, passivation layers, or substrate surface significantly influences the electrostatic Poisson potential and charge redistribution in the electrically contacted under-layer active graphene device. Furthermore, It's lateral transport characteristic even though there is no transport path assumed between interlayers in the verticle direction in the test device configuration. A more thorough investigation, incorporating interlayers scattering potential self-energy interaction in the non-equilibrium Green's functions formalism with Dyson's self-energy loop, is required to conclude stacking effect phenomena in the low-dimensional material devices. Furthermore, from a theoretical accounting point of view, for transport calculation in the low-dimensional layered device, we need the density of states of continuum systems and energy scattering rate to find the interlayer states between stacked layers, and a more in-depth description is required. Finally, we conclude the article by summarizing the observations, limitations in the physical model, mathematical validity of the governing equation in the present work, and future advancement proposal for the incoherent transport regime in low-dimensional materials. 

\section*{Summary}
\label{section:Summary}

An incoherent electronics transport regime at room temperature operating low-dimensional materials device is a formidable task. It incorporates multiple physically rich self-energy interaction phenomena while the system simultaneously is at a non-equilibrium state, exchanging carrier and heat with an external reservoir bath. Moreover, the material's low dimensionality and surface nature will further increase complexity in physics. Furthermore, employing a multi-scale framework by binding the atomistic tight-binding orbital wave-function bottom-up to experimentally measurand Drude mobility values makes it a full-scale end-to-end herculean task to compute. In addition to the aforementioned self-energy scattering potential, points like neutral impurity interaction center and grain boundaries will induce line defects in the roll-to-roll transfer-produced graphene devices. A physically transparent, suitable treatment should be incorporated in the future extension of the framework. Especially the line defects in a typically contaminated sample will increase the backscattering events as in single-crystalline ideal graphene backscattering of Dirac fermion suppressed due to unique lattice and electronic band structure. Nevertheless, at the nanoscale level, as the crystal domains' characteristic dimension reaches the length scale of the carrier mean free path $l_{mfp}$, backscattering events for the carriers at the grain boundaries are significantly enhanced to degrade the measurand mobility. Therefore, comparing the performance of the two-dimensional material devices and their reproducibility, scalability, and reliability are crucial inquiries for the industrial application of low-dimensional materials. Correspondingly, instead of using fictitious first-order phonons at the nanoscale's theoretical framework to catch the genesis of the dephasing process in the inelastic acoustic, optical, and polar optical phonon scattering. We have to incorporate and numerically solve the real phonon in the low-dimensional material with the self-heating interaction in the self-consistent Dyson loop, which will again increase the order of magnitude of computational cost at the device level investigation. Also, The trap states, vacancies, passivating atoms, screening effect treatment, monolayer bandgap fluctuation by an interface, and interlayer dynamics interplay in real devices and are challenging tasks for multi-scale implementation. Some of these challenging tasks at the device modelling level, like defects, are treated at the Density functional level for the small defective portion of the device. In such an approach, we remove or add some extra atoms and relax the structure to calculate the perturbation hamiltonian. We later include it in the total device hamiltonian and use a reliable statistical model to extract a scattering potential to use it in the non-equilibrium Green's function formalism for transport calculations to observe the influence of defects on electrical transport properties. The numerical correctness of such accounting is heavily tilted on the accuracy of underline Density functional theory relaxation calculation. However, The numerical efficiency of transfer of parameterization in the intermediate stage, e.g., tight-binding or Wannier functions to the perturbing potential and the statistical model of scattering potential for the transport calculation and each of these steps may introduce a systematic bias and error in calculation. Moreover, The lattice relaxation process is usually long-range, and extracting scattering potentials through the statistical model are nearest or next-nearest-neighbor couplings limited. Especially in the stacked-layered low-dimensional material systems, defect-induced lattice mismatch and relaxation process will build up a long-range strain field, and the scattering potential will alter in such a scenario. Another approach to treat the corrugated system is at the scattering self-energy level, which is numerical inefficient and challenging. The energy range of fluctuations is in order of ten to a hundred micro electron volts within the layer. The correlation length is an essential factor; for longer, the correlation length is slower is this pile-up of the fluctuation in the hamiltonian from one atom to the next.
Moreover, it depends upon the underline substrate and surrounding encapsulation layer. In most published works, nonlocality ignores without sufficient justification. However, most scattering mechanisms, even roughness scattering, are non-local to a reasonable extent. Nevertheless, the most famous Recursive Green's Function (RGF) algorithm numerical restricts up to the tri-diagonal block until recently. Therefore, the non-local treatment beyond the nearest neighbor atom truncate at the algorithm level in most calculations, and efficient algorithms are required to calculate Green's function. Though the bending and twisted fringing fields are fragile, as the interlayer coupling is also weak in the van der Waals low-dimensional materials, they are energetically comparable. As a result, systems may corrugate or twist and affect the scattering potential. The fringing-field impact on the electronic hamiltonian must be statistically parameterized in the tight-binding or Wannier functions procedure, especially for the multi-layer corrugated or twisted structure. 

\bibliographystyle{IEEEtran}
\bibliography{Article}

\end{document}